\documentclass[aps,prx,twocolumn,longbibliography]{revtex4-1}

\usepackage{times}
\usepackage{float}
\usepackage{amsmath}
\usepackage{amsthm}
\usepackage{amssymb}
\usepackage{amsbsy}
\usepackage{graphicx}
\usepackage{pstricks}
\usepackage{color}
\usepackage{subfigure} 
\usepackage{enumerate}
\usepackage{multirow} 
\usepackage{ulem}
\usepackage{wasysym}
\usepackage[pagebackref=false,pdfauthor={Meena}]{hyperref}

\begin{document}

\title{Classical and quantum theories of proton
disorder in hexagonal water ice}

\author{Owen Benton}
%
\affiliation{Okinawa Institute of Science and Technology
Graduate University, Onna, Okinawa, 904 0495, Japan}
\affiliation{H.\ H.\ Wills Physics Laboratory,
University of Bristol, Tyndall Av, Bristol BS8--1TL, UK}

\author{Olga Sikora}
%
\affiliation{Marian Smoluchowski Institute of Physics, Jagiellonian University, Prof. {\L}ojasiewicza 11, PL-30348 Krak{\'o}w, Poland}
\affiliation{Department of Physics, National Taiwan University, Taipei 10607, Taiwan}
\affiliation{Okinawa Institute of Science and Technology
Graduate University, Onna, Okinawa, 904 0495, Japan}
\affiliation{H.\ H.\ Wills Physics Laboratory,
University of Bristol, Tyndall Av, Bristol BS8--1TL, UK}

\author{Nic Shannon}
%
\affiliation{Okinawa Institute of Science and Technology
Graduate University, Onna, Okinawa, 904 0495, Japan}
\affiliation{Clarendon Laboratory, University of Oxford, Parks Rd., Oxford OX1 3PU, UK} 
\affiliation{H.\ H.\ Wills Physics Laboratory,
University of Bristol, Tyndall Av, Bristol BS8--1TL, UK}


\begin{abstract}

It has been known since the pioneering work of Bernal, Fowler and
Pauling that common, hexagonal (Ih) water ice is the archetype of a
frustrated material : a proton--bonded network in which protons satisfy
strong local constraints --- the ``ice rules'' --- but do not order.  
While this proton disorder is well established, there is now a growing
body of evidence that quantum effects may also have a role to play in
the physics of ice at low temperatures.  
In this Article we use a combination of numerical and analytic techniques 
to explore the nature of proton correlations in both
classical and quantum models of ice Ih.  
In the case of classical ice Ih, we find that the ice rules have two,
distinct, consequences for scattering experiments --- singular
``pinch points'', reflecting a zero--divergence condition on the uniform
polarization of the crystal, and broad, asymmetric features, coming 
from its staggered polarisation.
In the case of the quantum model, we find that the collective quantum
tunnelling of groups of protons can convert states obeying the ice rules
into a quantum liquid, whose excitations are birefringent, emergent photons.
We make explicit predictions for scattering experiments on both  
classical and quantum ice Ih, and show how the quantum theory 
can explain the ``wings'' of incoherent inelastic scattering 
observed in recent neutron scattering experiments 
[Bove {\it et al.}, Phys. Rev. Lett. {\bf 103}, 165901 (2009)].
These results raise the intriguing possibility that the protons in
ice Ih could form a quantum liquid at low temperatures, in
which protons are not merely disordered, but continually fluctuate 
between different configurations obeying the ice rules.

\end{abstract}

\maketitle

\section{Introduction}
\label{section:introduction}

\begin{figure*}
\centering
\subfigure[]{%
\includegraphics[width=.3\textwidth]{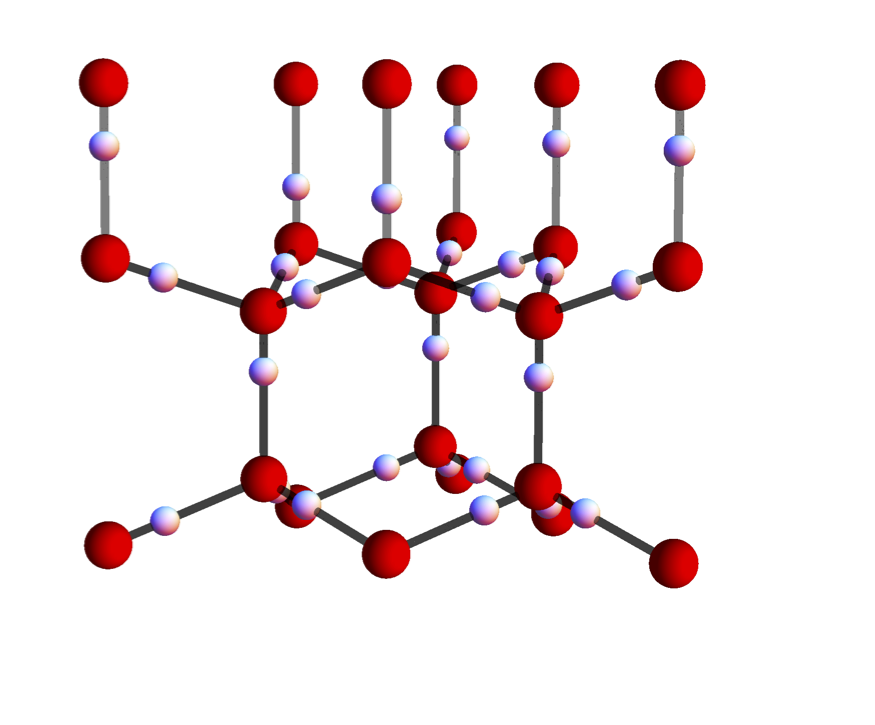}}
    \qquad
\subfigure[]{%
\includegraphics[width=.3\textwidth]{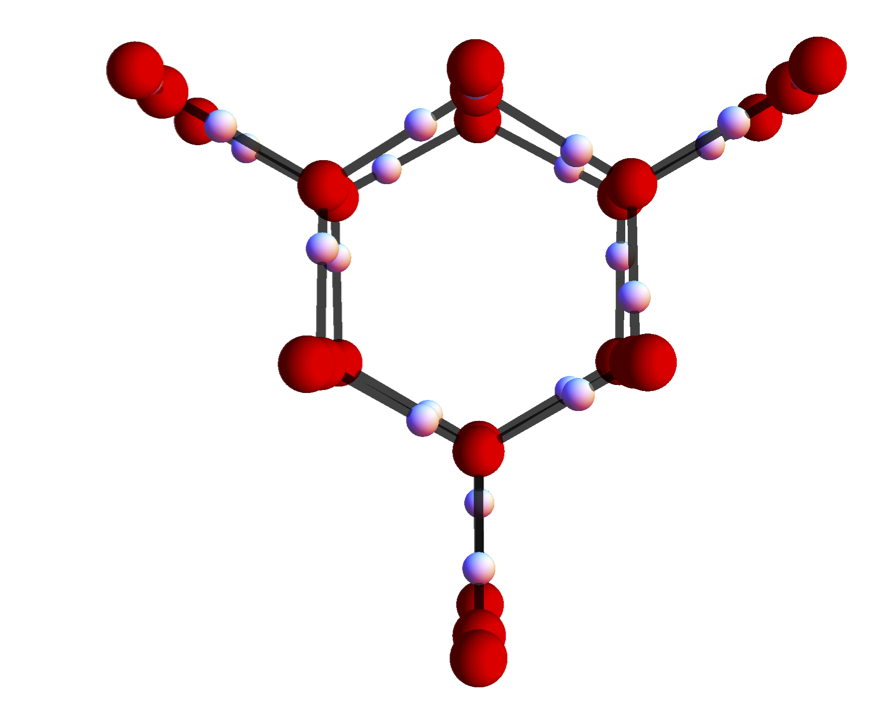}}
\caption{
(Color online).
Crystal structure of hexagonal (Ih) water ice.
Water ice can be viewed as a frozen configuration of water 
molecules, satisfying the Bernal--Fowler {\it ``ice rules''} \cite{bernal33,pauling35}, 
in which each oxygen (red sphere) forms two short, covalent bonds, 
and two long, hydrogen bonds with neighbouring protons (white spheres).   
Oxygen atoms form an ordered lattice, belonging to the hexagonal space 
group $P6_3/mmc$, with a 4--site primitive unit cell.
Protons do not show any long--range order.
(a) Structure viewed perpendicular to the hexagonal symmetry axis
(the crystallographic $c$-axis). 
(b) Structure viewed along the hexagonal symmetry axis.
}
\label{fig:1h_structure}
\end{figure*}

We learn as children 
that matter can exist in three different phases --- 
solid, liquid, and gas.
This concept is usually introduced through the example of water,  
familiar as a liquid (water), a gas (steam) and a solid (ice).
However, at least as far as its solid phase is concerned, water is a spectacularly 
unusual material.
At atmospheric pressure, water molecules freeze into a structure known 
as ``ice Ih'', illustrated in Fig.~\ref{fig:1h_structure}.
Ice Ih is remarkable in that the oxygen ions (O$^{2-}$) form an ordered lattice, 
while the protons (H$^+$) lack any kind of long--range order --- in flat contradiction with 
the usual paradigm for solids.


\begin{figure*}
\centering
\subfigure[\ Tunnelling on a plaquette within hexagonal plane.]{%
\includegraphics[width=.66\textwidth]{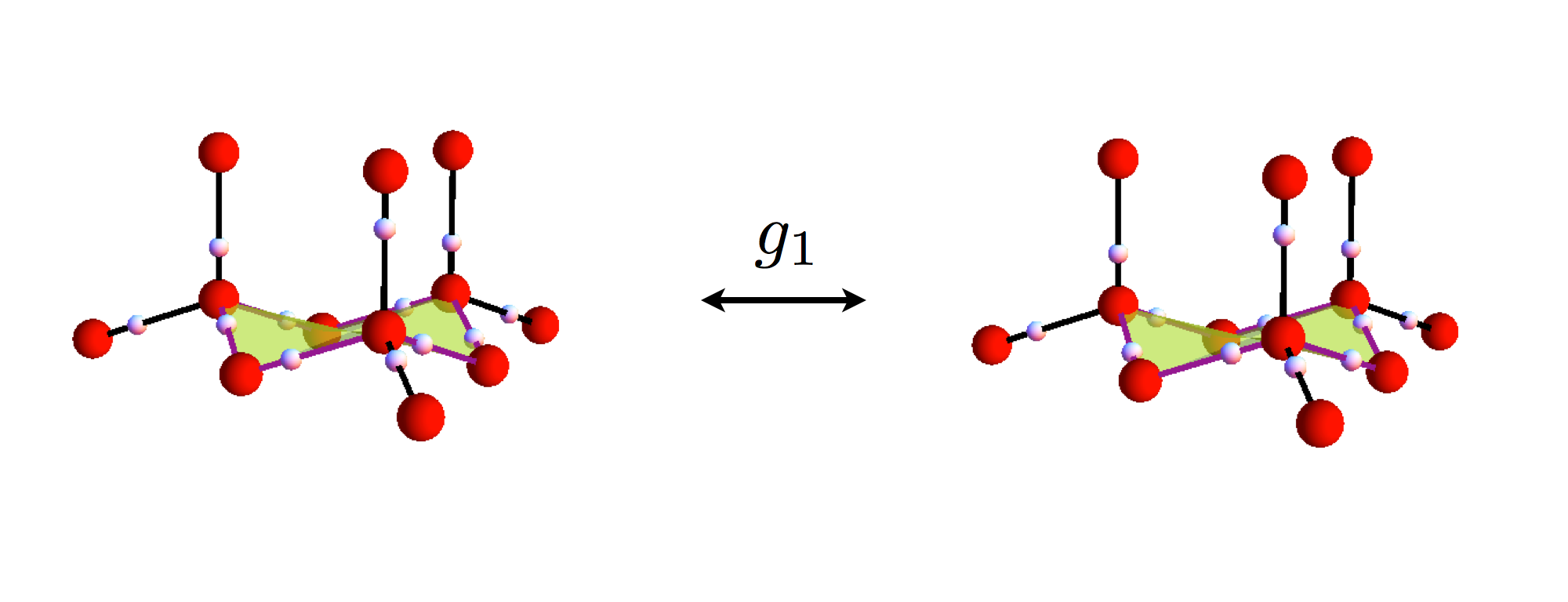}}%
    \\
\subfigure[\ Tunnelling on a plaquette connecting hexagonal planes.]{%
\includegraphics[width=.66\textwidth]{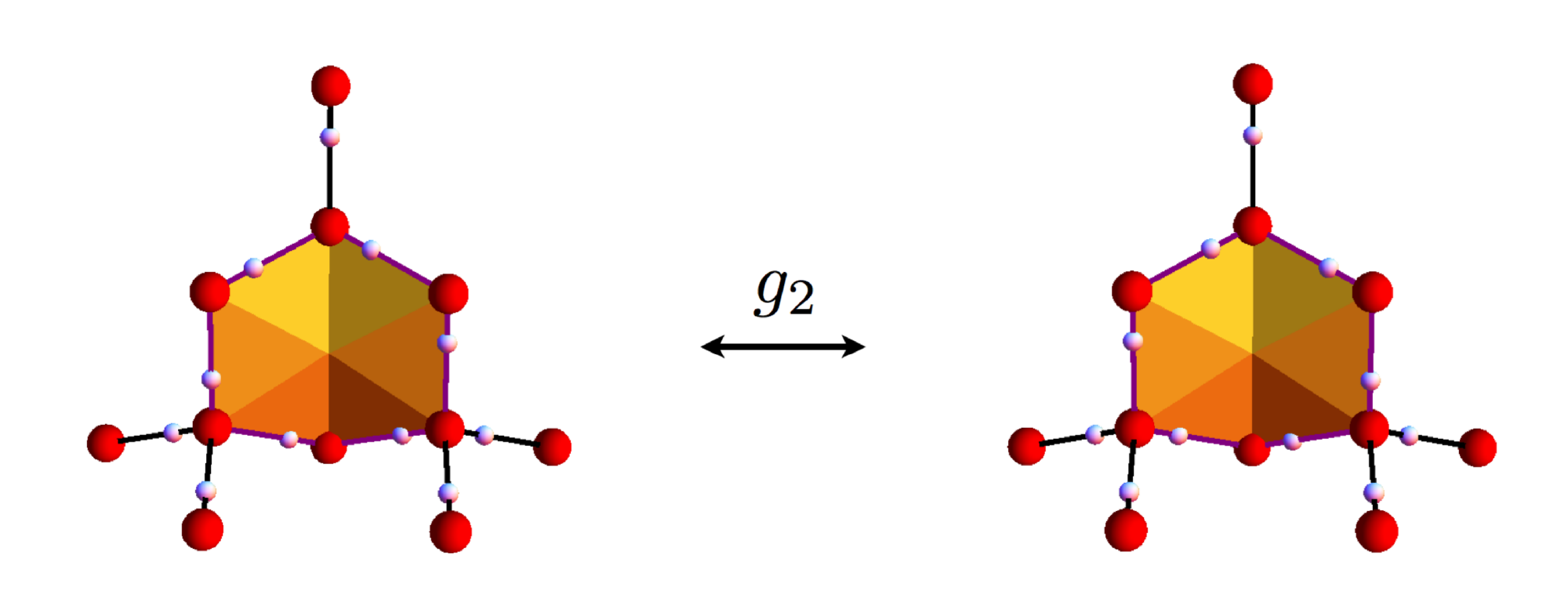}}%
\caption{
(Color online).
Collective quantum tunnelling between different proton configurations 
satisfying the ice rules in hexagonal (Ih) water ice.
The crystal structure of ice Ih contains two distinct types of hexagonal 
plaquette, each containing six oxygen atoms (red spheres). 
Tunnelling between different ice configurations is mediated the by   
correlated hopping of protons (white spheres) around such a plaquette.   
Where protons form an alternating sequence of short (S) and long (L) bonds 
(e.g. S--L--S--L\ldots --S--L), 
it is possible to tunnel to another, 
degenerate, ice configuration in which long and short bonds are interchanged, 
(i.e. L--S--L--S\ldots --L--S) 
\cite{bove09,ihm96, drechsel-grau14-PRL}.   
(a) Tunnnelling on hexagonal plaquette of ``type I'', in the plane perpendicular to the 
hexagonal symmetry axis.
This process has a matrix element $g_1$ in our model for proton dynamics, 
$\mathcal{H}_{\sf tunnelling}^{\sf hexagonal}$~[Eq.~(\ref{eq:htunnelling2})].
(b) Tunnnelling on hexagonal plaquette of ``type II', connecting different hexagonal planes. 
This process has a matrix element $g_2$.
}
\label{fig:loops}
\end{figure*}


This extraordinary property of water ice was first elucidated more than 80 years ago, 
by Bernal and Fowler~\cite{bernal33}.
Bernal and Fowler argued that ice should be viewed as a molecular solid, 
in which distinct water molecules are bound together by hydrogen bonds.
Each water molecule forms four such hydrogen bonds, and 
as a result, the proton configurations obey strong local constraints, 
commonly referred to as the ``{\it ice rules}'' \cite{bernal33,pauling35}.
The ice rules lead to strong correlations between protons, but can be satisified by 
an exponentially large number of different proton configurations~\cite{pauling35, nagle66}.
As a result, the protons remain disordered, and possess an extensive
residual entropy.
This ``ice entropy'' is observed in experiments on Ih 
water ice~\cite{giauque36}, and persists down to the lowest 
temperatures measured, in apparent defiance of the laws of thermodynamics.
Eighty years on, these striking discoveries continue to exert a profound 
influence on research into water ice~\cite{petrenko99}, 
and a wide range of other 
materials~\cite{harris97, bramwell01, ramirez99, klemke11, 
castelnovo12, wan12, anderson56, fulde02, youngblood80, 
chern14, meng15, algara-siller15}.  


Recent experiments by Bove~{\it et al.}~[\onlinecite{bove09}] suggest a new twist on
the behaviour of protons in ice Ih --- not only are protons disordered, but they 
remain mobile, even at temperatures as low as $5\, \text{K}$.  
This might seem surprising, since any attempt to move a proton will lead 
to a violation of the ice rules, at considerable cost in energy~\cite{petrenko99}.
This problem is avoided, however, if the proton dynamics consists of coherent
collective quantum 
tunnelling on hexagonal plaquettes, of the type 
illustrated in Fig.~\ref{fig:loops}.
This mechanism for proton dynamics in ice Ih finds support from {\it ab initio} 
calculations~\cite{ihm96, drechsel-grau14-PRL}, 
with recent results suggesting that, while the high--temperature dynamics proceeds 
via single--proton hopping,  collective motion
around loops becomes important at low temperatures \cite{drechsel-grau14-PRL}.
And an analogous correlated tunnelling of protons has recently been observed 
in an artificial assembly of four water molecules~\cite{meng15}.


There are, in fact, many classes of system whose low--temperature physics is subject 
to strong local constraints, similar to those found in water ice.
The most celebrated of these are the magnetic systems known as the 
``{\it spin ices}'', whose low--temperature spin configurations are in 
correspondence with the proton configurations in water 
ice~\cite{harris97, bramwell01, castelnovo12}.
Ice--like physics also arises in models of frustrated charge order \cite{anderson56, fulde02}, 
proton--bonded (anti-)ferroelectrics \cite{slater41, allen74, 
youngblood78, youngblood80, samara87, 
morimoto91, chern14}, 
dense polymer melts \cite{kondev98} 
and dimer models \cite{roberts35,henley10}.
In these systems, violations of the ice rules take
on the character of fractionalised charges~\cite{fulde02}.
This point has attracted particular attention in
the case of the spin ices, since these charges 
behave as effective magnetic monopoles 
\cite{ryzhkin05,castelnovo08, jaubert09,morris09}.
And, in fact, a direct analogy may be drawn between these 
magnetic monopoles and ionic defects in water ice
\cite{ryzhkin05, ryzhkin97, ryzhkin99}.


The effect of quantum tunnelling, of the type proposed by Bove~{\it et al.}~[\onlinecite{bove09}], 
has been studied in a range of other ice--like problems, with striking results.
Quantum tunnelling has been shown to give rise to a quantum ``spin--liquid'',  
comprising a coherent superposition of an exponentially large number of states 
obeying the ice rules, in models derived from 
spin ice~\cite{hermele04, banerjee08, shannon12, benton12,kato-arXiv,mcclarty-arXiv,gingras14}.   
Equivalent quantum liquids have also been found in three--dimensional quantum 
dimer models~\cite{moessner03, sikora09, sikora11}.
In both cases, the low--energy excitations of the liquid are gapped, fractionalised, 
``charges'' and gapless, linearly dispersing, ``photons'', in direct correspondence
with the theory of electromagnetism~\cite{moessner03,hermele04}.


A similar ``electromagnetic'' scenario has also been discussed in the context 
of a simplified model of water ice by Castro Neto {\it et al.} \cite{castro-neto06}.
Finding that quantum fluctuations drive the protons to order in a 
two--dimensional model of water ice~\cite{chern14, shannon04,syljusen06,henry14}, 
these authors argued, by extension, that quantum effects in a three--dimensional 
water ice could drive a finite--temperature phase transition between a low--temperature 
proton--ordered phase, and a high--temperature proton disordered phase. 
Within this scenario, ordered and disordered phases of protons 
in hexagonal water ice --- ice Ih and ice XI [\onlinecite{petrenko99}] --- 
would correspond to the confined, and deconfined phases of a compact, 
frustrated $U(1)$ lattice--gauge theory~\cite{kogut79,guth80}, similar to 
that proposed in the context of quantum spin ice~\cite{hermele04, benton12}.   


In the work which we present here we do not attempt to establish the
conditions under which the protons in hexagonal water ice order, 
but instead seek to characterise their deconfined, disordered phase.
To this end we develop theories which
describe the proton correlations in both 
classical and quantum models of hexagonal (Ih) water ice, making
explicit predictions for scattering experiments.


In the case of classical ice Ih, we find that the ice rules 
have two, distinct, consequences for proton correlations, directly 
visible in scattering experiments.
Firstly, algebraic correlations of the uniform polarisation lead to 
``pinch points'' --- singular features in scattering --- visible in 
a subset of Brillouin zones.
Secondly, exponential correlations of the staggered polarisation
lead to broad, assymetric features in 
a different subset 
of Brillouin zones.
This analysis provides new insight into diffuse scattering experiments
on ice Ih~\cite{li94, nield95}, and makes explicit the differences 
between ice Ih and spin ice, or (cubic) ice Ic.
It also provides the starting point needed to construct a theory
of ice Ih in the presence of quantum tunnelling.


In the case of quantum ice Ih, we find that it is possible to describe 
the proton configurations in terms of a lattice--gauge theory in which the 
connection with electromagnetism, long--implied by the ice rules, is made explicit.
The deconfined, proton--liquid phase of this theory is shown to support two 
types of excitation --- gapless, linearly--dispersing photons with birefringent character, 
and weakly--dispersing optical modes, corresponding to local fluctuations of the 
electric polarisation.
The predictions of this lattice gauge theory are shown to be in quantitative agreement 
with the results of variational quantum Monte Carlo simulations.


Throughout our analysis we will emphasise the ways in which proton--correlations in 
hexagonal water ice differ from spin--correlations in cubic spin ice, paying particular
attention to the new structure which arises in both the classical continuum field theory,
and the quantum lattice--gauge theory.


The remainder of the Article is structured as follows :


In Section \ref{section:definitions} we introduce a model of ice Ih
which includes tunnelling between different proton configurations obeying 
the ``ice rules'', and describe those aspects of the symmetry 
and geometry of the ice Ih lattice which are important for our discussion.

\begin{figure*}
\centering
\subfigure[]{%
\includegraphics[width=.2\textwidth]{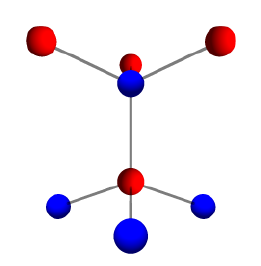}}%
    \qquad
\subfigure[]{%
\includegraphics[width=.2\textwidth]{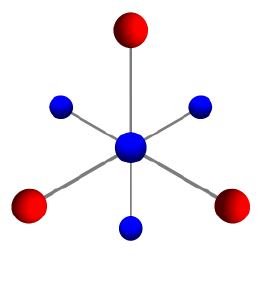}}%
\subfigure[]{%
 \includegraphics[width=.2\textwidth]{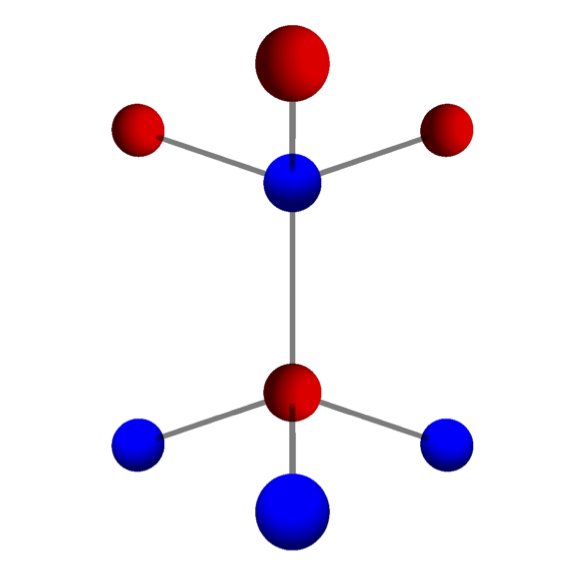}}%
    \qquad
\subfigure[]{%
\includegraphics[width=.2\textwidth]{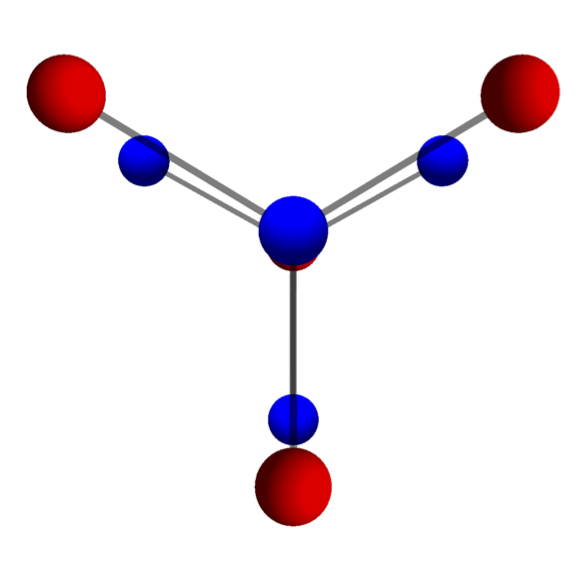}}%
\caption{
(Color online).
Different types of bond within the ordered oxygen lattice of hexagonal (Ih) water ice.
The Ih lattice can be viewed as a stack of buckled, honeycomb lattices, composed
of bonds which are symmetric under inversion about the centre of the bond.  
These honeycomb layers are connected by bonds running parallel to the 
hexagonal symmetry axis, which are symmetric under reflection in the plane 
perpendicular to the bond.
This lattice is bipartite --- i.e. it may be divided into two
sublattices --- here colored red and blue.
(a) Centre--symmetric bond, viewed from a direction perpendicular to the bond.
(b) Centre--symmetric bond, viewed along the bond direction.
(c) Mirror--symmetric bond, viewed from a direction perpendicular to the bond.
(d) Mirror--symmetric bond, viewed along the bond direction.
}
\label{fig:bondsymmetries}
\end{figure*} 


In Section \ref{section:classical} we develop a coarse--grained, classical field theory 
describing the correlations of protons in ice Ih, in the absence of quantum tunnelling, 
and compare this with the results of an equivalent, lattice--based calculation, developed in 
Appendix~\ref{appendix:projection}. 
The two approaches are shown to agree in the long-wavelength limit.
The lattice--based calculation is also used to make explicit predictions for 
scattering experiments on a ice Ih.
These are summarised in Fig.~\ref{fig:Sq.proton.classical}.


In Section~\ref{section:photons}, we develop a quantum $U(1)$ lattice gauge theory 
describing the correlations of protons in ice Ih, in the presence of quantum tunnelling. 
We construct the excitations of this lattice gauge theory, which include 
birefringent emergent photons, and make quantitative predictions for their signature 
in inelastic scattering experiments.
Key results are summarised in Fig.~\ref{fig:photon-dispersion} and Fig.~\ref{fig:XRDSq-quantum}.


In Section \ref{section:applications} we discuss our results in the
context of published experiments on Ih water ice. 
We consider in particular the long--wavelength features seen in diffuse neutron 
scattering, the incoherent inelastic neutron--scattering experiments of 
Bove~{\it et al.}~[\onlinecite{bove09}], and the thermodynamic properties of ice
at low temperatures.


We conclude in Section \ref{section:discussion}, with a summary of our results and a 
brief discussion of open questions.


Technical  aspects of this work, including comparison with classical 
and quantum Monte Carlo simulations, are developed in a series of Appendices.


Appendix~\ref{appendix:lattice} provides details of the ice Ih
lattice and of the coordinate system used to describe it.


Appendix~\ref{appendix:continuum.field.theory} provides details of the 
derivation of the classical, continuum field--theory introduced in Section~\ref{section:classical}.


Appendix~\ref{appendix:projection} provides details of the microscopic 
calculation of the proton--proton correlations in a classical ice Ih, 
described in Section~\ref{section:classical}.


Appendix~\ref{appendix:structure.factors.and.fields} provides
explicit relationships between the measurable proton correlations and
the correlations of the fields appearing in our continuum theory.

Appendix~\ref{appendix:photons} provides details of the quantum
$U(1)$ lattice--gauge theory introduced in Section~\ref{section:photons}.   


Appendix~\ref{appendix:numerics} provides a comparison 
between the analytic predictions for proton correlations 
described in Section~\ref{section:classical} and Section~\ref{section:photons}, 
with Monte Carlo simulation of both classical and quantum ice Ih.


Appendix~\ref{appendix:incoherent} provides the derivation of the
structure factor for inelastic, incoherent neutron scattering as measured by Bove {\it et al.}
in [\onlinecite{bove09}].

\section{Classical and quantum models of hexagonal (Ih) water ice}
\label{section:definitions}

The key to understanding the structure of water ice is the realization,
due to Bernal and Fowler, that water molecules retain their integrity on freezing~\cite{bernal33}.
It follows that each oxygen remains covalently bonded to two protons, while at the same 
time forming two, weaker, hydrogen bonds with protons on neighbouring water molecules.  
In the frozen state, the oxygen atoms form an ordered lattice, 
held together by intermediate protons, each of which forms 
one long (hydrogen) and one short (covalent) bond with a 
neighbouring oxygen.
This type of bonding favours a tetrahedral coordination of oxygen atoms, 
but does not select any one structure, with 17 different forms of ice crystal 
known to exist~\cite{petrenko99}.


Hexagonal (Ih) water ice, illustrated in Fig.~\ref{fig:1h_structure}, is the 
most common form of water ice, formed at ambient pressure.
In it, 
oxygen atoms form a crystal with the hexagonal space 
group $P6_3/mmc$.
This structure can be thought of as a set of buckled honeycomb lattices, 
composed of centre--symmetric bonds of the type shown in 
Fig.~\ref{fig:bondsymmetries}(a) and (b).
These honeycomb layers are linked by mirror--symmetric bonds, parallel
to the hexagonal symmetry axis, of the the type shown in 
Fig.~\ref{fig:bondsymmetries}(c) and (d).
The two types of bond have almost exactly the same length, leading to
a near--perfect tetrahedral coordination of oxygen atoms  
\cite{petrenko99, kuhs81, kuhs83, kuhs86, kuhs87}.
The primitive unit cell of ice Ih contains 4 oxygen atoms or, 
equivalently, 4 water molecules, with 8 associated protons.
This should be contrasted with the 2--site primitive unit cell needed 
to describe the diamond lattice of oxygen atoms in cubic (Ic) 
water ice (space group $Fd\overline{3}m$), or its magnetic analogue, 
spin--ice~\cite{henley05}.


While the oxygen atoms in ice Ih form an ordered lattice, protons
do not.
The way in which water molecules bond together does not select any one 
proton configuration~\cite{bernal33}, but rather an exponentially large set of 
\begin{eqnarray}
\Omega \sim \left(\frac{3}{2}\right)^N
\end{eqnarray}
proton configurations, where $N$ is the number of oxygen atoms 
(equivalently, water molecules) in the lattice~\cite{pauling35,nagle66}.  
As a consequence, the protons do not show any long--range order.
While unusual, extensive degeneracies of this type are by no means unique to 
 water ice, occurring in ``spin ice''~\cite{bramwell01, castelnovo12},  
problems of frustrated charge order
on the pyrochlore lattice~\cite{anderson56,fulde02}, and in a wide range 
of problems involving the hard--core dimer--coverings of 
two-- or three--dimensional lattices~\cite{roberts35, henley10}. 


The principles governing the arrangement of protons in water ice 
are neatly summarised in the Bernal--Fowler ``{\it ice rules}'' 
\cite{bernal33, pauling35, petrenko99}~:
\begin{enumerate}
\item{Each bond between oxygen atoms contains exactly one proton.}
\item{Each oxygen has exactly two protons adjacent to it.}
\end{enumerate}
Since the ice rules
define one of the simplest models with an extensive ground--state entropy, they have proved 
a rich source of inspiration for statistical studies~\cite{slater41, villain72,nagle66}, particularly 
in two dimensions, where the corresponding ``six--vertex model'' can be 
solved exactly~\cite{lieb67a, lieb67b, lieb67c, lieb67d, sutherland67, baxter71,baxter82}.


Violations of the ice rules cost finite energy, and fall into two types.  
Violations of the first ice rule, double--loaded or empty bonds, are known 
as Bjerrum defects~\cite{bjerrum52}.  
Violations of the second rule occur where a proton is transferred from
one water moleule to another, creating a pair of ``ionic'' defects, 
hydroxil (OH$^{-}$) and hydronium (H$_3${O}$^+$)~\cite{bernal33,ryzhkin97,petrenko99}.   
These ionic defects are in direct correspondence with the fractional charges found in 
models of frustrated charge order on the pyrochlore lattice~\cite{fulde02}, and the magnetic 
monopoles observed in spin ice~\cite{ryzhkin05,castelnovo08,morris09,jaubert09}.


\begin{figure*}
\centering
\includegraphics[width=.9\textwidth]{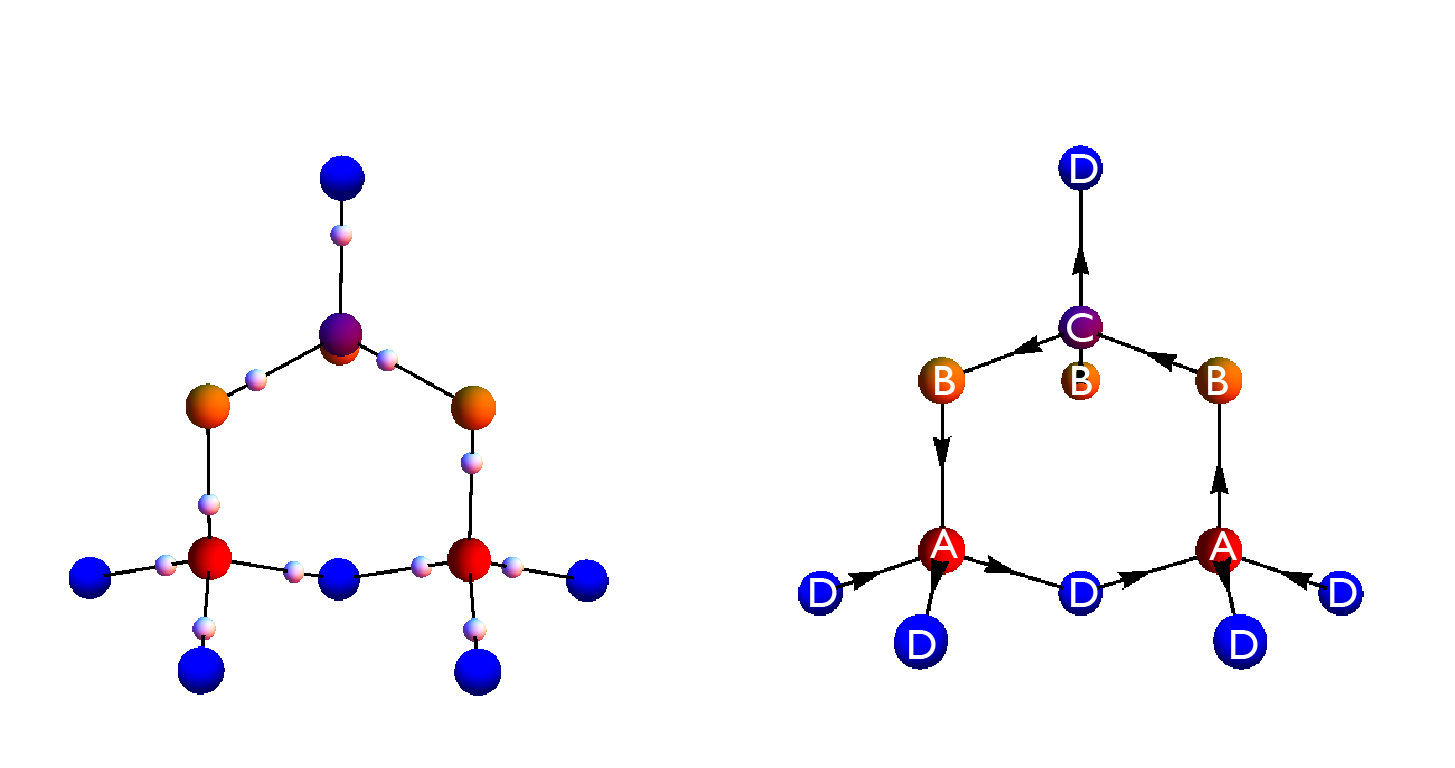}%
\caption{
(Color online).
Flux representation of proton configurations in hexagonal (Ih) water ice.
(a)  Proton configuration, including a hexagonal plaquette of Type II (cf.~Fig.~\ref{fig:loops}). 
(b)  Equivalent proton configuration represented using arrows.
The displacement of protons (white spheres) from the midpoint of each 
oxygen--oxygen bond can be mapped to an arrow on that bond.
The ice rules require that two ``in'' arrows and two ``out'' arrows meet at each 
vertex of the lattice (oxygen atom).
Where arrows form a closed loop, it is possible to tunnel between different proton
configurations satisfying the ice rules, by reversing the sense of all arrows on that loop.
The letters $A, B, C, D$, and colour coding, indicate the convention for labelling  
oxygen sublattices adopted in Section~\ref{section:classical}.
}
\label{fig:fluxrep}
\end{figure*}


While the proton configurations which satisfy the ice rules do not exhibit 
long--range order, they do possess a definite topological structure.   
This, and the connection between the ice rules
and electromagnetism, is most easily understood if proton configurations 
are viewed in terms of a conserved flux.


The mapping onto a flux representation starts with the observation that, 
in the absence of Bjerrum defects, all proton configurations can be represented
by a set of arrows on the bonds of the oxygen lattice.
Each arrow points in the direction of the displacement of the 
proton from the midpoint of that bond, as is illustrated in Fig. \ref{fig:fluxrep}.
Since the oxygen lattice is bipartite (i.e. it may be divided into two sublattices), each arrow 
can be thought of as the flux of a vector field ${\bf P}$ from one sublattice to the other.
The second ice rule then amounts to the condition that flux is conserved, i.e. 
there are two incoming arrows and two outgoing arrows at each oxygen vertex.
This in turn can be viewed as a zero--divergence condition on the flux, 
\begin{eqnarray}
\nabla \cdot {\bf P} = 0 \; ,
\label{eqn:zero-divergence}
\end{eqnarray}
true at every vertex of the lattice.


The flux representation of water ice and related systems
is an approach with a long history~\cite{youngblood80}, 
and is particularly useful in two dimensions, where ice states map onto the exactly--soluble 
six--vertex model~\cite{baxter82}.   
The existence of a zero--divergence condition, Eq.~(\ref{eqn:zero-divergence}), 
suggests a natural analogy with electromagnetism, which we will explore further 
in the remainder of this article.  
The flux representation also plays a crucial role in understanding scattering 
experiments~\cite{youngblood80}, since both the electric polarization of a given bond, 
and the distribution of mass on that bond, are determined uniquely by the flux of ${\bf P}$.


The topological structure of the ice states becomes evident once periodic boundary conditions 
are imposed on the lattice.
In this case, the local conservation of flux (second ice rule) gives rise to a distinct 
set of global topological sectors, with definite, quantised, flux through the (periodic) 
boundaries of the crystal.
In a real crystal, with open boundary conditions, these topological sectors correspond 
to the different possible values of the electrical polarisation of the crystal.
The very high dielectric constant of water ice~\cite{petrenko99}, can therefore
be interpreted as 
evidence of fluctuations between different topological sectors~\cite{ryzhkin05, 
ryzhkin97,  jaubert13, klyuev14}. 


Quantum mechanics enters into the physics of ice through the (quantum) tunnelling 
of protons from one configuration to another.
Since it is energetically expensive to violate the ice rules, tunnelling should 
occur between different configurations which satisfy the ice rules.
This can be accomplished through the collective tunnelling of a group of protons,
on any closed loop within the lattice, where the associated flux ``arrows'' 
also form a closed loop, as illustrated in Fig.~\ref{fig:fluxrep}.


Where such a loop exists, it is possible to generate a second proton configuration 
satisfying the ice rules, simply by interchanging long and short proton bonds.
This is equivalent to reversing the sense of all fluxes on the loop.  
The shortest loops for which this is possible in ice Ih consist of 
six oxygen--oxygen bonds, as illustrated in Fig. \ref{fig:loops}.
Quantum tunnelling of the form considered in this paper preserves the topological 
sector, since this is unchanged by any local rearrangements of protons which 
preserves the condition of local flux conservation.
As a consequence, the conserved flux is elevated to the role of a quantum number~\cite{sikora11}.  
We note that, since all of these properties follow from the topological structure 
of ice, they are independent of the mechanism by which quantum tunnelling
occurs \cite{footnote1}.


The effect of quantum tunnelling on loops of six bonds has 
previously been explored in the context of quantum dimer models on the diamond 
lattice~\cite{moessner03, sikora09, sikora11}, and of quantum effects in 
spin ice~\cite{hermele04, shannon12, benton12, mcclarty-arXiv}.   
These models have the same cubic symmetry, $Fd\overline{3}m$, as Ic water ice, 
for which the shortest closed loop of bonds defines the edge of an hexagonal plaquette.
Since all such plaquettes are related by lattice symmetries, a minimal model 
for quantum tunnelling can be obtained by introducing a single tunnelling 
matrix element $g$ 
\begin{eqnarray}  
\mathcal{H}_{\sf tunnelling}^{\sf cubic} =   
- g \sum_{\hexagon} 
   \big[
   |\! \circlearrowright \rangle\langle \circlearrowleft\! | + 
   |\! \circlearrowleft \rangle\langle \circlearrowright\! | 
   \big] \; , 
\label{eq:htunnelling1}
\end{eqnarray}
where the sum on $\hexagon$ runs over all hexagonal plaquettes in the lattice,
and tunnelling occurs between loops with the opposite sense of flux.
We emphasise operators in Eq. (\ref{eq:htunnelling1}) act only on those plaquettes on
which the flux arrows form a closed loop, as higlighted in Fig. \ref{fig:loops}.
These plaquettes are those described as ``proton-ordered rings'' in, e.g., 
Refs.~[\onlinecite{bove09}] and [\onlinecite{drechsel-grau14-PRL}].


In ice Ih, in contrast, there are two inequivalent types of six--sided plaquette, 
illustrated in Fig.~\ref{fig:loops}.   
The first is composed entirely of centre--symmetric bonds
[cf. Fig.~\ref{fig:bondsymmetries}], and comprises the buckled hexagonal plaquettes 
which make up the hexagonal--symmetry layers of the Ih structure.   
A crystal with $N$ oxygen atoms contains $N/2$ such plaquettes,
which we label as type I [cf.~Fig.~\ref{fig:loops}(a)].
The second is composed of four centre--symmetric 
and two mirror--symmetric bonds, linking neighbouring layers of the lattice.  
There are $3N/2$ such plaquettes, which we label as type II [cf.~Fig.~\ref{fig:loops}(b)].


Since there are two, inequivalent, types of six-sided plaquette in the 
Ih lattice, the minimal model for quantum effects in ice Ih
comprises two distinct matrix elements
\begin{eqnarray}
&&\mathcal{H}_{\sf tunnelling}^{\sf hexagonal} 
= - g_1 \sum_{\hexagon \in I}  
   \big[
   |\! \circlearrowright \rangle\langle \circlearrowleft\! | + 
   |\! \circlearrowleft \rangle\langle \circlearrowright\! | 
   \big] \nonumber \\
&&\qquad \qquad \quad  
   - g_2 \sum_{\hexagon \in II}  
   \big[
   |\! \circlearrowright \rangle\langle \circlearrowleft\! | + 
   |\! \circlearrowleft \rangle\langle \circlearrowright\! | 
   \big] \; ,
\label{eq:htunnelling2}
\end{eqnarray}
acting on the space of all possible proton configurations obeying 
the ice rules.   
Symmetry alone does not place any constraints on the values of 
$g_1$ and $g_2$, and {\it a priori}, these matrix elements can take 
on either sign.  

It is important to note that collective coherent quantum tunnelling
of the kind described by Eq. (\ref{eq:htunnelling2}) is distinct from 
incoherent single-proton tunnelling.
The study of proton tunnelling in ice has rather a long history.
The motion of single protons plays an important role in
theories of ice's dielectric properties \cite{jaccard64}
and it was believed for some time that this motion may proceed
by single particle quantum tunnelling \cite{petrenko99}.
However,
it is now generally believed that tunnelling of single protons
is not effective in ice Ih at low
temperature and ambient pressure \cite{cowin99}.
Collective tunnelling of protons, however, may still play a role
in ice Ih, as it does in many other H-bonded systems
\cite{brougham99, klein04, lopez07, meng15}.
The experimental results of
 Bove {\it et al}~\cite{bove09} indicate 
that collective tunnelling of protons on rings
of six H-bonds can indeed occur in ice Ih.
%

Various {\it ab initio} studies have considered the possibility of proton tunnelling 
in water ice.
Much of this work has been motivated by interest in the successive transformations 
between cubic, high--pressure phases of ice, and in particular by ice X, an extreme 
high--pressure phase where the Bernal--Fowler ice rules no longer apply \cite{holzapfel72}.
Benoit {\it et al.}~[\onlinecite{benoit98}] find that the transformation from ice VIII to ice VII 
(which do obey the ice rules) is driven by quantum tunnelling of individual protons.
This becomes more favourable under pressure, as the oxygen--oxygen distance decreases.
A later study by Lin {\it et al.}~[\onlinecite{lin11}] reinforces this conclusion, and highlights 
the emergence of collective, quantum tunnelling of groups of protons in cubic ice VII.
Lin {\it et al.} note that this type of tunnelling might also be effective in hexagonal ice Ih, 
albeit with a much smaller tunnelling matrix element.


The case considered in this Article  --- collective, quantum tunnelling of protons in ice Ih 
--- was recently studied in detail by Drechsel--Grau and Marx, using a combination of 
density-functional, molecular-dynamics and path-integral techniques~\cite{drechsel-grau14-PRL}.
Considering a cluster of 48 water molecules, containing a single ``proton--ordered'' 
ring of the type shown in Fig.~\ref{fig:loops}, Drechsel--Grau and Marx 
find that proton dynamics at low temperatures are dominated by the 
collective quantum tunnelling of protons from one ordered state of the ring 
to the other.
A further study by the same authors~\cite{drechsel-grau14-angewandte}
finds that partial deuteration of ice Ih will suppress this collective quantum 
tunnelling, in agreement with experimental results of Bove {\it et al.} [\onlinecite{bove09}].
Another recent {\it ab initio} study of nuclear quantum effects on the dielectric
constant of water ice also finds strong quantum fluctuations of the
proton system,  consistent with the findings of Drechsel-Grau and Marx
\cite{moreira15}.


If tunnelling of the form of Eq.~(\ref{eq:htunnelling2}) were to occur on a single isolated 
plaquette, the resulting ground state would be a quantum superposition of the two opposite 
senses of proton-ordering on the ring, in direct analogy with the resonating ground
state of single benzene ring.   
However in a typical state obeying the ice rules, approximately a {\it quarter} of the 
plaquettes are ``ordered'', and quantum tunnelling on these plaquettes allows the 
system to explore an exponentially large number of different states obeying the ice rules.


In related three-dimensional quantum dimer and quantum spin ice models, dynamics 
of this type have been shown to stabilise a quantum liquid state, formed by 
a coherent superposition of an exponentially large number of states
~\cite{moessner03, sikora09, sikora11,hermele04, shannon12, benton12, mcclarty-arXiv}.  
In water ice, such a state would be a massively entangled superposition of proton
configurations obeying the ice rules, in which the molecular character of 
water molecules was preserved, but individual protons could no longer be assigned
to a given water molecule.
This type of liquid should be contrasted with the ``plaquette--ordered'' phase 
found in simplified two--dimensional models of water ice, which are dominated 
by local resonances \cite{shannon04, syljusen06, castro-neto06, henry14}.


At this time, we are not aware of any attempt to independently determine the two 
different matrix elements $g_1$ and $g_2$, which define our model 
$\mathcal{H}_{\sf tunnelling}^{\sf hexagonal}$ [Eq.~(\ref{eq:htunnelling2})], 
from ab initio simulation of ice Ih.  
And to the best of knowledge, the only available estimate of the magnitude 
of collective tunnelling in ice Ih is 
\begin{eqnarray}
g \sim 1.46 \times 10^{-4}\ \text{eV} \approx 1.7\ \text{K} \; ,
\end{eqnarray}
taken from the density-functional calculations of Ihm [\onlinecite{ihm96}].
Intriguingly, this value is consistent with the energy scale of
proton dynamics observed in inelastic neutron scattering \cite{bove09}.
And it is somewhat larger than the corresponding estimates for the putative 
``quantum spin ice'' materials \cite{benton12}. 
Taken at face value, this would suggest that water ice is potentially 
a {\it more} favorable place to look for the formation of an exotic quantum liquid state 
than are the quantum spin ices.


With this in mind, in Section~\ref{section:photons} of this article, we develop a theory 
of disordered proton configurations in the presence of quantum 
tunnelling, based on the minimal quantum model of ice Ih, 
$\mathcal{H}_{\sf tunnelling}^{\sf hexagonal}$~[Eq.~(\ref{eq:htunnelling2})].
We make the assumption that
\begin{eqnarray}
g_1> 0 \quad  , \quad g_2 > 0 \; ,
\end{eqnarray}
so that the model is accessible to quantum Monte Carlo simulation.
Before examining the quantum model, however, it is necessary
to understand the classical proton correlations which arise simply 
from the ice--rule constraint.
This will form the subject of Section~\ref{section:classical}, below.   

\section{Proton correlations in a classical model of ice Ih}
\label{section:classical}

In what follows, we develop a theory of proton correlations in a classical model
of ice Ih, neglecting all quantum tunnelling between different proton
configurations.
This theory provides a detailed and microscopically--derivable
phenomenology to explain the diffuse scattering which is arises as 
a result of static proton disorder in ice Ih 
\cite{schneider80, li94, nield95, owston49, wehinger14, nield95-ActaCrystollagrA51, beverley97-JPCM9, 
beverley97-JPhysChemB101, villain73, villain72, descamps77}.


We start, in Section~\ref{subsec:classical.fields} by developing 
a long--wavelength, classical field theory of proton configurations
in ice Ih.
This field theory has parallels with those developed to explain 
pinch--points in proton--bonded ferroelectrics \cite{youngblood80}
and spin--ice~\cite{henley05}, but displays a number of 
new features, which will become important in the quantum case.
The details of these calculations are described in 
Appendix~\ref{appendix:continuum.field.theory}.


In Section~\ref{subsec:classical.lattice.theory}, we introduce a 
lattice theory of proton correlations in classical ice Ih, 
and show that this reproduces the predictions of Section~\ref{subsec:classical.fields}.
The details of these calculations, which are based on a generalisation 
of a method introduced for spin--ice by Henley~\cite{henley05},
are developed in Appendix~\ref{appendix:projection}.


Then, in Section~\ref{subsec:classical.predictions}, we use the lattice 
theory introduced in Section~\ref{subsec:classical.lattice.theory} to make explicit 
predictions for the structure factors measured in X--ray~\cite{wehinger14, owston49} 
and neutron scattering~\cite{li94, nield95, schneider80} experiments. 
These show a number of interesting features, which we interpret 
in terms of the classical field theory developed in 
Section~\ref{subsec:classical.fields}.



\subsection{Continuum field--theory for protons in classical ice Ih}
\label{subsec:classical.fields}

The natural place to start in constructing a theory of proton disorder
in ice Ih is from the ice rules~\cite{bernal33,pauling35,petrenko99}.
The first ice rule states that each bond of the oxygen lattice contains exactly 
one proton [cf.~Section~\ref{section:definitions}].  
This proton is displaced, relative to the centre of the bond, towards 
one of the two oxygen atoms which make up the bond 
[cf.~Fig~\ref{fig:1h_structure}].  
The displacement of this proton from the centre 
of the bond $(\mathbf r \mathbf{r}')$ can be described 
using the Ising variable 
\begin{eqnarray}
\sigma_{\mathbf{r} \mathbf{r}'} = \pm 1 \; ,
\label{eq:ising}
\end{eqnarray}
where $\sigma_{\mathbf{r} \mathbf{r}'} = +1$
if the proton is displaced towards $\mathbf{r}'$, 
and $\sigma_{\mathbf{r} \mathbf{r}'}=-1$
if it is displaced towards $\mathbf{r}$.
It follows that
\begin{eqnarray}
\sigma_{\mathbf{r} \mathbf{r}'}=-\sigma_{\mathbf{r}' \mathbf{r}}.
\label{eq:sigmadirectionality}
\end{eqnarray}


The second ice rule states that each oxygen must form exactly two short (covalent)
and two long (hydrogen) bonds with neighbouring protons.  
Written in terms of the Ising variable $\sigma_{\mathbf{r} \mathbf{r}'}$, this becomes 
a condition that, at every oxygen lattice site~$\mathbf{r}$
\begin{eqnarray}
m_{\mathbf{r}}\equiv\sum_{{\sf nn} \ i} \sigma_{\mathbf{r+\mathbf{d}_{ \mathbf{r} i}} \ \mathbf{r}}=0
\label{eq:vanishingm}
\end{eqnarray}
where the sum runs over the four nearest---neigbours within the oxygen lattice, 
located at sites $\mathbf{r}+\mathbf{d}_{\mathbf{r} i}$.
For this purpose, it is necessary to divide the lattice of oxygen sites into four
inequivalent sublattices as illustrated in Fig. \ref{fig:fluxrep}. 
These four sets of oxygen sites
have different associated vectors ${\bf d}_{\mathbf{r} i}$, 
defined in Appendix~\ref{appendix:lattice}.


Just as in proton-bonded ferroelectrics~\cite{youngblood80}, or spin ice~\cite{henley05},  
we can understand the proton correlations arising from Eq.~(\ref{eq:vanishingm}) by considering 
the spatial variation of the flux field represented by the arrows on the bonds in Fig.~(\ref{fig:loops}).
For each oxygen--oxygen bond $(\mathbf{r}\mathbf{r}')$, we assign a flux 
\begin{eqnarray}
{\bf P}_{\mathbf{r} \mathbf{r}'}=\frac{\mathbf{r}-\mathbf{r}'}{a_{\sf O}} \sigma_{\mathbf{r} \mathbf{r}'}
\label{eq:poldeff}
\end{eqnarray}
where $a_{\sf O}$ is the oxygen--oxygen bond distance.
This flux points in the direction of the polarisation 
of the associated H-bond, as illustrated in Fig.~\ref{fig:fluxrep}.
The total flux from the four arrows around a single oxygen site is thus
\begin{eqnarray}
{\bf P}_{\mathbf{r}}\equiv\sum_{{\sf nn} \ i} \frac{1}{a_{\sf O}}\mathbf{d}_{ \mathbf{r} i} \ \sigma_{\mathbf{r}+\mathbf{d}_{ \mathbf{r} i} \ \mathbf{r}}
\label{eq:tetrahedronflux}
\end{eqnarray}


Knowledge of the fields $m_{\mathbf{r}}$ [Eq. (\ref{eq:vanishingm})]
 and ${\bf P}_{\mathbf{r}}$ [Eq. (\ref{eq:tetrahedronflux})] 
on half of the oxygen sites (e.g. those on the A and C sublattices, shown in Fig. \ref{fig:fluxrep})
is sufficient to uniquely determine the proton configuration of the entire lattice.
We can see this as follows:
if, for a given oxygen site $\mathbf{r}$, we know both $m_{\mathbf{r}}$ (which must be zero for an ice rule state)
and ${\bf P}_{\mathbf{r}}$ we can determine the value of all of the surrounding bond
variables $\sigma_{\mathbf{r}+\mathbf{d}_{ \mathbf{r} i} \ \mathbf{r}}$ using the relation
\begin{eqnarray}
\sigma_{\mathbf{r}+\mathbf{d}_{ \mathbf{r} i} \ \mathbf{r}}=\frac{1}{4} \left( m_{\mathbf{r}}+\frac{3}{a_{\sf O}} 
\mathbf{d}_{ \mathbf{r} i} \cdot \mathbf{P}_{\mathbf{r}} \right).
\end{eqnarray}
Every bond belongs either to one $A$ oxygen site or one $C$ oxygen site [cf. Fig. \ref{fig:loops}(b)]
so knowing  $m_{\mathbf{r}}$ 
 and ${\bf P}_{\mathbf{r}}$ on just the $A$ and $C$ sites, (or equivalently just the $B$ and $D$
sites) is sufficient.


We may imagine generating a proton configuration by setting $m_{\mathbf{r}}=0$
on every $A$ and $C$ oxygen site and letting the flux ${\bf P}_{\mathbf{r}}$ vary between those sites.
For the configuration thus obtained to be consistent with the ice rules we would need
it to satisfy $m_{\mathbf{r}}=0$ on all of the $B$ and $D$ tetrahedra as well.
Naturally, this implies some constraints on the spatial variation of ${\bf P}_{\mathbf{r}}$.
These constraints control the form of the proton correlations.


The constraints on the spatial variation ${\bf P}_{\bf r}$ arising from the ice rules
may be understood by defining continuum fields (i.e. defined over all space, not just on the lattice)
$\bar{{\bf P}}_A(\mathbf{r})$ and $\bar{{\bf P}}_C(\mathbf{r})$ in such a way that evaluating them
at the lattice sites ${\bf r}_{A, C}$ returns the value of ${\bf P}_{\bf r}$ [Eq.  (\ref{eq:tetrahedronflux})].
If we then assume $\bar{{\bf P}}_{A, C}(\mathbf{r})$ to vary smoothly in space, we can use the condition
that $m_{\bf r}$ must vanish at the $B$ and $D$ sites to obtain a constraint on the fields 
$\bar{{\bf P}}_{A, C}(\mathbf{r})$ and their derivatives.
This procedure is described in more detail in Appendix \ref{appendix:continuum.field.theory}.


Neglecting terms beyond leading order in the bond distance $a_{\sf O}$, 
we obtain:
\begin{eqnarray}
&&-3\bar{P}_A^z(\mathbf{r})+3 \bar{P}_C^z(\mathbf{r})
-2 a_{\sf O} \nabla \cdot \bar{{\bf P}}_A
 \nonumber \\
&& \qquad\qquad\qquad+ 3 a_{\sf O}\partial_z \bar{P}_A^z -3 a_{\sf O} \partial_z \bar{P}_C^z 
= 0 \; ,
\label{eq:divconstraint1}
\\
&&-3\bar{P}_C^z(\mathbf{r})+3 \bar{P}_A^z(\mathbf{r})
-2 a_{\sf O} \nabla \cdot \bar{{\bf P}}_C
\nonumber \\
&& \qquad\qquad\qquad+ 3 a_{\sf O}\partial_z \bar{P}_C^z  -3 a_{\sf O} \partial_z \bar{P}_A^z 
=0 \; ,
\label{eq:divconstraint2}
\end{eqnarray}
where $z$ is the hexagonal symmetry axis of the crystal (i.e. the crystalographic c--axis).
These equations can be decoupled by introducing odd and even combinations
of the fields $\bar{{\bf P}}_{A, C}(\mathbf{r})$
\begin{eqnarray}
&&{\bf P}_{+}(\mathbf{r})= \frac{1}{\sqrt{2}}  (\bar{{\bf P}}^A(\mathbf{r})+\bar{{\bf P}}^C(\mathbf{r})) \, ,
\label{eq:p+def}\\
&&{\bf P}_{-}(\mathbf{r})= \frac{1}{\sqrt{2}}  (\bar{{\bf P}}^A(\mathbf{r})-\bar{{\bf P}}^C(\mathbf{r})) \; .
\label{eq:p-def}
\end{eqnarray}
It follows that the uniform polarisation ${\bf P}_{+}(\mathbf{r})$ satisfies a zero--divergence condition
\begin{eqnarray}
\nabla \cdot {\bf P}_{+} (\mathbf{r}) = 0 \; ,
\label{eq:p+constraint}
\end{eqnarray}
while the staggered polarisation ${\bf P}_{-}(\mathbf{r})$ is governed by the equation 
\begin{eqnarray}
P^z_-(\mathbf{r})+\frac{2}{3} a_{\sf O} \nabla \cdot {\bf P}_{-}(\mathbf{r})-
 a_{\sf O} \partial_z P^z_-(\mathbf{r}) = 0 \; .
\label{eq:p-constraint}
\end{eqnarray}


The very different form of the equations governing ${\bf P}_{+}$ and ${\bf P}_{-}$
suggest that these fields have qualitatively different correlations. 
Following \cite{henley05, henley10, huse03}, we can estimate these 
by assuming a free--energy of the form 
\begin{eqnarray}
{\mathcal F} = - \left( k_B T \right) 
       \int \frac{d^3{\bf r}}{V_{\sf u.c.}} 
        \sum_{\upsilon = \pm}
        \frac{\kappa_\upsilon {\bf P}_\upsilon^2}{2}     
\label{eq:gaussian.free.energy}
\end{eqnarray}
where $\kappa_\upsilon$ is an (unknown) 
constant of entropic origin, and $V_{\sf u.c.}$ is the volume of a unit cell.
Within this approximation, correlations of ${\bf P}_\pm$ are 
controlled by a Gaussian distribution of fields 
\begin{eqnarray}
p[{\bf P}_\upsilon]
   \propto \exp 
   \left[
   - \frac{\kappa_\upsilon}{2} 
   \int \frac{d^3{\bf r}}{V_{\sf u.c.}}  
   {\bf P}_\upsilon^2
              \right]
\label{eq:gaussianprob}
\end{eqnarray}
subject to the constraints Eq.~(\ref{eq:p+constraint}) and Eq.~(\ref{eq:p-constraint}).


We find that the correlations of ${\bf P}_{+}$ have the form 
of a singular ``pinch--point'' in reciprocal space 
\begin{eqnarray}
&&\langle P_+^{\alpha}  (-\mathbf{q})P_+^{\beta}(\mathbf{q}) \rangle
= \frac{1}{\kappa_+} \left( \delta_{\alpha \beta}-\frac{q_{\alpha} q_{\beta}}{q^2} \right) \; ,
\nonumber \\ 
&& \qquad \qquad \qquad  \qquad \qquad \qquad (\alpha, \beta =x, y, z)  \; .
\label{eq:p+corr}
\end{eqnarray}
Meanwhile, the correlations of ${\bf P}_{-}$ in reciprocal space 
show much broader, smoothly varying structure 
\begin{eqnarray}
&&\langle P_-^{\alpha}  (-\mathbf{q}) P_-^{\beta}(\mathbf{q}) \rangle
=\frac{1}{\kappa_-} \left( \delta_{\alpha \beta} 
- \frac{4 \zeta^2 q_{\alpha} q_{\beta}}{1+\zeta^2q_z^2+4 \zeta^2 {\bf q}_{\perp}^2} \right) \; ,
\nonumber \\ 
&& \qquad \qquad \qquad \qquad \qquad \qquad \qquad (\alpha, \beta =x, y)
\label{eq:p-perpcorr} \\
&&\langle P_-^{z}  (-\mathbf{q})P_-^{z}(\mathbf{q}) \rangle
=\frac{1}{\kappa_-} 
 \left( 1-\frac{1+\zeta^2 q_z^2}{1+\zeta^2q_z^2+4 \zeta^2 {\bf q}_{\perp}^2} \right) \; ,
 \nonumber \\
\label{eq:p-zcorr}
\end{eqnarray}
where
\begin{eqnarray}
{\bf q}_{\perp} &=& (q_x, q_y) \; 
\end{eqnarray}
and, for compactness, we have introduced the notation
\begin{eqnarray}
\zeta &=&\frac{a_{\sf O}}{3} \; .
\label{eq:shortlength}
\end{eqnarray}


Fourier--transforming Eq.~(\ref{eq:p+corr})--Eq.~(\ref{eq:p-zcorr}), 
we find that correlations of ${\bf P}_{+}$ decay algebraically in real space,
with the dipolar form
\begin{eqnarray}
 &&\langle P_+^{\alpha}  ({\bf r}) P_+^{\beta}(\mathbf{0}) \rangle 
= \frac{4 \pi}{\kappa_+} \left[ 
\delta({\bf r}) + \frac{\delta_{\alpha\beta}  r^2  - 3 r_\alpha r_\beta}{r^5}
\right]
\end{eqnarray}
Meanwhile correlations of ${\bf P}_-$ are very short--ranged, decaying over 
a length--scale $\zeta$~[Eq.~(\ref{eq:shortlength})].
It follows that proton correlations at large distances are controlled by the field 
${\bf P}_{+}$.


The algebraic correlations of ${\bf P}_{+}$ give rise to sharp pinch--point
singularities in structure factors, of the type observed by Li {\it et al.} 
in neutron scattering from ice Ih~\cite{li94}.
However in some Brilluouin zones, pinch--point singularities are suppressed
by the lattice form--factor, 
and scattering is instead dominated by 
broad, assymetric features coming from the correlations of ${\bf P}_{-}$.
We discuss this point further in Section III C, below, where we develop
an explicit theory for neutron and X--ray scattering experiments. 


The form of the constraint on ${\bf P}_{+}(\mathbf{r})$ [Eq.~(\ref{eq:p+constraint})]
strongly suggests an analogy with electromagnetism, 
where the zero--divergence condition on magnetic field
\begin{eqnarray}
\nabla \cdot {\bf B} = 0  \; , 
\end{eqnarray}
can be resolved as 
\begin{eqnarray} 
{\bf B} = \nabla \times {\bf A} \; , 
\end{eqnarray}
and the electric and magnetic fields are connected by an underlying $U(1)$ 
gauge symmetry.
This analogy, and the distinction between the two classical fields, 
${\bf P}_{+}$ and ${\bf P}_{-}$, become explicit once quantum 
effects are taken into account, as described in 
Section~\ref{section:photons}.


\begin{figure*}
\centering
\includegraphics[width=0.7\textwidth]{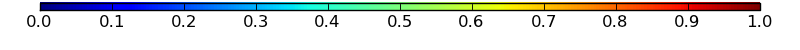} \\
\subfigure[\ \ Theory- hk0 plane]{%
  \includegraphics[width=.5195\textwidth]{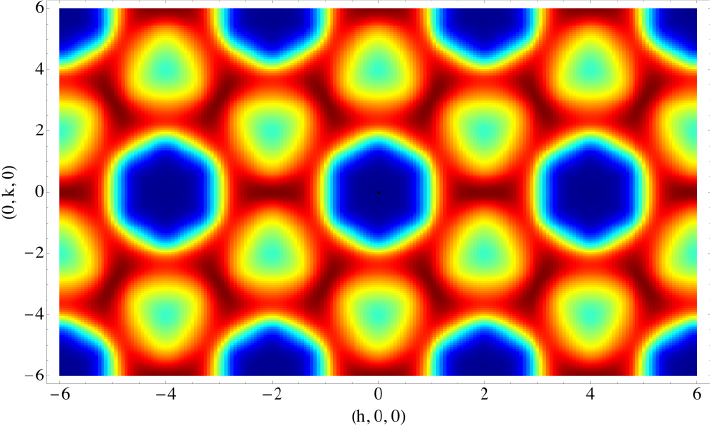}}%
\qquad
\subfigure[\ \ Theory- 0kl plane]{%
  \includegraphics[width=.3\textwidth]{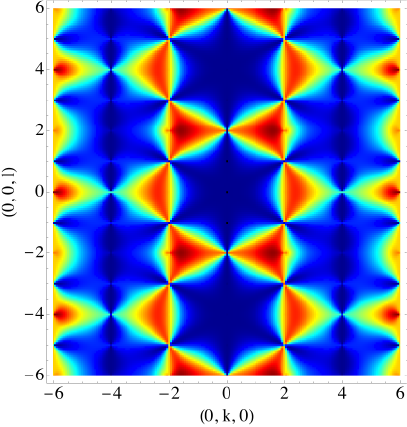}}%
\\
\subfigure[\ \ Theory- h0l plane]{%
  \includegraphics[width=.501962\textwidth]{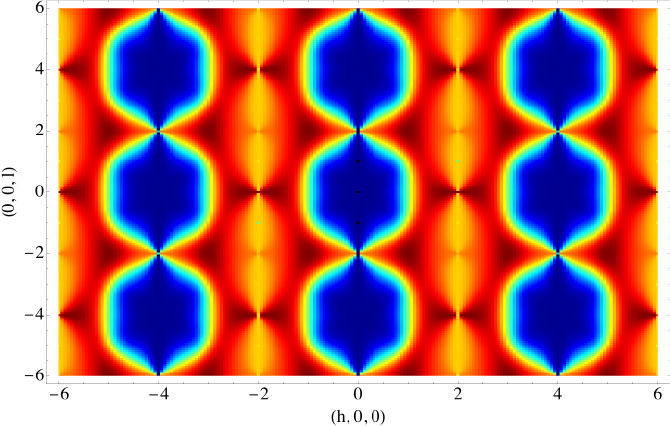}}%
\caption{
(Color online). 
Correlation of long and short proton bonds in a classical model of 
hexagonal (Ih) water ice.
The structure factor $S_{\sf Ising}(\mathbf{q})$~[Eq.~\ref{eq:S.of.q.Ising}], 
for the Ising variable $\sigma$~[Eqs.~(\ref{eq:ising})--(\ref{eq:poldeff})], 
is plotted in three orthogonal planes in reciprocal space.   
Near to zone centres, correlations are well--described by
a combination of pinch--point singularities, reflecting the algebraic correlations 
of the field ${\bf P}_+(\mathbf{q})$~[cf.~Eq.~(\ref{eq:p+corr})], and smooth features, 
reflecting the short--range correlations of the field 
${\bf P}_-(\mathbf{q})$~[cf.~Eq.~(\ref{eq:p-perpcorr},\ref{eq:p-zcorr})], 
as discussed in Section~\ref{subsec:classical.fields}.
Calculations were performed using the method outlined in Appendix~\ref{appendix:projection}, 
in which the ice--rule constraints are written as orthogonality conditions 
in Fourier space~\cite{henley05}.
Reciprocal--lattice vectors are indexed to the orthorhombic unit cell defined 
in Appendix~\ref{appendix:lattice}, following the conventions of 
Nield and Whitworth~\cite{nield95}.
} 
\label{fig:Sq.Ising.classical}
\end{figure*}

\subsection{Lattice theory of proton correlations in classical ice Ih}
\label{subsec:classical.lattice.theory}

The classical fields ${\bf P}_{+}$ and ${\bf P}_{-}$, introduced in Section~\ref{subsec:classical.fields}
provide a complete description of the correlations of protons in classical ice Ih 
at long--wavelength, i.e. near to zone--centers in reciprocal space.
However, X--ray and neutron scattering experiments on water ice measure 
proton correlations at all length--scales.
We have therefore developed a lattice--based theory of proton correlations 
in classical ice Ih, valid for all wave--numbers. 
The approach we take is a generalisation of the method developed for spin--ice 
by Henley~\cite{henley05}, in which the ice rules are expressed as a projection 
operator in reciprocal space. 
In what follows we explore how the predictions of this theory relate to the
those obtained from the continuum theory described in Section~\ref{subsec:classical.fields}.
We reserve all technical details for Appendix~\ref{appendix:projection}.


The underlying structure of proton correlations in classical ice Ih is most easily
understood throughout the correlations of the Ising variables 
$\sigma_{\mathbf{r} \mathbf{r}'}$~[Eq.~(\ref{eq:ising})], which 
describe the alternating long and short bonds  between protons and  
neighbouring oxygen atoms.
These are characterised by the structure factor
\begin{eqnarray}
S_{\sf Ising}(\mathbf{q}) 
= \sum_{\nu\nu^\prime} \langle \sigma_\nu (\mathbf{q}) \sigma_{\nu^\prime} (-\mathbf{q}) \rangle \; ,
\label{eq:S.of.q.Ising}
\end{eqnarray} 
where the sums on $\nu$ and $\nu^\prime$ run over the eight distinct bonds 
within the 4--site primitive unit cell, and 
\begin{eqnarray}
\sigma_\nu (\mathbf{q}) 
   &=& \sqrt{\frac{4}{N}}
   \sum_{\mathbf{r}\mathbf{r}^\prime \in \nu}  
   \exp \left( -i \mathbf{q} \cdot {\bf R}_{\mathbf{r}\mathbf{r}^\prime} \right) 
   \sigma_{\mathbf{r}\mathbf{r}^\prime} \; , \nonumber\\
   && \qquad  \qquad {\bf R}_{\mathbf{r}\mathbf{r}^\prime} = \frac{\mathbf{r} + \mathbf{r}^\prime}{2} \; .
\label{eq:sigma.q}
\end{eqnarray}
In calculating $\sigma_\nu (\mathbf{q})$, we label the 4 oxygen sublattices
within the unit cell A, B, C, D, (cf. Fig~\ref{fig:fluxrep}), and adopt a sign 
convention such that 
\begin{eqnarray}
\mathbf{r} \in \{A,C\} \; , \; \mathbf{r}^\prime \in \{B,D\} \; .
\label{eq:sgnconv}
\end{eqnarray}


In Fig.~\ref{fig:Sq.Ising.classical} we show results for 
$S_{\sf Ising}(\mathbf{q})$, 
calculated within the lattice--based theory.
The structure factor exhibits clear pinch--point singularities, characteristic of the 
ice rules~\cite{youngblood80,henley05}, at a subset of Brillouin--zone centers typified by
\begin{eqnarray}
\mathbf{Q}^*_{\sf p} =(0, 0, 2) \; ,
\end{eqnarray}
where, following Nield and Whitworth~\cite{nield95}, we index all reciprocal--lattice vectors 
to the 8-site orthorhombic unit cell defined in Appendix~\ref{appendix:lattice}.


Correlations near to reciprocal--lattice vectors (Brillouin--zone centres) are described 
by the classical field theory developed in Section~\ref{subsec:classical.fields}, 
with contributions from both fields, ${\bf P}_+$ and ${\bf P}_-$.
Near to a reciprocal lattice vector $\mathbf{Q}$, for $ | \tilde{\mathbf{q}} | \ll 1 $, 
the structure factor can be written 
\begin{eqnarray}
&&S_{\sf Ising }(\mathbf{Q} + \tilde{\mathbf{q}}) 
\nonumber\\
&&  \quad \approx 
     \sum_{\upsilon = \pm}
          F^{\sf Ising}_\upsilon (\mathbf{Q})
         \langle 
              | \boldsymbol{\hat{\lambda}}_{\mathbf{Q},\upsilon}^{\sf Ising} 
              \cdot {\bf P}_\upsilon (\tilde{\mathbf{q}}) |^2 
         \rangle \; ,
\label{eq:SIsing-Pcorr}
\end{eqnarray}
where the form--factor 
$F^{\sf Ising}_\upsilon (\mathbf{Q})$~[Eq.~(\ref{eq.F.Q.upslion})] 
and vectors $\boldsymbol{\hat{\lambda}}_{\mathbf{Q},\upsilon}^{\sf Ising}$~[Eq.~(\ref{eq:defn-Ising-lambda})]
are defined in Appendix~\ref{appendix:structure.factors.and.fields}, 
and the Fourier transform ${\bf P}_\upsilon (\tilde{\mathbf{q}})$~[Eq.~(\ref{eq:FT.P})] 
in Appendix~\ref{appendix:projection}.



Sharp pinch--points are seen for a subset of reciprocal lattice vectors 
$\mathbf{Q}_{\sf p}$, for which 
\begin{eqnarray}
F^{\sf Ising}_- ( \mathbf{Q}_{\sf p} ) = 0 \; .
\end{eqnarray}
In this case, correlations are controlled by the zero--divergence condition 
on ${\bf P}_{+}(\mathbf{q})$ [Eq.~(\ref{eq:p+constraint})], 
and it follows from Eq.~(\ref{eq:p+corr}) and Eq.~(\ref{eq:FT.P}) that
\begin{eqnarray}
S_{\sf Ising}( \mathbf{Q}_{\sf p} + \tilde{\mathbf{q}} ) 
   \approx
       \frac{F^{\sf Ising}_+ ( \mathbf{Q}_{\sf p} )}{\kappa_+}
            \left( 1 -  \frac{| \boldsymbol{\hat{\lambda}}_{\mathbf{Q},\upsilon}^{\sf Ising} 
            \cdot \tilde{\mathbf{q}} |^2}{| \tilde{\mathbf{q}} |^2} \right) \, .
\end{eqnarray}
Considering the specific example of 
\begin{eqnarray}
{\bf Q}^*_{\sf p} = (0, 0, 2)
\end{eqnarray}
for which 
\begin{eqnarray}
   \boldsymbol{\hat{\lambda}}_{{\bf Q}^*_{\sf p},+}^{\sf Ising} &=& (0,0,1) 
\end{eqnarray}
we have
\begin{eqnarray}
S_{\sf Ising}(\mathbf{Q}^*_{\sf p} + \tilde{\mathbf{q}}) 
   \propto \left( 1 -  \frac{\tilde{q}_z^2}{\tilde{q}^2} \right) \, .
\end{eqnarray}
This pinch--point, aligned with the hexagonal--symmetry axis (z--axis),
can be clearly resolved in Fig.~\ref{fig:Sq.Ising.classical}~(b) and (c).   


While there are no reciprocal lattice vectors for which 
$F^{\sf Ising}_+ ( \mathbf{Q} )$ vanishes identically with
$F^{\sf Ising}_- ( \mathbf{Q} )$ remaining finite, there are another 
set of lattice vectors $\mathbf{Q}_{\sf m}$, for which
\begin{eqnarray}
F^{\sf Ising}_+ ( \mathbf{Q}_{\sf m} ) \ll F^{\sf Ising}_- ( \mathbf{Q}_{\sf m} )
\end{eqnarray}
and the structure factor is dominated by the short--ranged correlations of 
${\bf P}_{-}$ [Eq.~(\ref{eq:p-constraint})].
Correlations of this type occurs for 
\begin{eqnarray}
\mathbf{Q}^*_{\sf m}=(1, 1, 0)  
\end{eqnarray}
and are visible as a broad feature centred on this reciprocal lattice vector 
in Fig.~\ref{fig:Sq.Ising.classical}~(a).  


For a general reciprocal lattice vector $\mathbf{Q}_{\sf pm}$
\begin{eqnarray}
F^{\sf Ising}_+ ( \mathbf{Q}_{\sf pm} ) \sim F^{\sf Ising}_- ( \mathbf{Q}_{\sf pm} )
\end{eqnarray}
and correlations reflect a combination of pinch--points originating in 
${\bf P}_{+}(\mathbf{q})$ and broad features originating in ${\bf P}_{-}(\mathbf{q})$.
An example of this occurs for
\begin{eqnarray}
\mathbf{Q}^*_{\sf pm} = (2, 0, 0)  
\end{eqnarray}
visible in Fig.~\ref{fig:Sq.Ising.classical}~(a) and (c), 
where a pinch--point has been superimposed on a featureless background.


Near to zone--centres, where they can be compared, the lattice--based theory is in complete
agreement with the predictions of the classical field--theory developed in 
Section~\ref{subsec:classical.fields}.  
In Appendix~\ref{appendix:projection} we show how the lattice--based theory reduces to the
continuum theory at long--wavelength.
We find the that entropic coefficient $\kappa_\upsilon$, which controls correlations of the fields
${\bf P}_\upsilon$ [Eq.~(\ref{eq:p+corr}--\ref{eq:p-zcorr})], 
is independent of $\upsilon$, i.e.
\begin{eqnarray}
\kappa = \kappa_+ = \kappa_- \, .
\end{eqnarray}


To confirm the validity of the lattice--based theory for more general ${\bf q}$, 
we have also performed classical Monte Carlo simulations of ice Ih, 
using local loop updates to sample proton configurations within the manifold 
of states obeying the ice rules.
The results, described in Appendix~\ref{appendix:numerics}, are in excellent agreement
with the predictions of the lattice theory.


\begin{figure*}
\centering
\includegraphics[width=0.7\textwidth]{colorbar.png} \\
\subfigure[\ \ hk0 plane \label{fig6a}]{%
\includegraphics[width=.5195\textwidth]{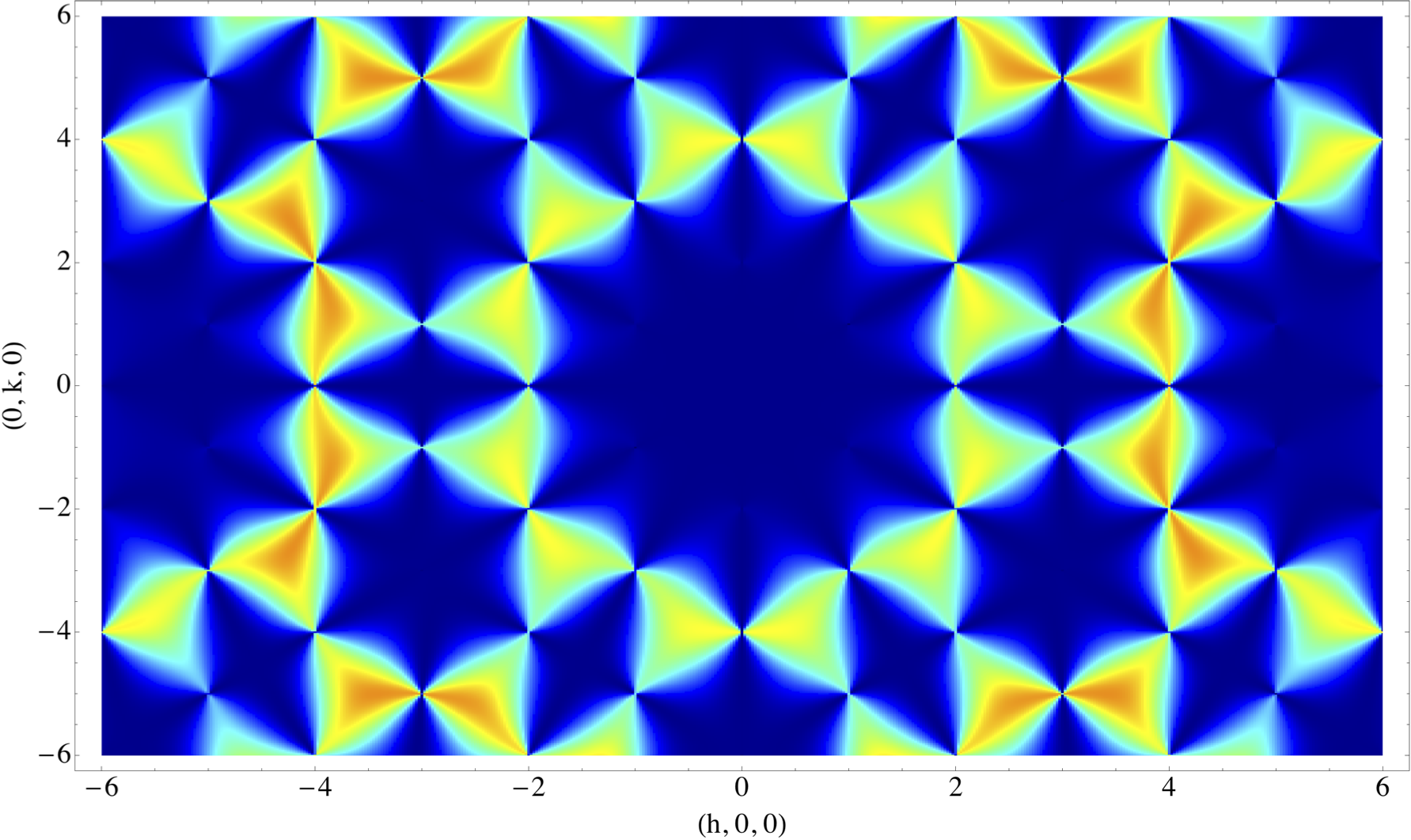}}%
    \qquad
\subfigure[\ \ 0kl plane \label{fig6b}]{%
\includegraphics[width=.3\textwidth]{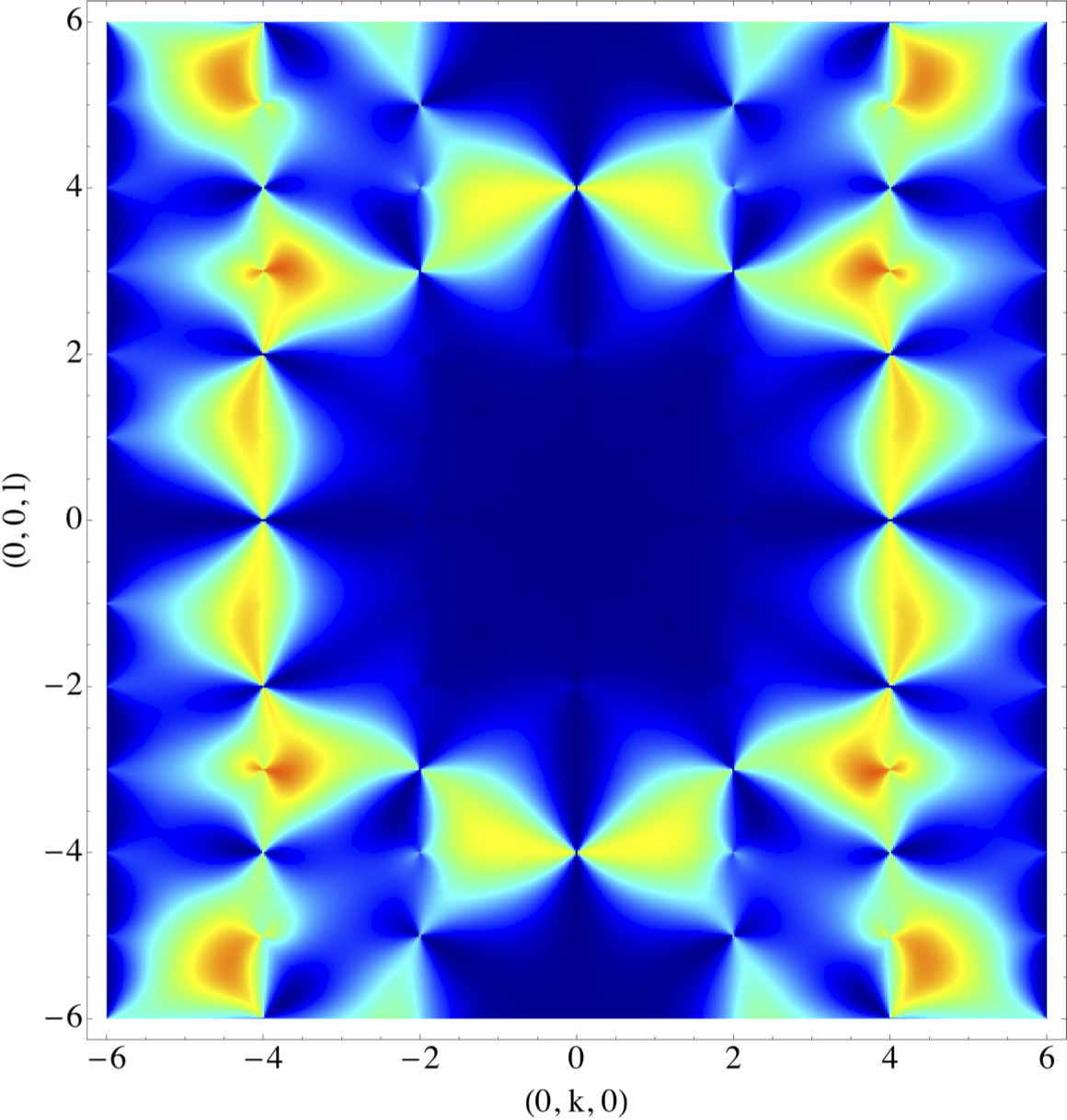}}%
\\
\subfigure[\ \ h0l plane \label{fig6c}]{%
\includegraphics[width=.501962\textwidth]{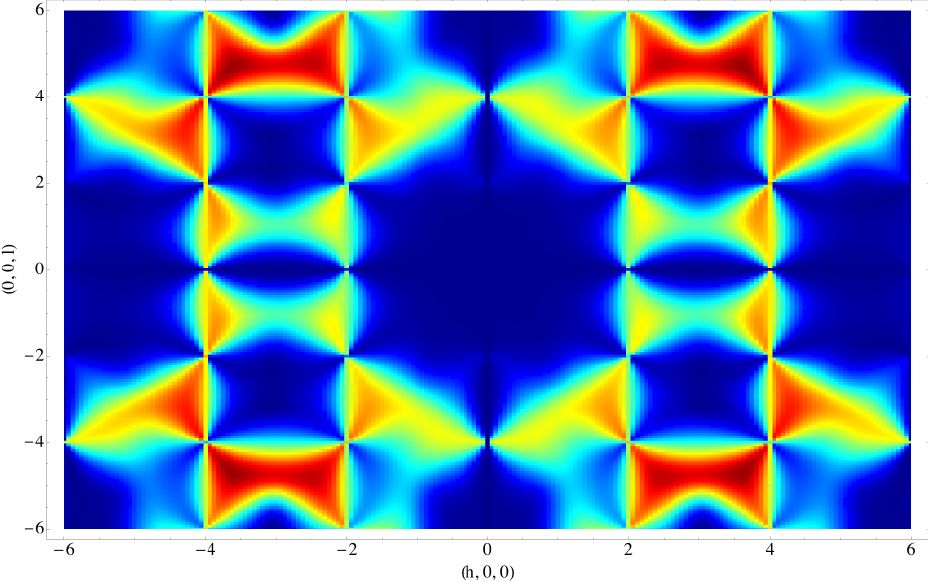}}%
\caption{
(Color online).
Prediction for diffuse scattering of neutrons or X--rays from protons in a classical
water ice described by the Bernal--Fowler ``ice rules''.  
The structure factor $S^{\sf diffuse}_{\sf proton}(\mathbf{q})$~[Eq. (\ref{eq:SHdiffuse})], 
is plotted in three orthogonal planes in reciprocal space.
The ice rules manifest themselves as both ``pinch--points'' --- singular features  
in scattering, visible at e.g. ${\bf Q}^*_{\sf p,H^+} = (0,0,4)$ in (c) 
--- and broad, asymmetric features at zone centers, visible at e.g. 
${\bf Q}^*_{\sf m,H^+} = (2,0,3)$ in (c). 
%
%
Results are shown for the theory described in Section~\ref{subsec:classical.predictions}
and Appendix~\ref{appendix:projection}, with reciprocal--lattice vectors  indexed to the orthorhombic 
unit cell defined in Appendix~\ref{appendix:lattice}, following the conventions of Nield and Whitworth~\cite{nield95}.
} 
\label{fig:Sq.proton.classical}
\end{figure*}

\subsection{Predictions for scattering from protons in a classical model of ice Ih}
\label{subsec:classical.predictions}

X--ray and neutron scattering experiments on water ice do not measure 
the structure factor for bond--variables $S_{\sf Ising}(\mathbf{q})$~[Eq.~(\ref{eq:S.of.q.Ising})], 
discussed in Section~\ref{subsec:classical.lattice.theory}, but rather the Fourier transform of the 
correlation function for the density of protons
\begin{eqnarray}
S_{\sf proton} (\mathbf{q}) = \langle n(-\mathbf{q}) n(\mathbf{q}) \rangle \; .
\end{eqnarray}
Information about the proton disorder is contained in
the diffuse part of this scattering, 
which is given by \cite{schneider80} 
\begin{eqnarray}
S^{\sf diffuse}_{\sf proton} (\mathbf{q})
  = \sum_{\nu\nu'} \langle \sigma_\nu (-\mathbf{q}) \sigma_{\nu'}(\mathbf{q}) \rangle
\sin(\mathbf{q} \cdot \mathbf{a}_\nu) \sin(\mathbf{q} \cdot \mathbf{a}_{\nu'}) \; ,
\nonumber\\
\label{eq:SHdiffuse}
\end{eqnarray}
where $\sigma_\nu (\mathbf{q})$ is given in Eq.~(\ref{eq:sigma.q}), and ${\bf a}_\nu$ 
is a set of vectors, defined in Appendix~\ref{appendix:lattice}, such that the 
displacement of a proton from the midpoint on any given bond $\nu$ is
\begin{eqnarray}
{\bf D}_\nu = \sigma_\nu {\bf a}_\nu \; ,
\end{eqnarray}
with
\begin{eqnarray}
| {\bf a}_\nu | \sim 0.15 \ a_{\sf O}   \; .
\end{eqnarray}


In Fig.~\ref{fig:Sq.proton.classical} we show the results for diffuse 
scattering from protons in a classical model of 
ice Ih.  
The structure factor $S^{\sf diffuse}_{\sf proton}(\mathbf{q})$~[Eq.~(\ref{eq:SHdiffuse})], 
was calculated using the lattice theory introduced in Section~\ref{subsec:classical.lattice.theory}.   
At small momentum transfers, the scattering 
is suppressed by the factors of $\sin(\mathbf{q}\cdot {\bf a}_\nu)$ in Eq.~(\ref{eq:SHdiffuse}), 
and as a result, there is essentially no scattering for $|\mathbf{q}| \lesssim 2$.  
For larger wave number, scattering shows a mixture of broad and 
sharp features, centered on two different sets of reciprocal lattice vectors. 
An example of broad feature can be seen near to ${\bf Q} = (2, 0, 3)$
in Fig.~\ref{fig:Sq.proton.classical}~(c).  
An example of a sharp feature --- a pinch--point --- can be seen near to ${\bf Q} = (4, 0, 0)$
in Fig.~\ref{fig:Sq.proton.classical}~(a) and (c).  


We can relate both broad and sharp features  
to the correlations of the 
classical fields ${\bf P}_+$ and ${\bf P}_-$, introduced in Section~\ref{subsec:classical.fields}.
Expanding the structure factor 
$S^{\sf diffuse}_{\sf proton} (\mathbf{q})$ [Eq.~(\ref{eq:SHdiffuse})]
about the reciprocal lattice vector ${\bf Q}$, for 
$ | \tilde{\mathbf{q}} | \ll 1 $, we find 
\begin{eqnarray}
&& S^{\sf diffuse}_{\sf proton} (\mathbf{Q} + \tilde{\mathbf{q}}) 
\nonumber\\
&&  \quad \approx 
     \sum_{\upsilon = \pm}
          F^{\sf proton}_\upsilon (\mathbf{Q})
         \langle 
              | \boldsymbol{\hat{\lambda}}_{\mathbf{Q},\upsilon}^{\sf proton} 
              \cdot {\bf P}_\upsilon(\tilde{\mathbf{q}}) |^2 
         \rangle \; ,
\label{eq:SIsing-Pcorr}
\end{eqnarray}
where the form--factors 
$F^{\sf proton}_{\mathbf{Q},\upsilon}$~[Eq.~(\ref{eq.F.Q.upslion})]
and vectors 
$\boldsymbol{\hat{\lambda}}_{\mathbf{Q},\upsilon}^{\sf proton}$~[Eq.~(\ref{eq:defn-proton-lambda})]
are defined in Appendix~\ref{appendix:structure.factors.and.fields},
and the Fourier transform 
${\bf P}_\upsilon(\tilde{\mathbf{q}})$ [Eq.~(\ref{eq:FT.P})] 
in Appendix~\ref{appendix:projection}.



Once again, there are a subset of reciprocal lattice vectors ${\bf Q}_{\sf p, H^+}$
for which 
\begin{eqnarray}
F^{\sf proton}_- ( \mathbf{Q}_{\sf p, H^+} ) \equiv 0 \; ,
\end{eqnarray}
and correlations are controlled by the zero--divergence condition 
on ${\bf P}_{+}$ [Eq.~(\ref{eq:p+constraint})].   
It follows from from Eq.~(\ref{eq:p+corr}) and Eq.~(\ref{eq:FT.P}) that 
\begin{eqnarray}
&&S^{\sf diffuse}_{\sf proton} ( \mathbf{Q}_{\sf p, H^+} + \tilde{\mathbf{q}} )
\nonumber \\
&& \quad \approx
   \frac{ F^{\sf proton}_+ ( \mathbf{Q}_{\sf p, H^+} )}{\kappa_+}
       \left( 1 -  \frac{ | \boldsymbol{\hat{\lambda}}_{\mathbf{Q}_{\sf p, H^+} , +}^{\sf proton}
              \cdot \tilde{\mathbf{q}} |^2}{|\tilde{\mathbf{q}}|^2 }
   \right) \nonumber \\
\end{eqnarray}


Considering the specific example of 
\begin{eqnarray}
{\bf Q}^*_{\sf p,H^+} = (4, 0, 0) \, 
\end{eqnarray}
for which 
\begin{eqnarray}
   \boldsymbol{\hat{\lambda}}_{\mathbf{Q}_{\sf p, H^+},\upsilon}^{\sf proton} &=& (1, 0, 0) \, ,
\end{eqnarray}
we have
\begin{eqnarray}
&& S^{\sf diffuse}_{\sf proton} ( \mathbf{Q}_{\sf p, H^+} + \tilde{\mathbf{q}} )
\propto \left( 1-\frac{\tilde{q}_x^2}{\tilde{q}^2} \right) \, .
\label{eq:pinchpointform}
\end{eqnarray}
A pinch--point singularity of this form is clearly visible near $ \mathbf{Q} = (4,0,0)$ 
in Fig.~\ref{fig:Sq.proton.classical}~(a) and (c).   


Similarly, while there are no reciprocal lattice vectors for which 
$F^{\sf proton}_+ ( \mathbf{Q} )$ vanishes identically while 
$F^{\sf proton}_- ( \mathbf{Q} )$ remains finite,
there are another 
set of lattice centers ${\bf Q}_{\sf m, H^+}$, for which
\begin{eqnarray}
F^{\sf proton}_+ ( \mathbf{Q}_{\sf m, H^+} ) \ll F^{\sf proton}_- ( \mathbf{Q}_{\sf m, H^+} )
\end{eqnarray}
and scattering from protons reflects the short--ranged correlations of ${\bf P}_-$ [Eq.~(\ref{eq:p-constraint})].
An example of this type scattering occurs for 
\begin{eqnarray}
{\bf Q}^*_{\sf m, H^+} = (0, 4, 1) \, .
\end{eqnarray}
A broad, asymmetric feature, centred on this reciprocal lattice vector, 
is clearly visible in Fig.~\ref{fig:Sq.proton.classical}~(b).   


For a more general choice of zone centre, $\mathbf{Q}_{\sf pm, H^+}$, 
\begin{eqnarray}
F^{\sf proton}_+ ( \mathbf{Q}_{\sf pm, H^+} ) 
   \approx F^{\sf proton}_- ( \mathbf{Q}_{\sf pm, H^+} ) \; ,
\end{eqnarray}
and scattering reflects the correlations of both ${\bf P}_+$ and ${\bf P}_-$.
An example of this type scattering occurs for 
\begin{eqnarray}
{\bf Q}^*_{\sf pm, H^+} = (0, 4, 3) \, .
\end{eqnarray}
A combination of pinch points and broad, assymetric features can be seen
near to this reciprocal lattice vector in Fig.~\ref{fig:Sq.proton.classical}~(b).  


We conclude this discussion with a brief word of caution ---  in some cases, the scattering from 
protons $S^{\sf diffuse}_{\sf proton}(\mathbf{q})$~[Eq.~(\ref{eq:SHdiffuse})] 
exhibits pinch--points at the same reciprocal lattice vectors as pinch--points in 
the structure factor $S_{\sf Ising}(\mathbf{q})$~[Eq. (\ref{eq:S.of.q.Ising})] --- as can be 
seen by comparing Fig.~\ref{fig:Sq.Ising.classical} and Fig.~\ref{fig:Sq.proton.classical}.
However, in general  
\begin{eqnarray}
F^{\sf proton}_- ( \mathbf{Q} ) = 0 
   \quad \not\Leftrightarrow \quad 
F^{\sf Ising}_- ( \mathbf{Q})  = 0 \, ,
\end{eqnarray}
and pinch--points in $S^{\sf diffuse}_{\sf proton}(\mathbf{q})$ 
do {\it not}, necessarily, occur at the same reciprocal lattice vectors as pinch--points 
in $S_{\sf Ising}(\mathbf{q})$.   

We will present a detailed comparison of these
results with diffuse neutron scattering 
experiments on ice Ih in Section \ref{subsection:experiment-classical}.


\section{Proton tunnelling and emergent photons in a quantum model of ice Ih}
\label{section:photons}

The analogy between the ice--rules and electromagnetism, underpinning
the classical analysis of Section~\ref{section:classical}, becomes 
complete once quantum effects are taken into account.   
In what follows, we show that the minimal model for quantum effects in ice Ih, 
$\mathcal{H}_{\sf tunnelling}^{\sf hexagonal}$~[Eq.~(\ref{eq:htunnelling2})], 
leads to a compact, frustrated quantum ${\sf U(1)}$ lattice--gauge theory, 
with precisely the form of electromagnetism on a lattice.   
We explore the new features of a proton liquid described by such a theory, 
and make explicit predictions for both inelastic and quasi--elastic 
(energy--integrated) scattering of X--rays or neutrons 
from disordered protons in a quantum ice Ih.
This discussion proceeds as follows~: 


Firstly, in Section~\ref{subsection:lattice.gauge.theory} we outline the derivation 
of this lattice--gauge theory.
Technical details of these calculations are provided 
in Appendix~\ref{appendix:photons}.   


Then, in Section~\ref{subsection:phenomenology} we explore some of the features of
lattice gauge theory in its deconfined (proton--disordered) phase.   
In particular, we show how its low--energy excitations can be thought of
as  the linearly---dispersing, birefringent ``photons'', and how these relate
to the classical fields ${\bf P}_{+}$ and ${\bf P}_{-}$ introduced 
in Section~\ref{section:classical}.


Finally, in Section~\ref{subsection.quantum.water.ice} we discuss the experimental 
signatures of quantum water ice, described by the deconfined phase of 
the lattice gauge theory.

\subsection{Lattice--gauge theory}
\label{subsection:lattice.gauge.theory}

Our route to a lattice--gauge theory of ice Ih closely parallels the 
cubic--symmetry case previously considered by Hermele~{\it et al.}~\cite{hermele04},  
and Benton~{\it et al.}~\cite{benton12}.  
The theory itself, however, contains  a number of new features.   


We begin introducing a set of pseudospin--$1/2$ operators $S^z, S^+, S^-$ 
defined on the bonds $\mathbf{r} \mathbf{r}'$ of the oxygen lattice.  
The $z$--component of the pseudo-spin is directly proportional
to the Ising variable $\sigma_{\mathbf{r} \mathbf{r}'}$ [Eq.~(\ref{eq:ising})], 
introduced to describe proton--correlations in the classical case
\begin{eqnarray}
S^z_{\mathbf{r} \mathbf{r}'}=\frac{1}{2} \sigma_{\mathbf{r} \mathbf{r}'} = -S^z_{\mathbf{r}' \mathbf{r} } \; .
\label{eq:Sz}
\end{eqnarray}
In keeping with the directedness of $S^z_{\mathbf{r} \mathbf{r}'}$ [Eq. (\ref{eq:Sz})]
the ladder operators obey the identity
\begin{eqnarray}
S_{\mathbf{r}\mathbf{r}'}^+=S_{\mathbf{r}'\mathbf{r}}^-
\end{eqnarray}


The minimal quantum model for ice Ih 
$\mathcal{H}_{\sf tunnelling}^{\sf hexagonal}$~[Eq.~(\ref{eq:htunnelling2})], 
can be expressed in terms of these operators as 
\begin{eqnarray}
&&\mathcal{H}_{\sf tunnelling}^{\sf hexagonal} 
   = - g_1 \sum_{\hexagon \in I}  
       [S^+_1 S^-_2 S^+_3S^-_4 S^+_5 S^-_6 + \text{h.c.}] 
       \nonumber \\
&& \qquad \qquad \quad  - g_2 \sum_{\hexagon \in II}  
     [S^+_1 S^-_2 S^+_3 S^-_4 S^+_5 S^-_6 + \text{h.c.}] \; . 
\label{eq:htunnelling3}
\end{eqnarray}
A mapping to a $U(1)$ lattice--gauge theory is then possible by writing the
spin-$1/2$ operators in a quantum rotor representation 
\cite{hermele04, castro-neto06, savary12}
\begin{eqnarray}
S^z_{\mathbf{r} \mathbf{r}'} \to E_{\mathbf{r} \mathbf{r}'} 
\quad S^{\pm}_{\mathbf{r} \mathbf{r}'} \to e^{\pm i A_{\mathbf{r} \mathbf{r}'}} \; ,
\label{eq:quantumrotor} 
\end{eqnarray}
subject to the canonical commutation relation 
\begin{eqnarray}
\left[ E_{\mathbf{r} \mathbf{r}'}, A_{\mathbf{r}'' \mathbf{r}'''} \right] 
   = i 
(\delta_{\mathbf{r} \mathbf{r}''}\delta_{\mathbf{r'} \mathbf{r}'''} 
-
\delta_{\mathbf{r} \mathbf{r}'''}\delta_{\mathbf{r'} \mathbf{r}''} 
)\; .
\label{eq:EAcomm}
\end{eqnarray}
The commutation relation Eq.~(\ref{eq:EAcomm}) is familiar
in quantum electromagnetism as the commutation between an electric
field ${\bf E}$ and a vector potential ${\bf A}$, and substituting
the rotor representation Eq.~(\ref{eq:quantumrotor}) into Eq.~(\ref{eq:htunnelling3})
results in a compact $U(1)$ gauge theory 
\begin{eqnarray}
&&\mathcal{H}_{\sf U(1)}^{\sf compact} = 
- 2g_1 \sum_{\hexagon \in I}  
     \cos( \left[ \nabla_{\scriptsize\hexagon} \times A \right]) \nonumber \\
&& \qquad \qquad \qquad
  - 2 g_2 \sum_{\hexagon \in II}  
   \cos( \left[ \nabla_{\scriptsize\hexagon} \times A \right]) \; . 
\end{eqnarray}
where the sum $\sum_{\langle \mathbf{r} \mathbf{r}' \rangle \in  {\sf CS}}$ runs over 
centre--symmetric oxygen--oxygen bonds, $\sum_{\langle \mathbf{r} \mathbf{r}' \rangle \in  {\sf MS}}$ 
runs over mirror--symmetric bonds [cf.~Fig.~\ref{fig:bondsymmetries}] 
and $\left[  \nabla_{\scriptsize\hexagon} \times A  \right]$ represents the lattice curl of 
$A_{\bf r r'}$ around a hexagonal plaquette, which may be of type--I or type--II [cf.~Fig.~\ref{fig:loops}].

The electric field $E_{\mathbf{r} \mathbf{r}'}$ subject to the constraint that 
\begin{eqnarray}
E_{\mathbf{r} \mathbf{r}'}=\pm\frac{1}{2} \quad \forall \ \ \ \text{bonds} \ \mathbf{r} \mathbf{r}' \; .
\label{eq:Econstraint}
\end{eqnarray}


Following~\cite{hermele04, benton12}, one may then argue that averaging over 
fast fluctuations of the gauge field softens the constraint Eq.~(\ref{eq:Econstraint}), 
and leads to a non--compact gauge theory on the links of the ice Ih lattice
\begin{eqnarray}
&&\mathcal{H}_{\sf U(1)}
 = \frac{\mathcal{U}}{2} \sum_{\langle \mathbf{r} \mathbf{r}' \rangle \in  {\sf CS}}
E_{\mathbf{r} \mathbf{r}'}^2 
 +\frac{\mathcal{U^{\prime}}}{2} \sum_{\langle \mathbf{r} \mathbf{r}' \rangle \in {\sf MS}}
E_{\mathbf{r} \mathbf{r}'}^2  \nonumber \\
 && \qquad + \frac{\mathcal{K}}{2} \sum_{\hexagon \in I}
   \left[  \nabla_{\scriptsize\hexagon} \times A  \right]^2 
   + \frac{\mathcal{K^{\prime}}}{2} \sum_{\hexagon \in II}
   \left[ \nabla_{\scriptsize\hexagon} \times A \right]^2 
\label{eq:HU1}
\end{eqnarray}
The parameters $\mathcal{U}$ and $\mathcal{U}'$ may be thought of as Lagrange multipliers
fixing the average value of $E_{\mathbf{r} \mathbf{r}'}^2$ on the two inequivalent types of bond.
The average over fast fluctuations will in general renormalise $\mathcal{K}$ and $\mathcal{K}'$
from their ``bare'' values 
\begin{eqnarray}
\mathcal{K}_0=2 g_1 \qquad \mathcal{K}_0'=2g_2.
\end{eqnarray}


On general grounds~\cite{kogut79,castro-neto06}, and by analogy with quantum spin 
ice~\cite{hermele04,banerjee08,shannon12}, we anticipate that this lattice--gauge theory 
will possess both a deconfined phase, in which the protons form a disordered quantum fluid, 
and confined phase(s), in which the protons order.   
In what follows we confine our discussion to the deconfined phase,
without attempting to characterise any competing fixed points.


Even with this restricted goal, 
the validity of $\mathcal{H}_{\sf U(1)}$~[Eq.~(\ref{eq:HU1})] depends critically on the 
assumptions made in passing to a non--compact gauge theory.
While these assumptions are reasonable, they can ultimately only be validated through 
quantum Monte Carlo simulation of the microscopic model 
$\mathcal{H}_{\sf tunnelling}^{\sf hexagonal}$~[Eq.~(\ref{eq:htunnelling3})] 
--- cf. [\onlinecite{sikora09,sikora11,shannon12,benton12}].
We have therefore used variational quantum Monte Carlo (VMC) 
simulation of $\mathcal{H}_{\sf U(1)}$ to establish that the deconfined phase 
of the lattice gauge theory, $\mathcal{H}_{\sf U(1)}$~[Eq.~(\ref{eq:HU1})], 
closely describes the correlations of the microscopic model 
$\mathcal{H}_{\sf tunnelling}^{\sf hexagonal}$~[Eq.~(\ref{eq:htunnelling3})], 
for the symmetric choice of parameters $g_1 = g_2$.   
These results are presented in Appendix~\ref{appendix:numerics}.


In principle, it is also possible to extract the parameters of 
of the lattice gauge theory --- $\mathcal{U}$, $\mathcal{U}^\prime$, 
$\mathcal{K}$ and $\mathcal{K}^\prime$ --- 
from detailed quantum Monte Carlo simulation of the microscopic model 
$\mathcal{H}_{\sf tunnelling}^{\sf hexagonal}$~[Eq.~(\ref{eq:htunnelling3})]
as a function of $g_1$ and $g_2$ --- cf. Ref.~[\onlinecite{benton12}].
However since the purpose of this Article is to explore the properties of the 
deconfined phase, and no reliable estimates are yet available for $g_1$ and $g_2$
in real water ice, we will continue to treat $\mathcal{U}$, $\mathcal{U}^\prime$, 
$\mathcal{K}$ and $\mathcal{K}^\prime$ as phenomenological parameters.


\begin{figure*} 
\centering
\includegraphics[width=\textwidth]{colorbar.png}
\\
\includegraphics[width=\textwidth]{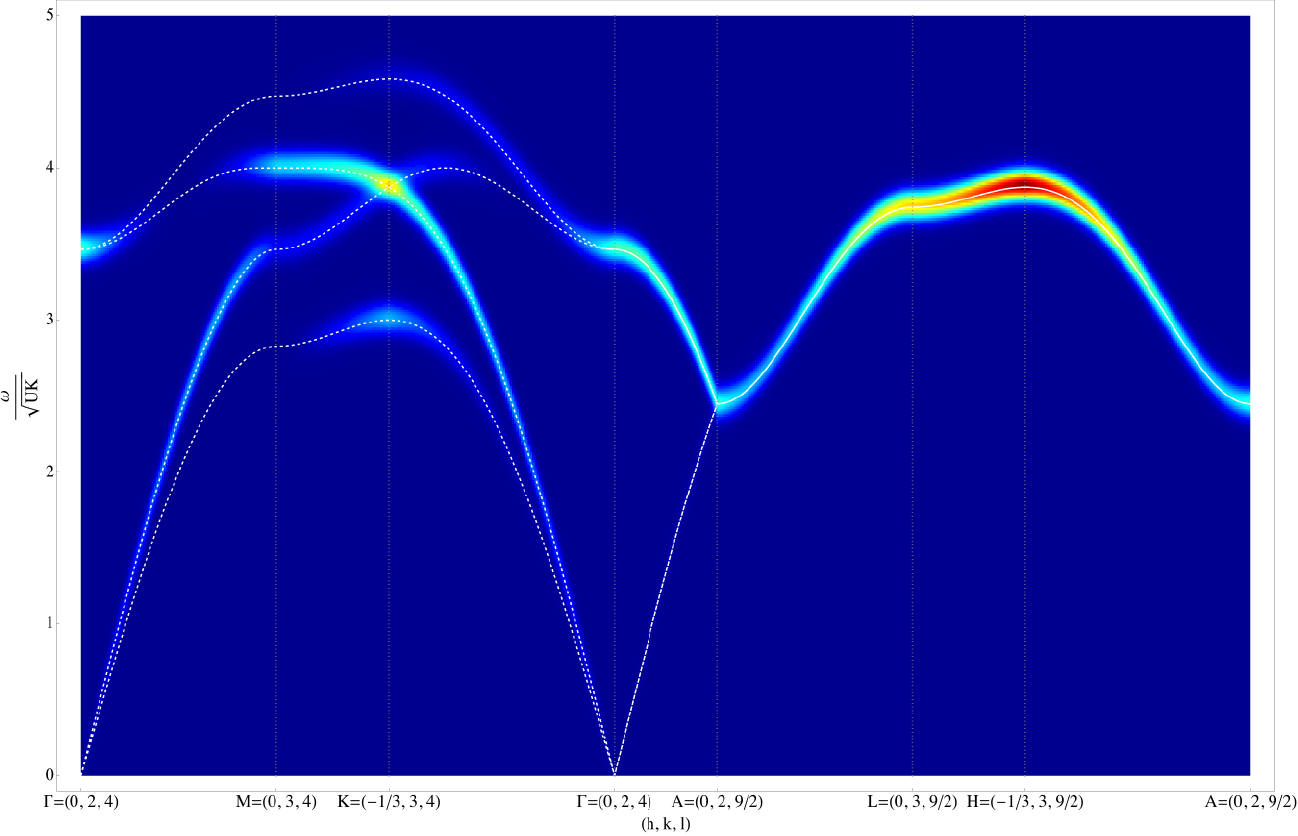}
\caption{
(Color online).
Prediction for inelastic scattering from disordered protons in a quantum model of ice Ih at 
zero temperature ($T=0$).   
Results for the dynamical structure factor 
$S_{\sf coh}^{\sf H^+}(\mathbf{q}, \omega)$~[Eq.~(\ref{eq:SH-definition})] 
are shown in three, orthogonal, planes in reciprocal space.
The ``photons'' of the lattice gauge theory $\mathcal{H}_{\sf U(1)}$~[Eq.~(\ref{eq:HU1})] 
are visible as gapless, linearly--dispersing excitations at long--wavelength.
These photons are birefringent, with a dispersion which depends 
on the polarisation of the photon, except along the optical axis, $\mathbf{z}$. 
As a result two bands of photons are visible on the path $\Gamma$--$M$--$K$--$\Gamma$, 
while only one appears on the path $\Gamma$--$A$--$L$--$H$--$A$.   
Additional spectral weight at higher energies is associated with 
gapped, exponentially--correlated excitations of the gauge--field.
The dynamical structure factor for coherent scattering from protons 
$S_{\sf coh}^{\sf H^+}(\mathbf{q}, \omega)$~[Eq.~(\ref{eq:SH-definition})] 
was calculated using the ${\sf U(1)}$ lattice gauge theory 
$\mathcal{H}_{\sf U(1)}$~[Eq.~(\ref{eq:HU1})], 
Results are shown for the parameter set given in Eq.~(\ref{eq:plain-vanilla-parameter-set}),  
and were convoluted with a Gaussian of $\text{FWHM} = 0.035 \sqrt{\mathcal{U}\mathcal{K}}$ 
to mimic the effect of experimental resolution.   
The color scale shows the intensity of the modes as they would appear in an
inelastic scattering experiment, in the Brillouin zone centred
on $(h, k, l)=(0, 2, 4)$.
}
\label{fig:photon-dispersion}
\end{figure*}

\subsection{Phenomenology of the deconfined phase : Why these photons ?}
\label{subsection:phenomenology}

In the absence of charges, the defining characteristic of the deconfined phase of the 
$U(1)$ lattice gauge theory, $\mathcal{H}_{\sf U(1)}$~[Eq.~(\ref{eq:HU1})], is its ``photon'', 
a transverse excitation of the gauge field $A$, with definite polarization and linear dispersion 
at long wavelength.   
Since $\mathcal{H}_{\sf U(1)}$ is quadratic in $A$, it can be solved by introducing 
a suitable basis for transverse fluctuations of the gauge--field.
This calculation is explained in detail in Appendix~\ref{appendix:photons}, 
following the methods described in Ref.~[\onlinecite{benton12}].
Here we concentrate instead on using this solution of $\mathcal{H}_{\sf U(1)}$
to describe the new features which arise from the tunnelling of protons in water ice.


In Fig.~\ref{fig:photon-dispersion}, we show the dispersion of the excitations of 
$\mathcal{H}_{\sf U(1)}$~[Eq.~(\ref{eq:HU1})], as they would appear in an inelastic 
neutron scattering experiment on ice Ih.
The dynamical structure factor for coherent scattering from protons,  
$S_{\sf coh}({\bf q},\omega)$~[Eq.~(\ref{eq:Scoh-definition})],  
was calculated for the symmetric choice of parameters 
\begin{eqnarray}
\mathcal{U}=\mathcal{U}' \; ,  \; \mathcal{K}=\mathcal{K}' \; ,
\label{eq:plain-vanilla-parameter-set}
\end{eqnarray}
and the dispersion has been normalised to the characteristic energy--scale of
the lattice--gauge theory, $\sqrt{\mathcal{U}\mathcal{K}}$.
Within this normalisation, the excitations have an overall bandwidth
\begin{eqnarray}
\frac{\Delta \omega}{\sqrt{\mathcal{UK}}} 
= f \left( \frac{\mathcal{U}'}{\mathcal{U}}, \frac{\mathcal{K}'}{\mathcal{K}}
 \right)  \; ,
\label{eq:bandwidth1}
\end{eqnarray}
where, for this parameter set
\begin{eqnarray}
f(1,1) \approx 4.56 \nonumber \; .
\end{eqnarray}
On the basis of published simulations for quantum 
spin-ice~\cite{banerjee08,benton12,kato-arXiv}, it is reasonable to expect that the bandwidth 
of the excitations of the gauge theory, $\Delta \omega$, should be of the same order of 
magnitude as the quantum tunnelling $g$.   


The excitations shown in Fig.~\ref{fig:photon-dispersion} possess 
a number of striking features, specific to ice Ih.
At low energies, the model supports two, linearly--dispersing modes, with 
intensity which vanishes linearly approaching zero energy.
These are the emergent ``photons'' of the lattice--gauge theory, and the fact 
that there are two such modes reflects the two possible polarisations of the photon.
The vanishing intensity of the photons at low energy is a feature
shared with the emergent photons of (cubic) quantum spin ice, and reflects the fact 
that neutrons scatter from fluctuations of the proton density, and not directly 
from the underlying gauge field~\cite{benton12}.
However, in contrast with the cubic case, 
the emergent photons of quantum 
ice Ih are {\it birefringent}, i.e. they have
 a dispersion which depends on
the polarisation of the photon.
The splitting of these two modes is clearly visible in Fig.~\ref{fig:photon-dispersion}, 
except for wavevectors parallel to the hexagonal symmetry axis of the crystal, 
where they are degenerate.


A second feature of note is the presence of gapped, optical modes.
These are clearly visible  in Fig.~\ref{fig:photon-dispersion} at the zone center, $\Gamma$, 
at energies above the photon dispersion.   
There two such modes, and away from high symmetry points, they are generally 
non--degenerate.  
The presence of these optical modes distinguishes quantum ice Ih from quantum spin ice, 
where the pure gauge theory only supports photons~\cite{benton12}.  


\begin{figure*}
\centering
\includegraphics[width=0.7\textwidth]{colorbar.png} \\
\subfigure[\ \ hk0 plane]{
\includegraphics[width=.5195\textwidth]{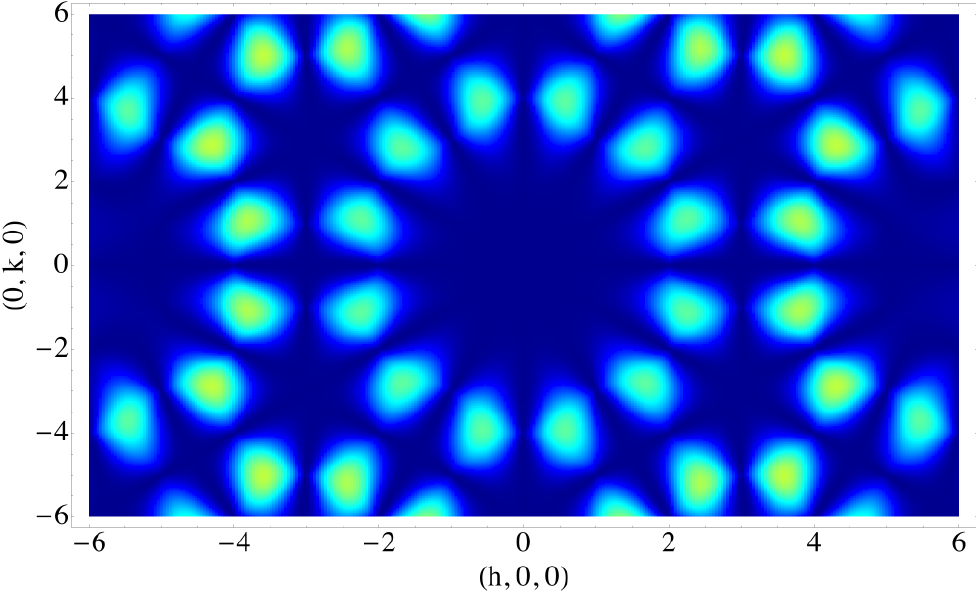}}%
    \qquad
\subfigure[\ \ 0kl plane]{%
\includegraphics[width=.3\textwidth]{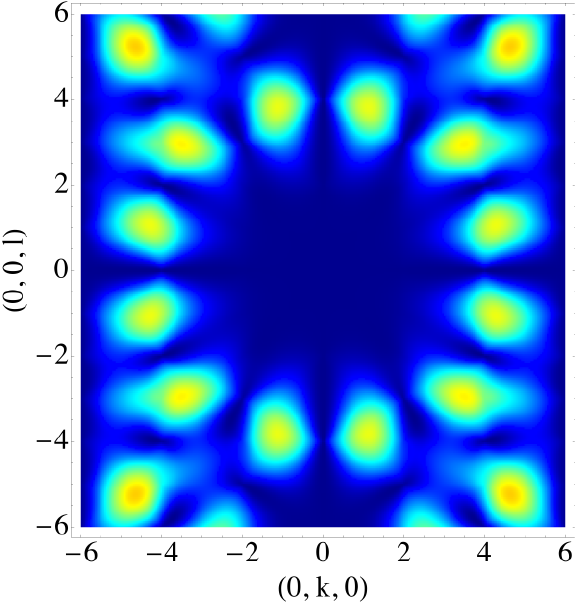}}%
\qquad
\subfigure[\ \ h0l plane]{%
\includegraphics[width=.501962\textwidth]{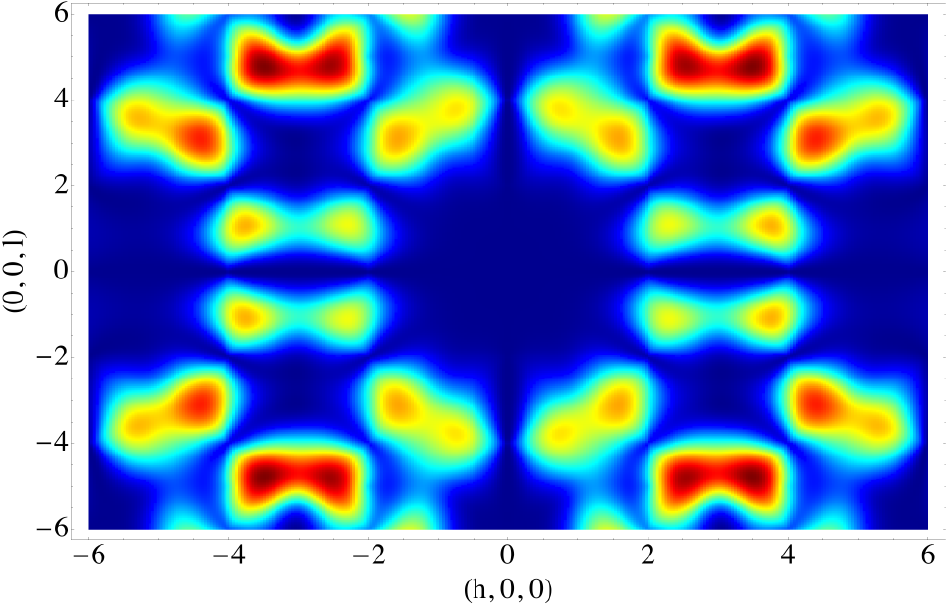}}%
\caption{
(Color online).
Prediction for coherent, quasi-elastic scattering from protons in a quantum 
model of ice Ih at zero temperature ($T=0$).  
The ``pinch--points'' associated with the ice--rules [cf. Fig.~\ref{fig:Sq.proton.classical}], 
are eliminated by quantum fluctuations [cf.~Ref.~\cite{shannon12}].
At finite temperatures, these pinch--points will be restored with a weight 
linear in $T$~[\onlinecite{benton12}].  
Results are shown for the energy--integrated, equal--time structure factor 
\mbox{$S^{\sf H^+}_{\sf coh}(\mathbf{q}, t=0)$}~[Eq.~\ref{eq:equaltimeprotoncorr}], 
calculated within the lattice--gauge theory $\mathcal{H}_{\sf U(1)}$~[Eq.~(\ref{eq:HU1})], 
for the same parameter set [Eq.~(\ref{eq:plain-vanilla-parameter-set})] 
as Fig.~\ref{fig:photon-dispersion}.  
Reciprocal--lattice vectors are indexed to the orthorhombic unit cell 
defined in Appendix~\ref{appendix:lattice}, following the conventions 
of Nield and Whitworth~\cite{nield95}.
} 
\label{fig:XRDSq-quantum}
\end{figure*}


We wish to emphasize that the symmetric choice of parameters, 
Eq.~(\ref{eq:plain-vanilla-parameter-set}), is made purely for illustrative purposes, 
and that --- as long as the lattice--gauge theory remains in its deconfined 
phase --- a more general choice of parameters 
will lead to qualitatively the same behaviour.
In fact, as we shall show, the signature features of quantum ice Ih 
--- the birefringence of the emergent photons, and the presence of gapped 
optical modes --- are strongly constrained by symmetry, and follow naturally 
from the quantisation of the two classical fields introduced 
in Section~\ref{section:classical}, ${\bf P}_+$ and ${\bf P}_-$.


The correspondence between these classical fields, and the 
excitations of the lattice--gauge theory $\mathcal{H}_{\sf U(1)}$~[Eq.~(\ref{eq:HU1})], 
is a consequence of the fact that both theories automatically respect the ice--rules.
It follows that constraints on the classical fields, 
Eq. (\ref{eq:p+constraint}) and Eq. (\ref{eq:p-constraint}), remain valid in the 
presence of quantum tunnelling $\mathcal{H}_{\sf tunnelling}^{\sf hexagonal}$~[Eq.~(\ref{eq:htunnelling2})], 
and that the long--wavelength dynamics of the quantum model can be found
by quantizing the fluctuations of the classical fields.


Let us consider first the case of ${\bf P}_+$.
The constraint, Eq.~(\ref{eq:p+constraint}), can be enforced by writing
\begin{eqnarray}
{\bf P}_+=\nabla \times {\bf A}'
\end{eqnarray}
where it is important to distinguish the course--grained field ${\bf A}'$ 
from the microscopic field $A_{\mathbf{r} \mathbf{r}'}$, entering into  
the lattice gauge theory $\mathcal{H}_{\sf U(1)}$.
The form of the Lagrangian describing
fluctuations of ${\bf A}'$ is then the Maxwell Lagrangian,
subject to the hexagonal symmetry of the lattice.
Choosing the Coulomb gauge
\begin{eqnarray}
\nabla \cdot {\bf A}'=0
\end{eqnarray}
this is given by 
\begin{eqnarray}
\mathcal{L}_{{\bf P}_+}
&=&\frac{1}{2}
\int dt \int d^3{\bf r} \sum_{\alpha \beta} \nonumber \\
&&\left[
   \rho_{\alpha \beta} \partial_t A^{'\alpha} \partial_t A^{'\beta}
   - \nu_{\alpha \beta} (\nabla \times A')^{\alpha} (\nabla \times A')^{\beta}
\right] \; , \nonumber \\
\label{eq:LMaxwell-hex}
\end{eqnarray}
where the tensors $\rho_{\alpha \beta}$ and $\nu_{\alpha \beta}$ are diagonal 
in the crystal basis, and have the form
\begin{eqnarray}
\rho &=&
\begin{pmatrix}
\rho_{\perp} & 0 & 0 \\
0 & \rho_{\perp} & 0 \\
0 & 0 & \rho_{z} 
\end{pmatrix}  \; ,\\
\nu &=&
\begin{pmatrix}
\nu_{\perp} & 0 & 0 \\
0 & \nu_{\perp} & 0 \\
0 & 0 & \nu_{z}
\end{pmatrix} \; .
\end{eqnarray}
As a consequence, the photons described by $\mathcal{L}_{{\bf P}_+}$ [Eq.~(\ref{eq:LMaxwell-hex})] 
are degenerate when propagating with momentum parallel to the crystallographic ${\bf z}$-axis, with dispersion 
\begin{eqnarray}
\omega = \sqrt{\frac{\nu_{\perp}}{\rho_{\perp}}} |k| \; .
\end{eqnarray}
However, they are non-degenerate when propagating in the plane perpendicular 
to the crystallographic ${\bf z}$-axis, with dispersion
\begin{eqnarray}
\omega_1 = \sqrt{\frac{\nu_{\perp}}{\rho_{z}}} |k| \; , \quad
\omega_2 = \sqrt{\frac{\nu_{z}}{\rho_{\perp}}} |k| \; .
\end{eqnarray}
which depends on the polarisation of the photon.  


We now turn to the field ${\bf P}_-$.
This has exponentially, rather than algebraically, decaying correlations 
in the classical limit [Eq.~(\ref{eq:p-perpcorr}) and Eq.~(\ref{eq:p-zcorr})], 
and no associated gauge symmetry.
However, because of the form of the constraint Eq. (\ref{eq:p-constraint}), fluctuations 
of ${\bf P}_-$ can described by a Lagrangian written in terms of its planar component ${\bf P}^\perp_-$~:
\begin{eqnarray}
\mathcal{L}_{{\bf P}_-}
&=&
\frac{1}{2}
\int dt \int d^3\mathbf{r} \left[
\gamma (\partial_t {\bf P}^{\perp}_-)^2 - \Delta ( {\bf P}^{\perp}_-)^2 \right. \nonumber \\
&& \left.
- \sum_{\mu \nu=x,y,z} \sum_{\alpha \beta= x,y} \epsilon_{\alpha \beta}^{\mu \nu}
\partial_{\mu} P^{\perp, \alpha}_- \partial_{\nu} P^{\perp, \beta}_-
\right] \; ,
\end{eqnarray}
with the $z$-component of ${\bf P}_-$ being fixed by Eq.~(\ref{eq:p-constraint}).  
Once again, the form of the tensor $\epsilon_{\alpha \beta}^{\mu \nu}$ is 
dictated by the symmetry of the lattice.  
For a general choice of $\epsilon_{\alpha \beta}^{\mu \nu}$,  we find that Lagrangian $\mathcal{L}_{{\bf P}_-}$ 
supports two gapped modes, which become degenerate approaching $\mathbf{q}=0$ 
\begin{eqnarray}
\omega(\mathbf{q} \to 0) = \sqrt{\frac{\Delta}{\gamma}} \; .
\end{eqnarray}


In the light of this analysis, at long wavelength, we can identify the gapless, birefringent 
emergent photons of $\mathcal{H}_{\sf U(1)}$~[Eq.~(\ref{eq:HU1})] with quantised fluctuations 
of ${\bf P}_+$, and its gapped optical modes with quantised fluctuations of  ${\bf P}_-$.
This correspondence is demonstrated explicitly in Appendix \ref{appendix:photons}.

\subsection{Predictions for diffuse, coherent inelastic neutron scattering
at low temperatures}
\label{subsection.quantum.water.ice}

The formation of the quantum liquid state at $T=0$ would have
profound consequences for scattering experiments.
In this section we will briefly comment on what one could
expect to observe in a measurement of the coherent scattering
from water ice, in the scenario where quantum tunnelling of
protons leads to the formation of a fluctuating $U(1)$ liquid state.
We shall postpone a consideration of the consequences for
measurements of {\it incoherent}
scattering, such as that performed by Bove~{\it et al}~[\onlinecite{bove09}] 
until our discussion of existing experiments in Section \ref{section:applications}.


Fig. \ref{fig:XRDSq-quantum} shows the equal time (energy-integrated)
structure factor for the diffuse coherent scattering from disordered protons
at $T=0$
\begin{eqnarray}
&&S^{\sf H^+}_{\sf coh}(\mathbf{q}, t=0)=4\sum_{ij} 
\sin(\mathbf{q} \cdot {\bf a}_i)\sin(\mathbf{q} \cdot {\bf a}_j) \nonumber \\
&& \qquad \qquad \langle {\sf S}_i^z(-\mathbf{q}, t=0) {\sf S}_j^z(\mathbf{q}, t=0)  \rangle.
\label{eq:equaltimeprotoncorr}
\end{eqnarray}
The pinch points are absent at $T=0$, replaced by suppressions of the scattering
around Brillion zone centers \cite{shannon12, benton12, savary12}. 


This effect is most clearly understood by comparing Fig.~\ref{fig:XRDSq-quantum}
with the corresponding classical result shown in  Fig.~\ref{fig:Sq.proton.classical}.
Around certain reciprocal lattice vectors, e.g.
\begin{eqnarray}
{\bf Q}_{\sf p}=(2, 0, 0)
\end{eqnarray}
the classical scattering is directly proportional to a correlation function of 
the uniform polarisation ${\bf P}_+$, 
with no contribution from the staggered polarisation ${\bf P}_-$, and takes
the form given in Eq.~(\ref{eq:pinchpointform}), i.e. 
\begin{eqnarray}
S^{\sf diffuse}_{\sf proton}(\mathbf{Q}_{\sf p}+\mathbf{q})\propto
\left( 1-\frac{(\mathbf{q} \cdot\mathbf{Q}_{\sf p})^2}{q^2 Q_{\sf p}^2} \right) \; . 
\nonumber
\end{eqnarray}
In the quantum case the pinch point form of Eq. (\ref{eq:pinchpointform})
becomes modified by a factor of $q$, suppressing the pinch point
\begin{eqnarray}
S_{\sf H^+}(\mathbf{Q}_{\sf p}+\mathbf{q}, t=0)\propto q
\left( 1-\frac{(\mathbf{q} \cdot \mathbf{Q}_{\sf p}^2)}{q^2 Q_{\sf p}^2} \right).
\label{eq:suppressedpinchpointform}
\end{eqnarray}


At finite temperature these pinch points are restored with a weight
linear in $T$~[\onlinecite{benton12}].  
This being the case, the clearest signature of the formation of a quantum liquid
which could be obtained from energy integrated scattering is the observation
of a pinch point at high temperature, the intensity of which reduces in as the
system cooled, heading towards a linear suppression of the scattering of the
form of Eq. (\ref{eq:suppressedpinchpointform}) as $T \to 0$.
The nodal lines which are predicted in the classical scattering [Fig.~\ref{fig:Sq.proton.classical}]
remain nodal in the quantum case [Fig.~\ref{fig:XRDSq-quantum}].


Around other reciprocal lattice vectors, e.g
\begin{eqnarray}
{\bf Q}_{\sf m}=(2, 0, 3)
\end{eqnarray}
there is a large contribution to the scattering from the fluctuations of ${\bf P}^{\perp}_-$.
In these cases the broad, assymetric features present in the classical scattering remain present
in the quantum case at $T=0$, but are now shifted to finite energy in accordance with
the gapped nature of the fluctuations of  ${\bf P}^{\perp}_-$.


Around reciprocal lattice vectors such as 
\begin{eqnarray}
{\bf Q}_{\sf pm}=(0, 4, 3)
\end{eqnarray}
where the classical scattering shows a combination of pinch points and broad, assymetric features,
the quantum theory predicts that the pinch--point contribution will be linearly suppressed $T=0$,
as in Eq. (\ref{eq:suppressedpinchpointform}), while the broad feature will remain, albeit shifted
to a finite energy.


This separation of the fluctuations of ${\bf P}_+$ and ${\bf P}_-$ as a function of energy 
would be clearly manifested in a measurement of the inelastic scattering.
This is illustrated in Fig. \ref{fig:photon-dispersion} which 
shows a prediction for the inelastic scattering around the
reciprocal lattice vector ${\bf Q}_{\sf in}=(0, 2, 4)$ at $T=0$.


The linearly dispersing photon modes are visible with intensity 
$I \propto q \propto \omega$,
vanishing as they approach $\omega=0$ at the zone center.
The gapped modes have finite weight approaching the zone center.
Observation of these modes in an inelastic scattering experiment would
represent convincing evidence for the formation of a protonic quantum liquid in ice Ih.

\section{Application to experiment}
\label{section:applications}

In this Article we have developed a comprehensive theory of the disordered proton correlations 
in hexagonal (Ih) water ice, considering both a classical model based on the 
``ice rules'' [Section~\ref{section:classical}], and a quantum model allowing for 
coherent quantum tunnelling of protons [Section~\ref{section:photons}].  
We anticipate that the classical theory should accurately describe proton
correlations at temperatures where quantum effects are unimportant, 
while the quantum theory becomes of interest at low temperatures, 
i.e. temperatures comparable with the tunnelling matrix elements $g_{1,2}$
in $\mathcal{H}_{\sf tunnelling}^{\sf hexagonal}$ [Eq.~(\ref{eq:htunnelling2})].


In what follows, we place these results in the context of
published experiment, exploring three particular themes :
ice--rule correlations, as revealed by diffuse neutron scattering 
at relatively high temperatures [Sec.~\ref{subsection:experiment-classical}]; 
low-energy excitations of protons at low temperatures, as revealed by the  
incoherent inelastic neutron--scattering experiments of Bove {\it et al.}~[\onlinecite{bove09}] 
[Sec.~\ref{subsection:bove}]; and the thermodynamics of ice at low 
temperatures [Sec.~\ref{subsection:thermodynamics}].


\begin{figure*}
\centering
\subfigure[\label{fig:9a}]{%
\includegraphics[width=0.6\columnwidth]{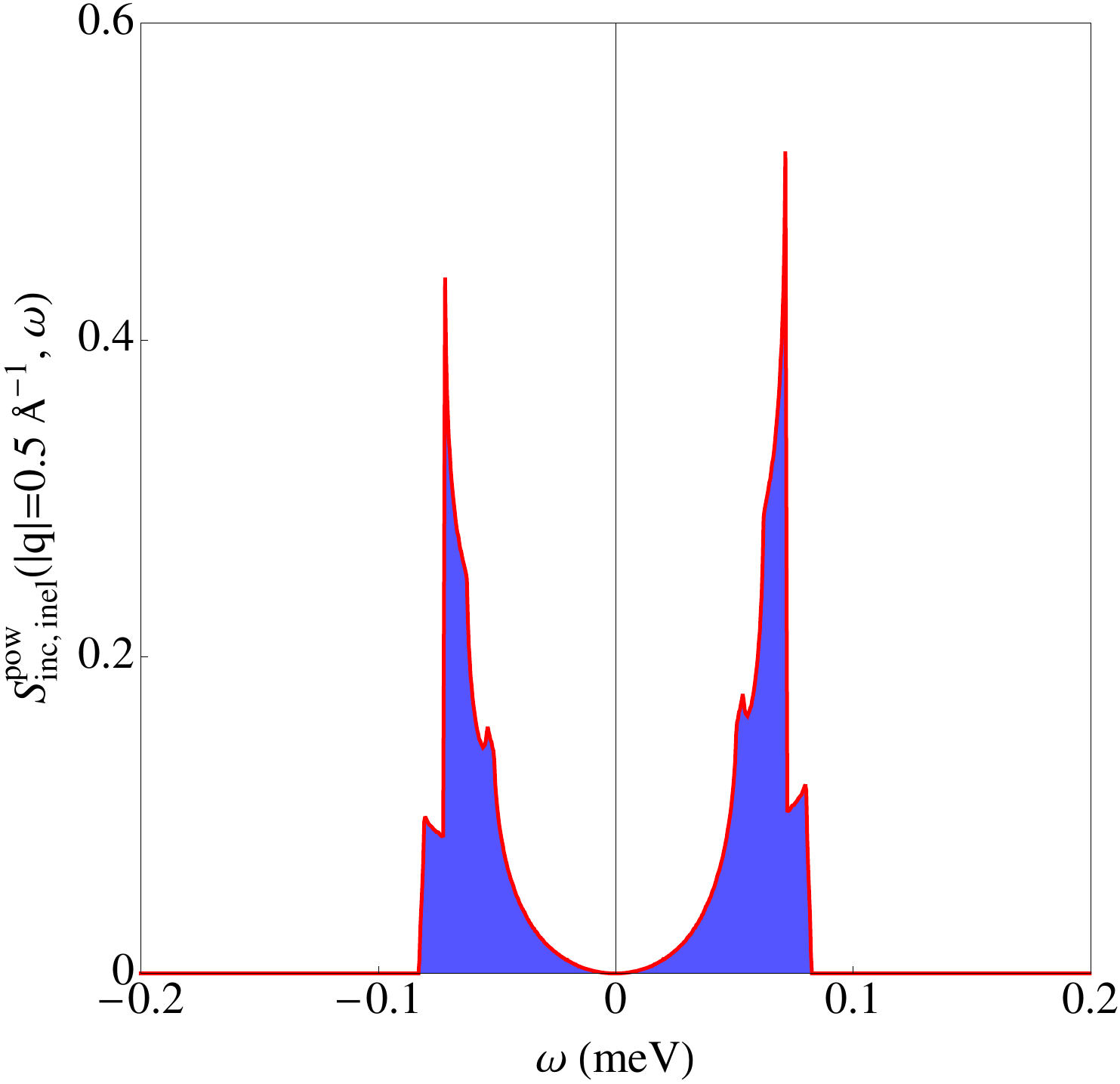}}
    \qquad
\subfigure[\label{fig:9b}]{%
\includegraphics[width=0.6\columnwidth]{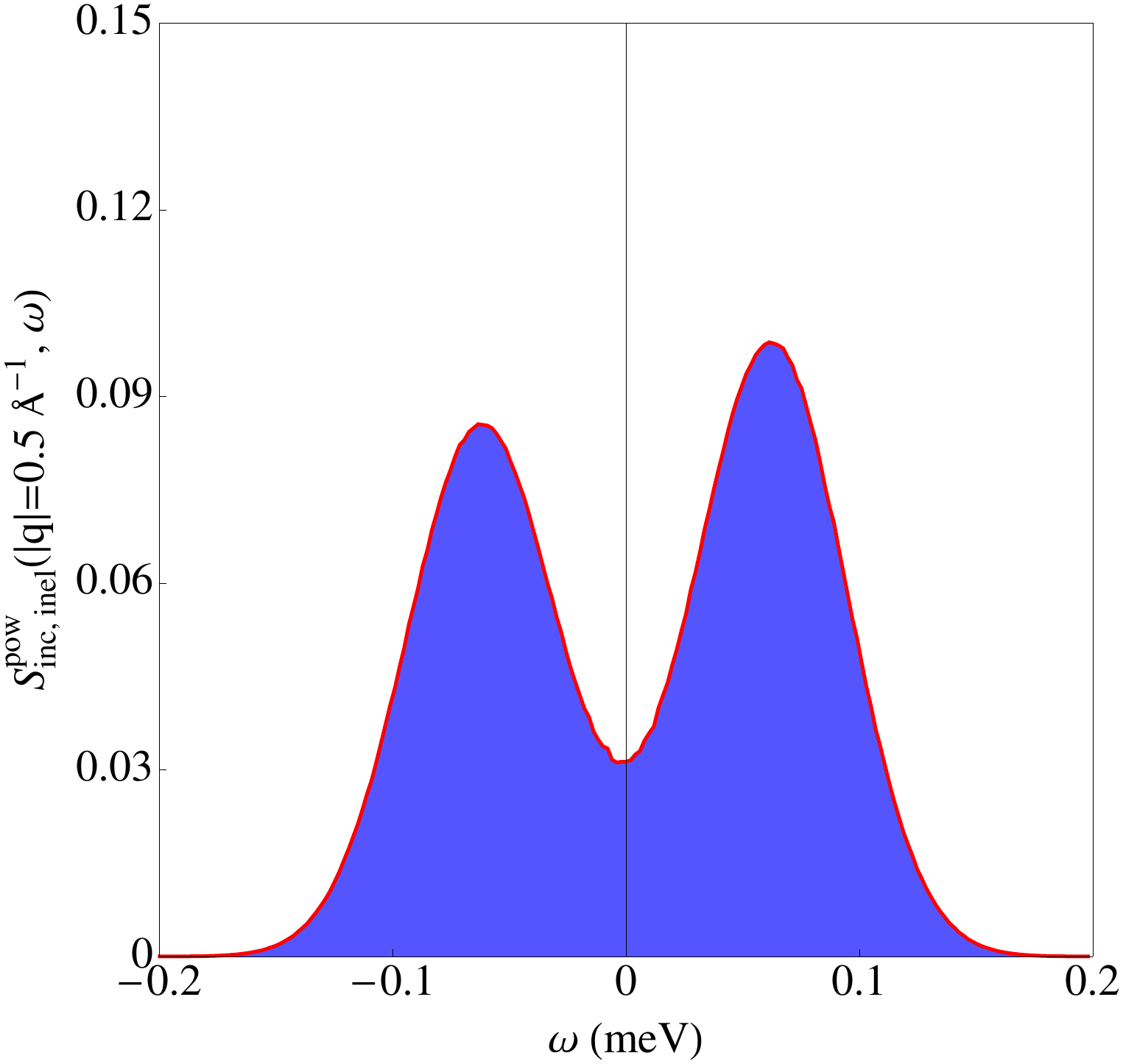}}
    \qquad
\subfigure[[\label{fig:9c}]{%
\includegraphics[width=0.6\columnwidth]{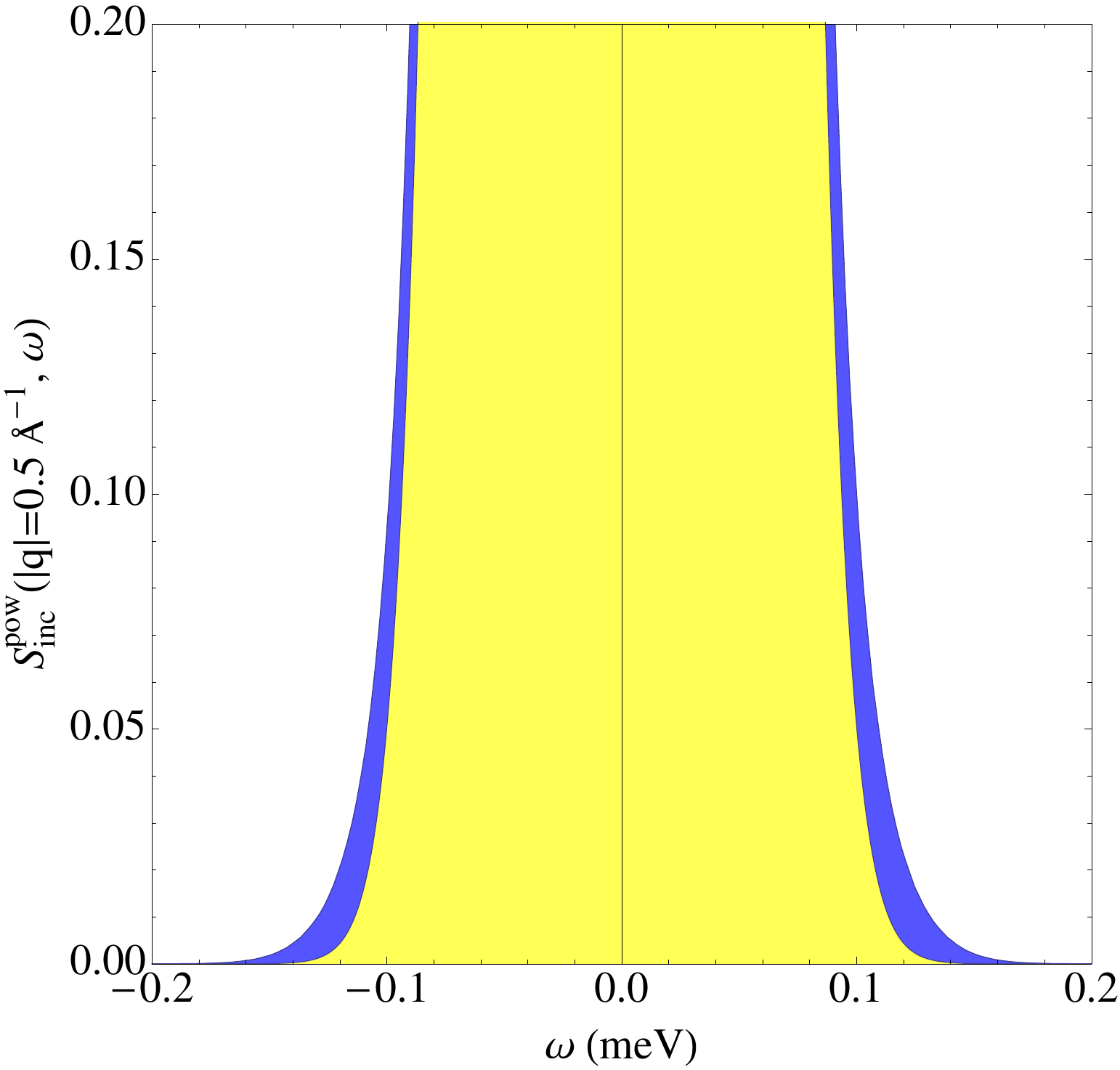}}
\caption{
(Color online).  
Prediction for inelastic, incoherent scattering from protons in quantum ice Ih.  
(a) Prediction of the lattice--gauge theory, $\mathcal{H}_{\sf U(1)}$~[Eq.~(\ref{eq:HU1})], 
at a temperature of $T=5\ \text{K}$, for the same parameters set used in 
discussing coherent inelastic scattering [cf. Fig.~\ref{fig:photon-dispersion} 
and Fig.~\ref{fig:XRDSq-quantum}].
(b) Prediction of the lattice--gauge theory, convoluted with a Gaussian 
of FHWM $0.07\ \text{meV}$ to represent finite experimental resolution.
(c) Prediction of the lattice--gauge theory (blue), combined with the elastic contribution
to the incoherent scattering (yellow).
Both have been convoluted with a Gaussian to represent finite experimental resolution.   
The combined line--shape shows inelastic ``wings'', similar to those observed in 
experiments on ice Ih by Bove {\it et al.}~[\onlinecite{bove09}].   
Parameters are given in Eq.~(\ref{eq:plain-vanilla-parameter-set}) 
and Eq.~(\ref{eq:symmetric-parameter-set}), with details of calculations  
in Appendix~\ref{appendix:incoherent}.
}
\label{fig:powderfig}
\end{figure*}

\subsection{Diffuse, coherent neutron scattering in the classical regime}
\label{subsection:experiment-classical}

In Section~\ref{subsec:classical.predictions} of this Article we developed a 
theory of diffuse neutron scattering from hexagonal water ice based entirely 
on the Bernal--Fowler ice rules~\cite{bernal33,pauling35,petrenko99}.    
We found that the ice rules give rise to both pinch--point
singularities, arising from the algebraic correlations of the uniform polarisation, 
${\bf P}_+$ [Eq.~(\ref{eq:p+def})], and broad, assymetric features, arising 
from the short--ranged correlations of the staggered polarisation, 
${\bf P}_-$  [Eq.~(\ref{eq:p-def})].
The way in which these two fields contribute is controlled by form factors,
and depends on the Brillouin zone in question --- cf. 
Appendix~\ref{appendix:structure.factors.and.fields}.
We now consider how these predictions, summarised in Fig.~\ref{fig:Sq.proton.classical}, 
compare with experiment.


We consider first the pinch points, originating in the uniform polarisation, ${\bf P}_+$.
That the ice rules give rise to pinch--point singularities in the structure factor is a very 
general result and widely known from the study of ice-like systems 
\cite{youngblood80, henley05, henley10}.   
In the context of water ice, the prediction that the proton correlation function should 
be singular at Brillouin zone centres going back as far as Villain in 1972 
[\onlinecite{villain72}].  
These pinch--point singularities are also visible in Monte Carlo simulations of the 
structure factor, based on the ice rules \cite{wehinger14} and in reverse Monte Carlo 
fits to neutron scattering data 
\cite{li94, nield95,nield95-ActaCrystollagrA51, beverley97-JPCM9, beverley97-JPhysChemB101}.
Other theoretical studies, which have utilised random walk 
approximations \cite{schneider80, villain73} and graph series 
expansions \cite{descamps77}, while not specifically identifying the pinch points,  
have noted the presence of nodal lines in the structure factor.


Experimental observation of pinch points in water ice is challenging, 
since coherent scattering of both neutrons and X-rays from protons is 
very weak, and the pinch points are located at reciprocal lattice vectors, 
where scattering is dominated by Bragg peaks associated with the ordered 
lattice of oxygen atoms.
Nevertheless, pinch--point structures {\it are} visible in the neutron--scattering 
from deuterated water ice, D$_2$O, where coherent scattering is much 
stronger\cite{li94,nield95,beverley97-JPhysChemB101,wehinger14}.  


Clear examples of pinch points, occuring at the zone--centers predicted
by our theory, can be seen in e.g. Fig.~2(a) of Li~{\it et al.}~[\onlinecite{li94}], 
or equivalently, Fig.~5(a) of Wehinger~{\it et al.}~[\onlinecite{wehinger14}].   
These should be compared with Fig.~\ref{fig6c} of this Article, noting 
that these authors have indexed their reciprocal lattice vectors to 
an hexagonal unit cell, such that~:
\begin{eqnarray}
H\ H\ L\ \; && \text{(hexagonal)} \nonumber\\
   && \rightarrow 2H\ 0\ L\  \; \text{(orthorhombic)} 
\end{eqnarray}
The pinch--points present in the data
are well reproduced by the theory developed in 
Section~\ref{subsec:classical.predictions}.
Nodal lines, connecting the zone--centers where there are pinch points, 
can also be seen as marked suppressions of the scattering along certain 
high--symmetry directions in both theory and experiment.


We now turn to the broad, assymetric features, originating in the 
staggered polarisation, ${\bf P}_-$.
As well as pinch points, the neutron--scattering data of 
Li~{\it et al.}~[\onlinecite{li94}] also shows zone--center 
scattering with a broad, assymetric character.
An example of a broad feature, arising from the 
correlations of ${\bf P}_-$
can be seen near to $\tilde{{\bf Q}}_{\sf Li} = (1, 1, 5)$ in 
Fig.~2(a) of Ref.~\onlinecite{li94}, or equivalently,  
Fig.~5(b) of Ref.~\onlinecite{wehinger14}.  
This should be compared the scattering around ${\bf Q} = (2, 0, 5)$ 
in Fig.~\ref{fig6c}.  
The broad zone-center features present in the data are well 
reproduced by theory.


More generally, correlations at zone centers are described by a combination
of pinch--points, originating in ${\bf P}_+$, and broad features, 
originating in ${\bf P}_-$.
An example of a this type of feature, can be seen near to 
$\tilde{{\bf Q}}_{\sf Li} = (2, 0,3)$ in 
Fig.~3(a) of Ref.~\onlinecite{li94}, or equivalently,  
Fig.~5(c) of Ref.~\onlinecite{wehinger14}.  
This should be compared the scattering around ${\bf Q} = (0, 4, 3)$ 
in Fig.~\ref{fig6b}, noting that 
\begin{eqnarray}
H\ 0\ L\ \; && \text{(hexagonal)} \nonumber\\
   && \rightarrow 0\ 2K\ L\  \; \text{(orthorhombic)} 
\end{eqnarray}
Once again, there is good agreement between theory and experiment.


Taking a broader view of reciprocal space, the proton correlations measured
by neutron scattering from D$_2$0 at \mbox{$20\ \text{K}$} are generally well--described 
by the ice-rules, differing only in a stronger diffuse background, 
and ``streaks'' of diffuse scattering in certain Brillouin zones where 
static proton correlations should not be visible~\cite{owston49, wehinger14}.
Both effects can be explained in terms of the thermal excitation of phonons~\cite{wehinger14}.


X--ray diffraction from protons does not suffer from the same problems 
as neutron scattering, and can be performed directly on H$_2$0.
However X--ray measurements present problems of cooling, 
and to date all experiments on ice Ih have been carried out relatively 
high temperatures.  
In this case, diffuse scattering reveals relatively little about the ice rules, 
being dominated by a pattern of line--like ``streaks'', 
characteristic of thermally excited phorons~\cite{wehinger14}.  

\subsection{Incoherent inelastic neutron scattering at low temperatures}
\label{subsection:bove}

At present, the strongest experimental evidence in support of the
collective quantum tunnelling of protons in ice Ih comes from 
the recent neutron scattering experiments of Bove~{\it et al.}~[\onlinecite{bove09}].
In these experiments, inelastic, incoherent scattering from protons 
in ice Ih at $T=5\ \text{K}$, was observed for a range of energies 
$\Delta \omega \sim 0.1\ \text{meV}$, outside the experimental 
width of the elastic line.
The authors found that the momentum dependence of this incoherent signal
was consistent with a ``double--well'' model in which there is one proton
on each bond, tunnelling between two sites.
Since moving a single proton within a state obeying the ice rules 
has a prohibitive energy cost, they interpreted their 
result in terms of correlated tunnelling of protons on hexagonal plaquettes, 
of exactly the type described by $\mathcal{H}_{\sf tunnelling}^{\sf hexagonal}$~[Eq.~(\ref{eq:htunnelling2})]
--- cf.~Fig.~\ref{fig:loops}.
This interpretation finds support in {\it ab inito} calculations for Ih 
water ice~\cite{ihm96,drechsel-grau14-PRL}, 
and the energy--scale of the dynamics observed is broadly 
consistent with the single published estimate of the scale of 
quantum tunnelling, $g \sim 0.1\ \text{meV}$~[\onlinecite{ihm96}].


Given this, it is interesting to compare the predictions of the theory of water ice with 
quantum tunnelling developed in Section~\ref{section:photons} of this Article, with the 
neutron--scattering experiments of Bove {\it et al.}~[\onlinecite{bove09}].
Since incoherent scattering probes only the local proton correlations, 
it is not capable of distinguishing the signal long-wavelength features of 
the lattice-gauge theory, such as birefringent photons, or 
temperature--dependent pinch--points.
Nonetheless, the results of the comparison remain very intriguing.    


In Fig.~\ref{fig:powderfig} we present the results of a calculation
of the incoherent scattering at finite temperature $T= 5\ \text{K}$, 
within the lattice--gauge theory $\mathcal{H}_{\sf U(1)}$~[Eq.~(\ref{eq:HU1})].  
Calculations were carried out for the for the symmetric choice of 
parameters 
\begin{eqnarray}
\mathcal{U}=\mathcal{U}' \; ,  \; \mathcal{K}=\mathcal{K}'
\nonumber
\end{eqnarray}
[cf. Eq.~(\ref{eq:plain-vanilla-parameter-set})].   
In the absence of further constraints on parameters, we set
\begin{eqnarray}
\sqrt{\mathcal{U}\mathcal{K}} = 0.018 \  \text{meV} \; ,
\label{eq:symmetric-parameter-set}
\end{eqnarray}
to give an overall bandwidth of excitations [Eq. (\ref{eq:bandwidth1})]
\begin{eqnarray}
\Delta \omega \sim 0.1\ \text{meV} \sim 1\ \text{K} \; , 
\end{eqnarray}
consistent with {\it ab inito} estimates of $g$~\cite{ihm96}.
The details of the calculation are given in Appendix~\ref{appendix:incoherent}.


We find that the quantum tunnelling of protons does indeed produce ``wings'' 
of inelastic scattering which extend appreciably beyond the experimental 
width of the elastic line, as observed by Bove {\it et al.}~[\onlinecite{bove09}].
Fine--structure in the incoherent inelastic scattering [Fig.~\ref{fig:9a}], coming from 
the details of the dispersion [Fig.~\ref{fig:photon-dispersion}], is obscured by finite 
experimental resolution [Fig.~\ref{fig:9b}], leading to broad wings on 
the elastic line [cf.~Fig.~\ref{fig:9c}].
The energy--width of these wings is controlled by the bandwidth of collective 
excitations of protons.
Within the lattice--gauge theory this is set by $\sqrt{\mathcal{U}\mathcal{K}}$
and, for the parameters chosen, is a little less than that observed
in experiment.
  

The good, qualitative, agreement between theory and experiment is very encouraging.
None the less, it is hard to draw a definitive conclusion on the nature of ice Ih 
from incoherent scattering alone.
A much cleaner test would be a measurement of 
the dispersion of the emergent photons, and associated gapped modes, as 
a function of wave vector.
This would require {\it coherent} inelastic scattering,
which is rendered rather challenging by the fact that
incoherent neutron scattering cross section for protons 
is approximately $50$ times greater than the coherent cross section \cite{sears92}.
However, in some circumstances, it is possible to separate coherent and
incoherent scattering using polarisation analysis \cite{moon69}.


It is possible to enhance the ratio of coherent to incoherent scattering
by using deuterated ice, D$_2$O~[\onlinecite{li94}].  
However, some caution is called for.
One of the key findings of the experiments of Bove {\it et al.} [\onlinecite{bove09}] 
was that {\it partial} deuteration suppressed proton dynamics.
This conclusion is supported by subsequent {\it ab initio} simulations, which find  
that partial deuteration inhibits collective quantum 
tunnelling on ``ordered'' plaquettes~\cite{drechsel-grau14-angewandte}.


It follows from the structure of the lattice gauge theory $\mathcal{H}_{\sf U(1)}$ [Eq.~\ref{eq:HU1}], 
that the quantum liquid is stable against all small perturbations which do not
violate the ice rules~\cite{hermele04}.
Consequently, the loss of quantum tunnelling on a very small proportion of 
plaquettes, through the natural abundance of deuterium in water, 
should not injure the quantum liquid state.
Nonetheless, the loss of tunnelling on a macroscopic proportion of plaquettes
is a different proposition, and could easily lead the protons to freeze.
It is plausible that proton the dynamics is restored in the case of full (or very large)
deuteration.
But even so the change from protons to deuterons will alter the relevant energy scale, 
presumably forcing quantum effects to lower temperatures, and changing the balance 
with any other interactions which favour proton order.


A possible alternative to neutron scattering is X--ray diffraction.
In this it has been shown that at high temperatures 
the diffuse scattering is dominated by thermally excited phonons \cite{wehinger14},
making it difficult to observe the diffuse scattering which comes from the proton disorder.
However, the quantum effects described in this article should manifest themselves
at temperatures far below the phonon Debye temperature.
At these temperatures the contribution of thermally excited phonons should
be substantially reduced and one may hope to observe the correlations associated with
the onset of a quantum proton liquid regime. 


It is also interesting to speculate that the optical properties of ice Ih,
which so closely resemble those of the emergent photons in the lattice gauge 
theory, might be sensitive to a proton-liquid at low temperatures.

\subsection{Thermodynamics}
\label{subsection:thermodynamics}

One of the most famous experimental
results in the study of water ice is the demonstration by Giauque and
Stout \cite{giauque36} that it retains a residual entropy down to $T=15$K, 
and that the size of the residual entropy is very close to Pauling's estimate of the
entropy arising from disordered proton configurations obeying the 
Bernal--Fowler ice rules~\cite{pauling35}.    
Subsequent experiments \cite{flubacher60, smith07}
have measured down to temperatures as low as  $T=0.5$K
and find only a very small change in the entropy below $15$K and no new
features in the heat capacity.


The quantum proton--liquid discussed in this article is a coherent superposition 
of an exponentially large number of proton configurations obeying the ice rules.  
At $T=0$, it provides a unique quantum ground state, with vanishing entropy, 
in accordance with the third law of thermodynamics\cite{shannon12}.  
It follows the residual entropy associated with ice must be lost in cooling from
the high--temperature, classical, regime, to the low--temperature quantum 
regime.
It is therefore natural to ask how this entropy would be lost, and whether this is
consistent with published results for the heat capacity of ice Ih.


In principle, entropy could be lost through either a sharp phase transition, or 
a smooth crossover.
Establishing which of these two possibilities occurs would require 
finite--temperature quantum Monte Carlo simulations of 
$\mathcal{H}_{\sf tunnelling}^{\sf hexagonal}$~[Eq.~(\ref{eq:htunnelling2})], 
which lie beyond the scope of the current Article.
None the less, we can gain some insight by analogy with quantum spin--ice, 
where related simulations have already been carried out.


The nature of a thermal crossover between a classical and a quantum spin ice 
was explored in Ref.~\onlinecite{benton12}.
The crossover was found to be controlled by a single length--scale $\lambda_{\sf T}$, 
the thermal de Broglie wavelength set by the thermal excitation of (emergent) photons.
At small, but finite temperatures, the correlations return to their classical form at length scales
\mbox{$r \gtrsim \lambda_{\sf T}$} where
\begin{eqnarray}
 \lambda_{\sf T} \sim \frac{1}{T}
\end{eqnarray}
The corollary of this result for scattering experiments is that the
pinch point singularities, suppressed by quantum fluctuations, are restored 
with a weight linear in $T$ [\onlinecite{benton12}].  


The thermal excitation of photons also has consequences for the thermodynamics
of a quantum ice.
Since the photons of the lattice gauge theory are linearly dispersing, they give a 
contribution
\begin{eqnarray}
C \sim T^3
\end{eqnarray}
to the heat capacity at low temperature \cite{hermele04,benton12,savary12}.
In the absence of a phase transition, this $T^3$ behaviour would be expected to 
merge into a Schottky--like peak in the heat capacity, at a 
temperature of order the quantum tunnelling matrix element, $g$. 


The thermal--crossover scenario for quantum spin ice, developed in 
Ref.~\onlinecite{benton12}, finds strong support in finite--temperature quantum 
Monte Carlo simulations~\cite{banerjee08, kato-arXiv}.  
In particular, a recent numerical study by Kato and Onoda \cite{kato-arXiv}
presents a detailed analysis of the thermodynamics of a quantum spin--ice 
model as the temperature is lowered from a ``classical ice'' into a quantum
spin--liquid state with emergent photon excitations.
They found that the ice entropy is lost in a smooth crossover occuring at a temperature
$T \lesssim g$, where $g$ is the leading tunnelling matrix element between ice states.
This crossover is observable in the heat capacity as a low--temperature peak.
At temperatures lower than the peak, the heat capacity behaves as 
$C \sim T^3$, as expected for linearly--dispersing photons.


Assuming that similar considerations apply, quantum effects should begin to influence 
the thermodynamics of water ice at a temperature a little smaller than the tunnelling 
matrix elements  $g_{1, 2}$.  
Below this temperature, we would anticipate a peak in the heat capacity
signalling the loss of the majority of the Pauling ice--entropy.  
Below this peak there would be a substantial $T^3$ contribution to the specific heat 
from thermal--excitation of photons, in addition to usual the $T^3$ contribution coming 
from acoustic phonons.


To date, to the best of our knowledge, measurements of ice Ih have not revealed 
any anomaly in the heat capacity which could be interpreted as the onset of quantum 
correlations of protons, down to \mbox{$T = 500\ \text{mK}$}~[\onlinecite{smith07}].  
However, for a number of reasons, it is hard to draw any definitive conclusion 
from these experiments.  


Firstly, considerable ambiguity remains about the temperature at which quantum 
effects should be expected to occur in ice Ih.
The single published {\it ab initio} estimate of quantum tunnelling 
$g \sim 0.1\ \text{meV} \sim 1\ \text{K}$~[\onlinecite{ihm96}], is very similar 
to the lowest temperatures achieved in experiment~\cite{smith07}.  
Moreover, in the absence of finite--temperature quantum Monte Carlo 
simulations, it is not known at what fraction of $g_1$ and $g_2$ 
a heat--capacity anomaly should be expected to appear.
It may therefore be that quantum tunnelling of protons {\it does} lead to quantum--liquid
state in ice Ih, but that the associated loss of entropy occurs at a  
temperature lower than that currently measured.
%


The second problem associated with the interpretation of theremodynamic
measurements low temperatures is the extreme difficulty of performing 
experiments, in equilibrium, on a system which retains an extensive entropy.
Here, parallel studies of spin ice provide a stark warning :
recent experiments on the spin ice Dy$_2$Ti$_2$O$_7$ 
by Pomaranski {\it et al.} revealed thermal equilibriation times in excess 
of a week at temperatures of order $300\ \text{mK}$ [\onlinecite{pomaranski13}].  
Once their sample had equilibriated, \mbox{Pomaranski~{\it et al.}}
found an upturn in the heat capacity at low temperatures, in contrast 
with earlier experiments, where the Pauling ice--entropy was reported to persist
down to $200\ \text{mK}$~[\onlinecite{ramirez99},\onlinecite{klemke11}].
If similar problems of equillibriation occur, it is possible to envisage that a specific 
heat peak associated with the loss of Pauling's ice--entropy could yet be observed 
in ice Ih at temperatures of order $1\ \text{K}$.


While thermodynamic measurements at very low temperatures 
will always be challenging, it is important to note that the consequences of 
quantum tunnelling in water ice should be observable at temperatures 
considerably greater than that associated with the loss of the Pauling ice--entropy.
This is particularly true of dynamical properties, measured at wavelength shorter 
than the thermal de Broglie wavelength, $\lambda_{\sf T}$.  
As a result, Quantum Monte Carlo simulation of quantum spin ices show clear 
signs of quantum effects in dynamical structure factors at relatively 
high temperatures \cite{banerjee08, kato-arXiv, henry14}.
It therefore remains reasonable to discuss the inelastic neutron scattering of 
Bove {\it et al.}~[\onlinecite{bove09}], carried out at $5\ \text{K}$, 
in terms of a lattice gauge theory with characteristic energy scale 
$\sqrt{\mathcal{U}\mathcal{K}} \sim 1\ \text{K}$ [cf. Section~\ref{subsection:bove}].
%

\section{Conclusions}
\label{section:discussion}

Common, hexagonal, water ice is a wholly remarkable substance --- a proton--bonded 
network of water molecules in which oxygen atoms form a regular crystal, but protons 
need never order.
In this Article we have explored the nature of proton disorder in hexagonal (Ih) 
water ice, considering both a classical model based on the Bernal--Fowler ``ice-rules''
[\onlinecite{bernal33},\onlinecite{pauling35}], and a quantum model which respects
the ice rules, but also incorporates collective quantum tunnelling of protons on 
hexagonal plaquettes [cf. Fig.~\ref{fig:loops}].   
Quantum tunnelling of this type is known to have profound consequences in models 
of quantum spin ice~\cite{hermele04, banerjee08, shannon12, benton12,kato-arXiv,mcclarty-arXiv}, 
and there is a growing body of evidence that it also plays a role in water ice, 
including {\it ab initio} calculations~\cite{ihm96,drechsel-grau14-PRL},
inelastic neutron scattering~\cite{bove09}, and parallel 
studies of other proton--bonded systems~\cite{meng15,brougham99}.  


In the case of classical ice Ih, we have developed a comprehensive 
theory of diffuse scattering from protons [Section~\ref{section:classical}].
We find that the ice--rules have two distinct signatures in scattering~: 
singular ``pinch--points'', originating in a zero--divergence 
condition on the uniform polarisation ${\bf P}_+$, and broad, assymetric, 
zone--center features, coming from the staggered polarisation ${\bf P}_-$
[Fig.~\ref{fig:Sq.proton.classical}].   
Both of these features have previously been observed in experiment~\cite{li94}.


In the case of the quantum model, we have obtained a description of a low--temperature
quantum liquid --- in which protons resonate between an exponentially large number 
of configurations satisfying the ice rules --- in terms of a quantum $U(1)$ lattice--gauge theory 
[Section~\ref{section:photons}].  
%
%
The long--wavelength excitations of this quantum liquid take the form of gapless, 
emergent photons, originating in the uniform polarisation ${\bf P}_+$,
and gapped, optical modes originating in the staggered polarisation ${\bf P}_-$.  


We have used this lattice--gauge theory to make concrete predictions for 
inelastic scattering experiments on a ice Ih [Fig.~\ref{fig:photon-dispersion}].
We find that both the emergent photons and the optical modes can be 
clearly resolved at finite energy.  
Much like real light in water ice \cite{brewster1814, brewster1818}, 
the emergent photons are birefringent, 
exhibiting a dispersion which depends on both their polarisation and their 
direction of propagation.   
We have also explored how quantum tunnelling of protons modifies 
diffuse scattering at low temperatures [Fig.~\ref{fig:XRDSq-quantum}].
We find that the ``pinch-points'', characteristic of the ice--rules, are progressively
eliminated as the system is cooled toward \mbox{$T=0$}.



The assertion, in Ref. [\onlinecite{castro-neto06}], that the lattice gauge 
theory describing a two--dimensional quantum water ice should be confining 
at $T=0$, is consistent with numerical results for two--dimensional quantum 
ice \cite{chern14, shannon04,syljusen06,henry14}.    
However a number of three--dimensional ice-like models are known 
to support deconfined, quantum--liquid ground states\cite{sikora09,shannon12}.
In the absence of detailed simulations it is not clear for which parameters the 
ground state of $\mathcal{H}_{\sf tunnelling}^{\sf hexagonal}$~[Eq.~(\ref{eq:htunnelling2})] 
should be ordered. 
However the incoherent inelastic neutron scattering experiments 
of Bove {\it et al.}~[\onlinecite{bove09}] provide {\it prima facie} 
evidence of collective proton dynamics at a temperature ($5\ \text{K}$) 
comparable with {\it ab initio estimates} of quantum tunnelling 
in ice Ih [\onlinecite{ihm96}], and is in good, qualitative, agreement 
with the predictions of our theory [Fig.~\ref{fig:powderfig}].  


The observation of a birefringent, emergent, photon in coherent inelastic scattering 
from protons at low temperatures would provide very strong evidence for the existence 
of quantum fluid of protons in ice Ih. 
At present, however it is not possible to compare these predictions directly with 
experiment, since coherent inelastic scattering data is unavailable for the 
temperatures 
where quantum tunnelling is expected to relevant.
Experimental evidence for a quantum fluid of protons from thermodynamic 
measurements is also lacking [cf. Section~\ref{section:applications}].
We hope that further experiment will provide a definitive answer to these questions.


Experiments probing the excitation spectrum of other water ices, 
where protons order at low temperatures, could also be of interest, 
since experience with other ice models suggests that a quantum liquid may 
still be observable at finite temperature \cite{banerjee08,henry14,kato-arXiv}.
And in the absence of evidence to the contrary, a quantum--liquid ground state 
in ice Ih remains an intriguing possibility.


In conclusion, more than 80 years after the pioneering work of 
Bernal and Fowler, the behaviour of protons in common, hexagonal 
water ice at low temperatures remains a problem of great 
fundamental interest.
And, on the basis recent experiments~\cite{bove09}, a 
quantum liquid of protons, of the type explored in this Article, 
remains a tantalising possibility.

\section*{Acknowledgments}

The authors are pleased to acknowledge helpful discussions with 
Paul McClarty and Karlo Penc.
We thank Livia Bove and Henrik Ronnow for constructive comments
on the manuscript.
This work was supported by the Theory of Quantum Matter Unit of the 
Okinawa Institute of Science and Technology, and through 
EPSRC Grants EP/C539974/1 and EP/G031460/1. 


Since this work was completed, two papers of direct relevance have appeared.   
The first, by Isakov {\it et al.} [\onlinecite{isakov15}], also examines the proton 
correlations in a classical model of ice Ih.
The second, by Yen and Gao [\onlinecite{yen15}], reports evidence for 
quantum coherence in water ice, from measurements of its dieletric constant.

\appendix

\section{Details of the lattice and coordinate system}
\label{appendix:lattice}

In this Appendix we give details of the
lattice and coordinate system which we have used in our
calculations.
We will write all lattice length scales in terms
of the oxygen-oxygen bond distance, $a_{\sf O}$, which in
ice Ih is approximately \cite{petrenko99}
\begin{eqnarray}
a_{\sf O}\approx2.75 \ \text{\AA}.
\end{eqnarray}


We use a coordinate system in which the hexagonal
symmetry axis is the ${\bf z}$-axis and the repeat 
vectors of the hexagonal unit cell
\begin{eqnarray}
&&{\bf G}_1= a_{\sf O} \left( \frac{2 \sqrt{2}}{\sqrt{3}}, 0, 0 \right) \\
&&{\bf G}_2= a_{\sf O} \left( \frac{\sqrt{2}}{\sqrt{3}}, \sqrt{2}, 0 \right)\\
&&{\bf G}_3= a_{\sf O} \left( 0, 0, \frac{8}{3} \right).
\end{eqnarray}
The primitive unit cell contains four oxygen atoms and eight protons.


One can also define an orthorhombic unit cell, containing eight
oxygen atoms with orthogonal repeat vectors
\begin{eqnarray}
&&{\bf G}_X= {\bf G}_1 
\label{eq:GX}\\
&&{\bf G}_Y= 2 {\bf G}_2 -{\bf G}_1
\label{eq:GY}\\
&&{\bf G}_Z={\bf G}_3.
\label{eq:GZ}
\end{eqnarray}
This is the unit cell used by Nield and Whitworth in Ref. \cite{nield95}.


\begin{figure*}
\centering
\subfigure[\ \ Bond labelling convention]{
\includegraphics[width=.3\textwidth]{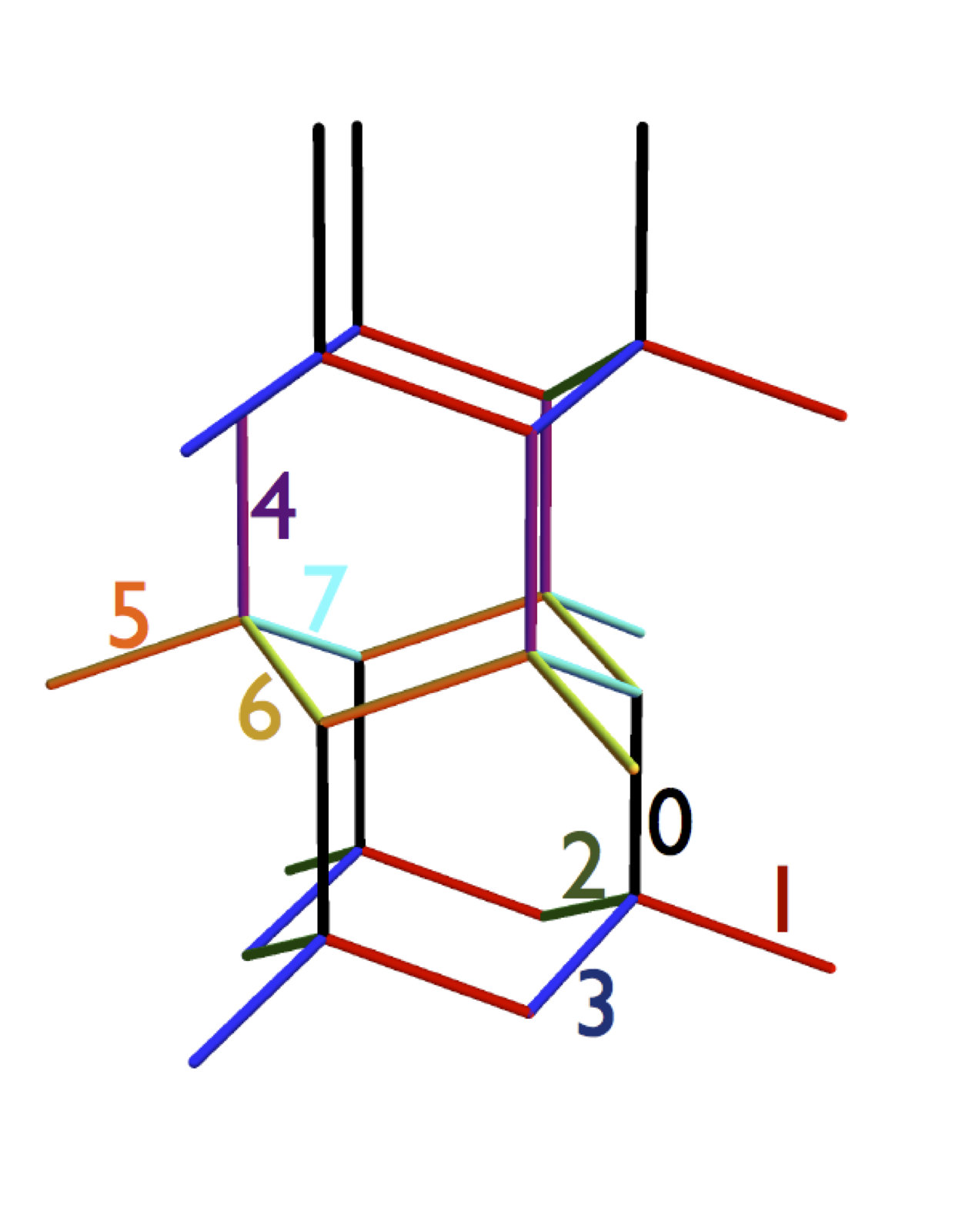}}%
    \qquad
\subfigure[\ \ Plaquettes labelled `0' and '4']{%
\includegraphics[width=.3\textwidth]{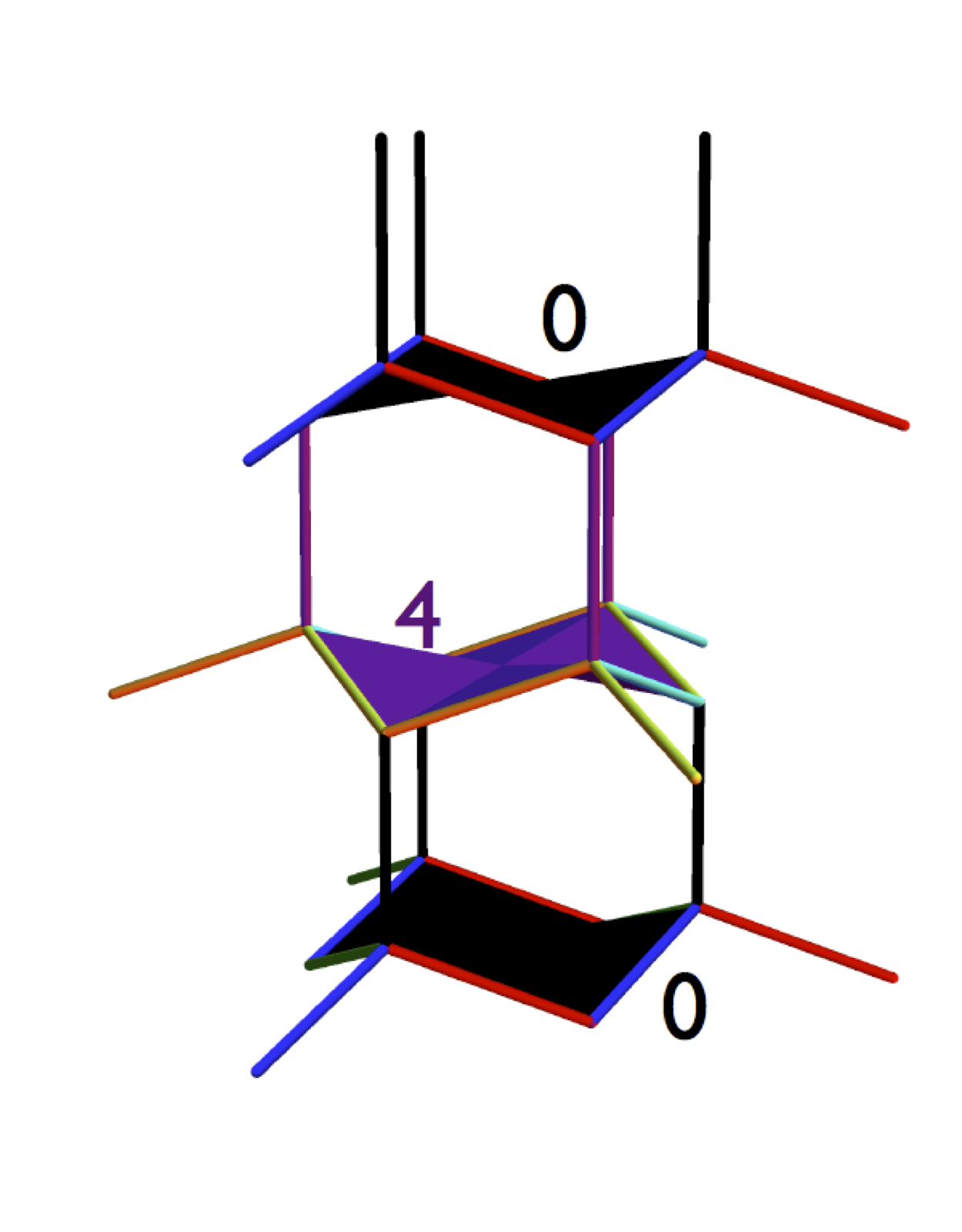}}%
    \qquad
\subfigure[\ \ Plaquettes labelled `1' and '5']{%
\includegraphics[width=.3\textwidth]{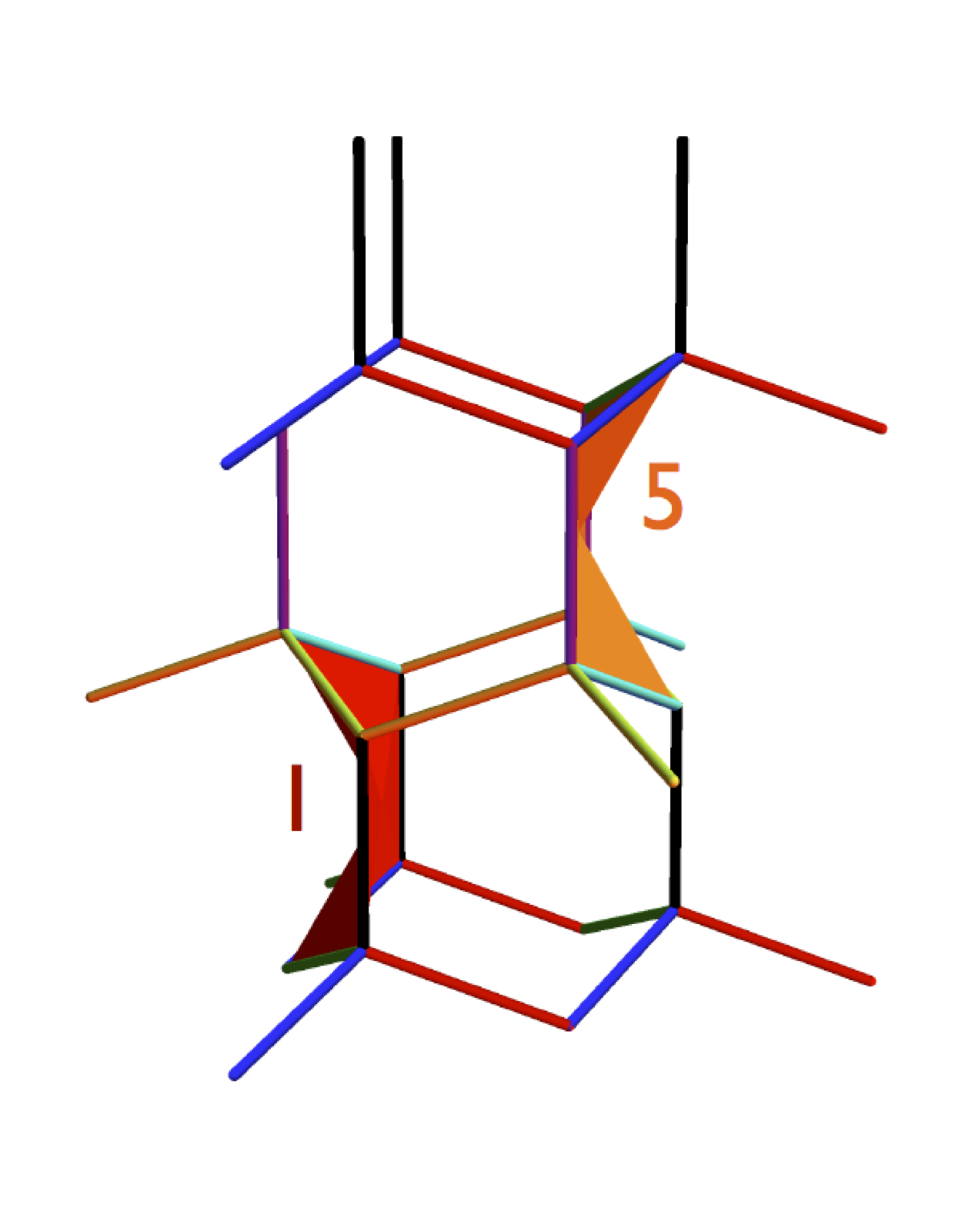}}
\\
\subfigure[\ \ Plaquettes labelled `2' and '6']{%
\includegraphics[width=.3\textwidth]{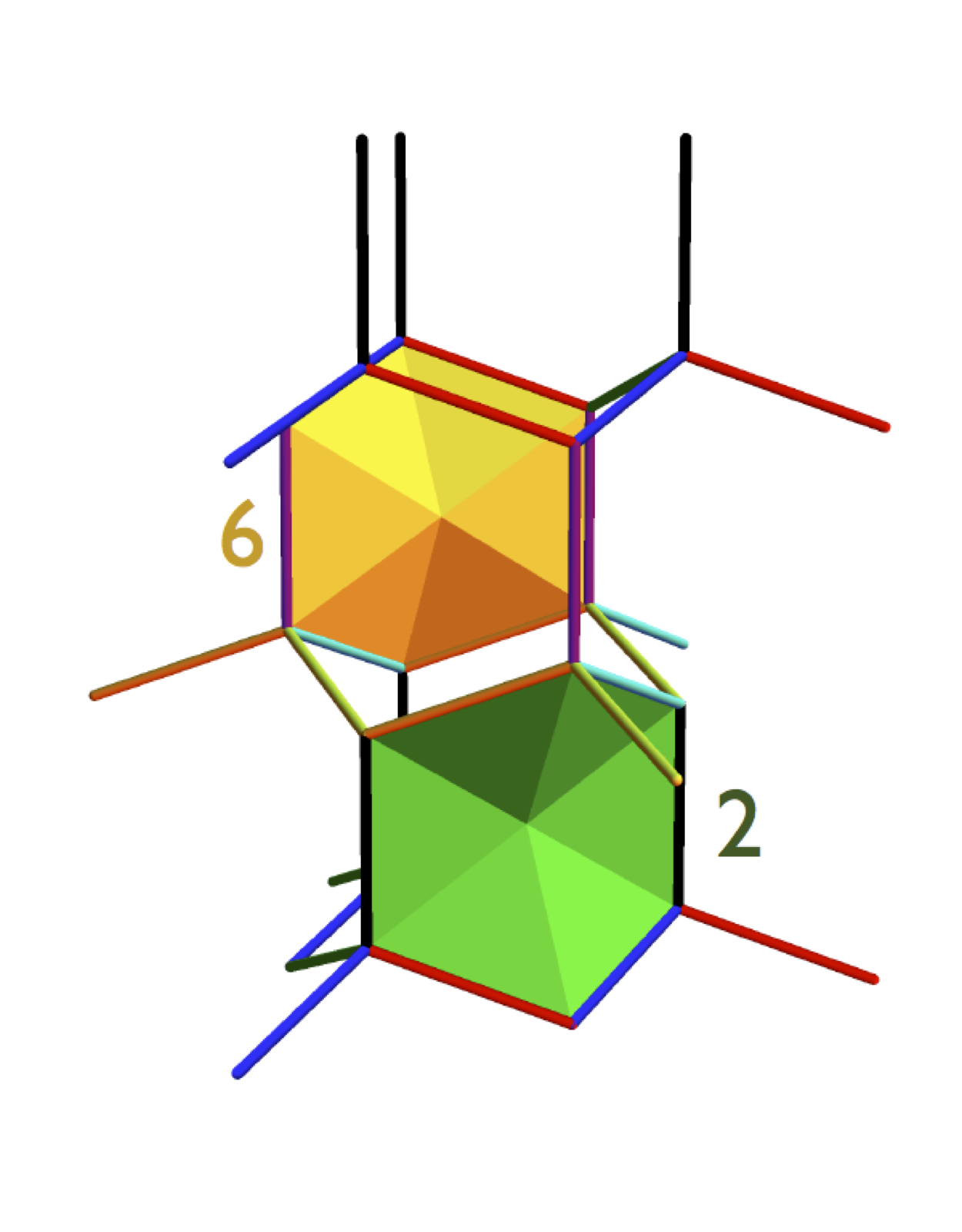}}%
    \qquad
\subfigure[\ \ Plaquettes labelled `3' and '7']{
\includegraphics[width=.3\textwidth]{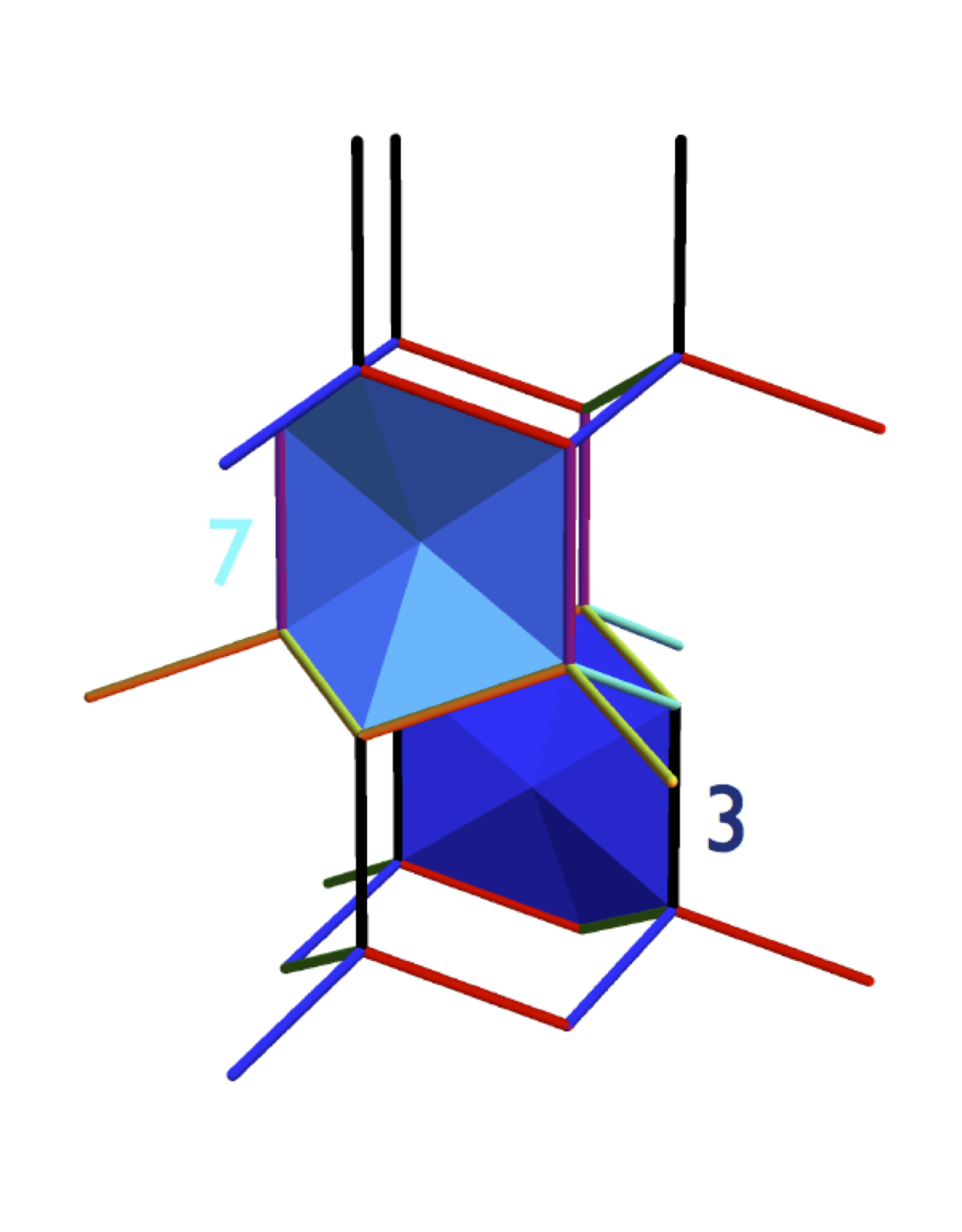}}
\\
\caption{
(Color online).
Labelling convention for the eight sets of oxygen--oxygen bonds and eight sets
of six-link plaquettes unrelated by translational symmetry.
This is the convention employed in diagonalising the lattice gauge theory
described in Section \ref{section:photons} and Appendix \ref{appendix:photons}.
(a) is color-coded such that bonds related by translational symmetry are the same color.
The bonds labelled $0$ and $4$ are mirror symmetric bonds aligned along the ${\bf z}$-axis
and the remaining bonds are centre symmetric [cf. Fig. \ref{fig:bondsymmetries}].
(b) shows plaquettes of type I with matrix element $g_1$.
(c)-(e) shows plaquettes of type II with matrix element $g_2$.
 } 
\label{fig:labelling}
\end{figure*}


In the scattering patterns shown in Figs. \ref{fig:Sq.Ising.classical}, \ref{fig:Sq.proton.classical},
\ref{fig:isingSq-quantum} and \ref{fig:XRDSq-quantum}, the reciprocal
lattice units used for the momentum scale are with reference to the orthorhombic
unit cell, as in Ref.  \cite{nield95}.
Thus, $h$, $k$ and $l$ [cf.  Figs. \ref{fig:Sq.Ising.classical}, \ref{fig:Sq.proton.classical},
\ref{fig:isingSq-quantum} and \ref{fig:XRDSq-quantum}] relate to the
momentum transfer ${\bf q}$ via
\begin{eqnarray}
{\bf q}= \left( \frac{2 \pi h}{|{\bf G}_X|},  \frac{2 \pi k}{|{\bf G}_Y|},  \frac{2 \pi l}{|{\bf G}_Z|} \right)
\end{eqnarray}
where ${\bf G}_{XYZ}$ are defined in Eqs. (\ref{eq:GX})-(\ref{eq:GZ}).


The bond vectors ${\bf d}^{\alpha}_i$ connecting an oxygen
to it's four neighbours [cf. Eq. (\ref{eq:vanishingm})] are given by
\begin{eqnarray}
{\bf d}^A_0&=&{\bf d}^C_0=a_{\sf O} \left( 0, 0, 1 \right)
\label{eq:dA0} \\
{\bf d}^A_1&=&a_{\sf O}\left( \sqrt{\frac{2}{3}}, \frac{\sqrt{2}}{3}, -\frac{1}{3} \right)
\label{eq:dA1}
\\
{\bf d}^A_2&=&a_{\sf O}\left( -\sqrt{\frac{2}{3}}, \frac{\sqrt{2}}{3}, -\frac{1}{3} \right)
\label{eq:dA2}
 \\
{\bf d}^A_3&=&a_{\sf O}\left(0, \frac{-2\sqrt{2}}{3}, -\frac{1}{3} \right)
\label{eq:dA3} \\
{\bf d}^C_1&=&a_{\sf O}\left( -\sqrt{\frac{2}{3}}, -\frac{\sqrt{2}}{3}, -\frac{1}{3} \right)
\label{eq:dC1}
\end{eqnarray}
\begin{eqnarray}
{\bf d}^C_2&=&a_{\sf O}\left( \sqrt{\frac{2}{3}}, -\frac{\sqrt{2}}{3}, -\frac{1}{3} \right)
\label{eq:dC2}
 \\
{\bf d}^C_3&=&a_{\sf O}\left(0, \frac{2\sqrt{2}}{3}, -\frac{1}{3} \right) 
\label{eq:dC3}
\\
{\bf d}^B_i&=&-{\bf d}^C_i \ \ \forall \ i 
\label{eq:dBi}
\\
{\bf d}^D_i&=&-{\bf d}^A_i \ \ \forall \ i .
\label{eq:dDi}
\end{eqnarray}

The displacement of protons from the bond midpoint on the 8 sublattices of bonds are given
by $\pm {\bf a}_i$ where
\begin{eqnarray}
&&{\bf a}_0=\phi {\bf d}_0^A \quad
{\bf a}_1=\phi {\bf d}_1^A  \quad
{\bf a}_2=\phi {\bf d}_2^A \quad
{\bf a}_3=\phi {\bf d}_3^A \nonumber \\
&&{\bf a}_4=\phi {\bf d}_0^C \quad
{\bf a}_5=\phi {\bf d}_1^C \quad
{\bf a}_6=\phi {\bf d}_2^C \quad
{\bf a}_7=\phi {\bf d}_3^C  \; . \nonumber\\
\label{eq:dispvectors}
\end{eqnarray}
For the purpose of our scattering calculations we take the size of the
proton displacement from the bond midpoint
relative to the bond length to be
\begin{eqnarray}
\phi = 0.15
\label{eq:displacement.parameter}
\end{eqnarray}
assuming an O-H covalent bond length \cite{wehinger14}
\begin{eqnarray}
a_{\sf OH}=0.95 \ \ \text{\AA}.
\end{eqnarray}


For the calculation of the dispersion of the lattice gauge theory,
presented in Section \ref{section:photons} and Appendix \ref{appendix:photons},
it is necessary to define a labelling convention for the eight sets of
oxygen--oxygen bonds not related by translational symmetries and for the
eight sets six-link plaquettes.
Our convention is defined in Fig. \ref{fig:labelling}.


Of particular are importance are the set of vectors which link the bond midpoints
around the edge of a plaquette to the center of the plaquette.
We denote these vectors ${\bf c}_{nm}$ where two bonds of sublattice $m$
belong to a plaquette of sublattice $n$ and are located at ${\bf p}_n \pm {\bf c}_{nm}$
with the plaquette center being at ${\bf p}_n$.
Where only one bond of sublattice $m$ belongs to a plaquette of sublattice $n$ we denote
it's position relative to the centre of the plaquette by ${\bf C}_{nm}$.
These vectors are
\begin{eqnarray}
&&{\bf c}_{01}=-{\bf c}_{10}={\bf c}_{45}=-{\bf c}_{54}=a_{\sf O} \left( \frac{-1}{\sqrt{6}}, \frac{1}{\sqrt{2}}, 0 \right)  \nonumber \\
&&{\bf c}_{02}=-{\bf c}_{20}={\bf c}_{46}=-{\bf c}_{64}=a_{\sf O} \left( \frac{-1}{\sqrt{6}}, -\frac{1}{\sqrt{2}}, 0 \right) \nonumber \\
&&{\bf c}_{03}=-{\bf c}_{30}={\bf c}_{47}=-{\bf c}_{74}=a_{\sf O} \left( \sqrt{\frac{2}{3}}, 0, 0 \right) \nonumber \\
&&{\bf C}_{12}={\bf C}_{21}=a_{\sf O}\left(0, \frac{\sqrt{2}}{3}, \frac{2}{3} \right) \nonumber \\
&&{\bf C}_{13}={\bf C}_{31}=a_{\sf O}\left(\frac{1}{\sqrt{6}}, -\frac{1}{3\sqrt{2}}, \frac{2}{3} \right) \nonumber 
\end{eqnarray}
\begin{eqnarray}
&&{\bf C}_{23}={\bf C}_{32}=a_{\sf O}\left(-\frac{1}{\sqrt{6}}, -\frac{1}{3\sqrt{2}}, \frac{2}{3} \right) \nonumber\\
&&{\bf C}_{56}={\bf C}_{65}=a_{\sf O}\left(0, -\frac{\sqrt{2}}{3}, \frac{2}{3} \right) \nonumber  \\
&&{\bf C}_{57}={\bf C}_{75}=a_{\sf O}\left(-\frac{1}{\sqrt{6}}, \frac{1}{3\sqrt{2}}, \frac{2}{3} \right)  \nonumber \\
&&{\bf C}_{67}={\bf C}_{76}=a_{\sf O}\left(\frac{1}{\sqrt{6}}, \frac{1}{3\sqrt{2}}, \frac{2}{3} \right) \nonumber \\
&&{\bf C}_{16}={\bf C}_{25}=a_{\sf O}\left(0, \frac{\sqrt{2}}{3}, -\frac{2}{3} \right) \nonumber \\
&&{\bf C}_{17}={\bf C}_{35}=a_{\sf O}\left(\frac{1}{\sqrt{6}}, -\frac{1}{3\sqrt{2}}, -\frac{2}{3} \right) \nonumber \\
&&{\bf C}_{27}={\bf C}_{36}=a_{\sf O}\left(-\frac{1}{\sqrt{6}}, -\frac{1}{3\sqrt{2}}, -\frac{2}{3} \right) \nonumber \\
&&{\bf C}_{53}={\bf C}_{71}=a_{\sf O}\left(-\frac{1}{\sqrt{6}}, \frac{1}{3\sqrt{2}}, -\frac{2}{3} \right) \nonumber \\
&&{\bf C}_{63}={\bf C}_{72}=a_{\sf O}\left(\frac{1}{\sqrt{6}}, \frac{1}{3\sqrt{2}}, -\frac{2}{3} \right) \nonumber \\
&&{\bf C}_{61}={\bf C}_{52}=a_{\sf O}\left(0, -\frac{\sqrt{2}}{3}, -\frac{2}{3} \right) \nonumber
\end{eqnarray}

\section{Derivation of constraints within in continuum field--theory}
\label{appendix:continuum.field.theory}

In this Appendix we describe in detail the coarse--graining procedure
used in Section \ref{subsec:classical.fields} to obtain the
constraint Eqs.~(\ref{eq:p+constraint})-(\ref{eq:p-constraint}).


The argument begins by setting $m_{\bf r}=0$ for all $A$ and $C$
vertices [cf. Fig. \ref{fig:loops}(b)].
If we then set the value of the total flux ${\bf P}_{\bf r}$ around
every $A$ and $C$ vertex, we have specified the H-bond configuration of
the entire lattice.
The necessity that this configuration must also obey $m_{\bf r}=0$ at the $B$
and $D$ vertices induces constraints on how  ${\bf P}_{\bf r}$ varies in space.


To derive those constraints in the long wavelength limit we introduce
two coarse--grained fields $\bar{{\bf P}}^A(\mathbf{r})$, $\bar{{\bf P}}^C(\mathbf{r})$
defined at every point in space (not just on the lattice).
We define $\bar{{\bf P}}^A(\mathbf{r})$ in such way that when it is evaluated at the position of an
oxygen vertex of the $A$ sublattice, it returns the precise value of the total
flux around that vertex
\begin{eqnarray}
\bar{{\bf P}}^A(\mathbf{r}=\mathbf{r}_A)={\bf P}_{\mathbf{r}_A}.
\end{eqnarray}
Similarly
\begin{eqnarray}
\bar{{\bf P}}^C(\mathbf{r}=\mathbf{r}_C)={\bf P}_{\mathbf{r}_C}.
\end{eqnarray}


To obtain our constraints on $\bar{{\bf P}}^A(\mathbf{r})$, $\bar{{\bf P}}^C(\mathbf{r})$
we write the scalar field $m_{\bf r}$ at the $B$ and $D$ vertices in terms of the degrees
of freedom at the four surrounding vertices.


A vertex of the $B$ sublattice [cf. Fig. \ref{fig:fluxrep}], located at ${\bf r}_B$,
neighbours one $A$ vertex located at
\begin{eqnarray}
{\bf r}_A^{(0)}={\bf r}_B-{\bf d}_0^A \\
\end{eqnarray}
and three $C$ vertices located at
\begin{eqnarray}
{\bf r}_C^{(i)}={\bf r}_B-{\bf d}_i^C \quad i=1,2,3
\end{eqnarray}
[cf. Eqs. (\ref{eq:dA0})-(\ref{eq:dDi})].


Using the vanishing of $m_{\bf r}$ at the $A$ and $C$ vertices
\begin{eqnarray}
m_{\bf r_B}=\frac{3}{4} \left\{ \sum_{i=1}^3
\frac{1}{a_{\sf O}}
 {\bf d}_i^C \cdot {\bf P}_{\mathbf{r}_B-{\bf d}_i^C}
+
P^z_{\mathbf{r}_B-{\bf d}_0^A}
 \right\}
\end{eqnarray}
and similarly
\begin{eqnarray}
m_{\bf r_D}=\frac{3}{4} \left\{ \sum_{i=1}^3
\frac{1}{a_{\sf O}}
  {\bf d}_i^A \cdot {\bf P}_{\mathbf{r}_D-{\bf d}_i^A}
+
P^z_{\mathbf{r}_D-{\bf d}_0^C}
 \right\}.
\end{eqnarray}


Assuming smooth variation of $\bar{{\bf P}}^A(\mathbf{r})$, $\bar{{\bf P}}^C(\mathbf{r})$
we can use a Taylor expansion to write
\begin{eqnarray}
P^{\alpha}_{{\bf r}_{A, C}+\delta {\bf r}}
\approx
\bar{P}^{\alpha}_{A, C}(\mathbf{r})
+
\delta \mathbf{r} \cdot \nabla \bar{P}^{\alpha}_{A, C}(\mathbf{r}).
\end{eqnarray}


Enforcing $m_{\bf r_B} = m_{\bf r_D}=0$ throughout the lattice, we obtain the constraints 
on our continuum field--theory, Eq.~(\ref{eq:divconstraint1})--(\ref{eq:divconstraint2}).

\section{Projection operator calculation of structure factor
for classical water ice}
\label{appendix:projection}

Here we show how to calculate the structure factor
for the Ising bond variables $\sigma_{\mathbf{r}\mathbf{r}^\prime}$ [Eq.~(\ref{eq:S.of.q.Ising})]
via a generalisation of the lattice calculation in Ref. [\onlinecite{henley05}].
In this approximation the Ising nature of the variables $\sigma_{\mathbf{r}\mathbf{r}^\prime}$ 
is relaxed, such that they can take on any real value and their normalisation is enforced on average
\begin{eqnarray}
\langle \sigma_{\mathbf{r}\mathbf{r}^\prime}^2 \rangle = 1 \; .
\label{eq:avenorm}
\end{eqnarray} 
The calculation proceeds by constructing a projection operator 
which acts on the Fourier transform of an arbitrary proton configuration, 
to remove all states which do not satisfy the ice rules.


We begin by defining the Fourier transform over the bond variables 
as in Section~\ref{subsec:classical.lattice.theory}  [Eq.~(\ref{eq:sgnconv})], i.e.
\begin{eqnarray}
\sigma_\nu (\mathbf{q}) 
   &=& \sqrt{\frac{4}{N}}
   \sum_{\mathbf{r}\mathbf{r}^\prime \in \nu}  
   \exp \left( -i \mathbf{q} \cdot {\bf R}_{\mathbf{r}\mathbf{r}^\prime} \right) 
   \sigma_{\mathbf{r}\mathbf{r}^\prime} \; , 
   \nonumber\\
   && \qquad  \qquad {\bf R}_{\mathbf{r}\mathbf{r}^\prime} = \frac{\mathbf{r} + \mathbf{r}^\prime}{2} \; , 
   \nonumber
\end{eqnarray}
where
$\nu$ indexes one of the eight sublattices of bonds which are not related
by a translational symmetry of the lattice, and $N$ counts the number of 
oxygen atoms.
The sign of $\sigma_{\mathbf{r}\mathbf{r}^\prime}$ is fixed by the convention 
[Eq.~(\ref{eq:sgnconv})]
\begin{eqnarray}
\mathbf{r} \in \{A,C\} \; , \; \mathbf{r}^\prime \in \{B,D\} \; .
\nonumber
\end{eqnarray}
where $\alpha=A, B, C, D$ indexes the four oxygen sublattices (cf. Fig. \ref{fig:loops}(b)).


From these Fourier transforms we define an 8-component vector 
\begin{eqnarray}
&&\tilde{\sigma}(\mathbf{q})= \nonumber \\
&& (\sigma_1(\mathbf{q}),
\sigma_2(\mathbf{q}),
\sigma_3(\mathbf{q}),
\sigma_4(\mathbf{q}),
\sigma_5(\mathbf{q}),
\sigma_6(\mathbf{q}),
\sigma_7(\mathbf{q}),
\sigma_8(\mathbf{q})
) \; . \nonumber\\
\label{eq:sigmavector}
\end{eqnarray}


We also introduce a net Ising polarisation for each oxygen sublattice $\alpha=A, B, C, D$, 
\begin{eqnarray}
m_{\alpha}(\mathbf{q})
   = \sqrt{\frac{4}{N}}\sum_{\mathbf{r} \in \alpha} 
   \exp(-i \mathbf{q} \cdot \mathbf{r}) m (\mathbf{r}) \; .
\end{eqnarray}
It follows from Eq.~(\ref{eq:vanishingm}) that 
\begin{eqnarray}
m_{\alpha}(\mathbf{q}) = 0 \quad \forall \ \alpha, \mathbf{q} 
\label{eq:qspaceicerule}
\end{eqnarray}
in any state obeying the ice rules.


The constraint Eq. (\ref{eq:qspaceicerule}) can be written as 
a set of orthogonality conditions on $\tilde{\sigma}(\mathbf{q})$
\begin{eqnarray}
(\tau_{\alpha}(\mathbf{q}), \tilde{\sigma}(\mathbf{q}))=0, \ \ \alpha=A, B, C, D
\label{eq:tauconstraints}
\end{eqnarray}
where the inner product
\begin{eqnarray}
(A, B)=\sum_i A_i^{\ast} B_i
\end{eqnarray}
and the 8-component vectors $\tau_{\alpha}(\mathbf{q})$ are defined by 
\begin{eqnarray}
&&\tau_{A}= \nonumber \\
&& (e^{i \mathbf{q} \cdot \mathbf{d}_0^A/2},
e^{i \mathbf{q} \cdot \mathbf{d}_1^A/2},
e^{i \mathbf{q} \cdot \mathbf{d}_2^A/2},
e^{i \mathbf{q} \cdot \mathbf{d}_3^A/2},
0, 
0,
0,
0) \nonumber \\
\label{eq:tauA}\\
&&\tau_{B}=
\nonumber \\
&&  (e^{i \mathbf{q} \cdot \mathbf{d}_0^B/2}, 0, 0, 0,
0,
e^{i \mathbf{q} \cdot \mathbf{d}_1^B/2},
e^{i \mathbf{q} \cdot \mathbf{d}_2^B/2},
e^{i \mathbf{q} \cdot \mathbf{d}_3^B/2})
\label{eq:tauB}
 \nonumber \\
\\
&&\tau_{C}= \nonumber \\
&& (0, 0, 0, 0,
e^{i \mathbf{q} \cdot \mathbf{d}_0^C/2},
e^{i \mathbf{q} \cdot \mathbf{d}_1^C/2},
e^{i \mathbf{q} \cdot \mathbf{d}_2^C/2},
e^{i \mathbf{q} \cdot \mathbf{d}_3^C/2}) \nonumber \\
\label{eq:tauC}\\
&&\tau_{D}= \nonumber \\
&& (0,
e^{i \mathbf{q} \cdot \mathbf{d}_1^D/2},
e^{i \mathbf{q} \cdot \mathbf{d}_2^D/2},
e^{i \mathbf{q} \cdot \mathbf{d}_3^D/2},
e^{i \mathbf{q} \cdot \mathbf{d}_0^D/2},
0,
0
0
) \nonumber \\ \nonumber 
\label{eq:tauD}\\
\end{eqnarray}
with the vector $\mathbf{d}_i^{\alpha}$ defined in Eqs.~(\ref{eq:dA0})--(\ref{eq:dDi}).


At long wavelength (i.e. in the vicinity of a Brillouin zone center), 
the orthogonality conditions Eqs.~(\ref{eq:tauA})--(\ref{eq:tauD}) 
must reduce to the constraints on the classical fields ${\bf P}_{+}$ and ${\bf P}_{-}$, 
discussed Section~\ref{subsec:classical.fields} and Appendix~\ref{appendix:continuum.field.theory}.
The connection between the two approaches can be made explicit by rewriting 
the constraints on bond variables, Eqs.~(\ref{eq:tauA})--(\ref{eq:tauD}), in terms of 
the lattice variable $m_{\bf r}$ and ${\bf P}_{\bf r}$.
To this end we introduce the Fourier transform of the polarisation 
${\bf P}_{\bf r}$ on subblatice $\alpha$
\begin{eqnarray}
{\bf P}_{\alpha}(\mathbf{q})
   = \sqrt{\frac{4}{N}}\sum_{\mathbf{r} \in \alpha} 
   \exp(-i \mathbf{q} \cdot \mathbf{r}) {\bf P}_{\bf r} \; ,
\end{eqnarray}

For the moment, we restrict our discussion to
wavevectors ${\bf q}$ which are close to a
reciprocal lattice vector ${\bf Q}$ and write
\begin{eqnarray}
\mathbf{q}=\mathbf{Q}+\tilde{\mathbf{q}}
\end{eqnarray}

By analogy with Eqs. (\ref{eq:p+def})--(\ref{eq:p-def}) of the main text, 
we write
\begin{eqnarray}
&& m_{\pm}(\tilde{\mathbf{q}}) =
   \frac{1}{\sqrt{2}} 
   \left[ 
        \exp(i \mathbf{Q} \cdot \mathbf{r}_A)\ m_A (\mathbf{q}) 
  \right. 
 \nonumber \\
&& \left. 
 \qquad \qquad \qquad \qquad 
 \pm 
     \exp(i \mathbf{Q} \cdot \mathbf{r}_C)\ m_{C}(\mathbf{Q}+\tilde{\mathbf{q}}) 
 \right]  \nonumber \\
\label{eq:FT.m}
 \\
&& {\bf P}_{\pm}(\tilde{\mathbf{q}}) =
\frac{1}{\sqrt{2}} \left[   \exp(i \mathbf{Q} \cdot \mathbf{r}_A)\ {\bf P}_A (\mathbf{Q}+\tilde{\mathbf{q}}) 
\right. \nonumber \\
&& \left. 
\qquad \qquad \qquad \qquad \pm 
 \exp(i \mathbf{Q} \cdot \mathbf{r}_C)\ 
{\bf P}_{C} (\mathbf{Q}+\tilde{\mathbf{q}}) \right] \, . \nonumber \\
\label{eq:FT.P}. 
\end{eqnarray}
The vectors $\mathbf{r}_A$ and $\mathbf{r}_C$ are the positions of the oxygen sites labelled
$A$ and $C$ within a primitive unit cell.
Note that we have defined the fields in Eqs. (\ref{eq:FT.m}) and (\ref{eq:FT.P})
in such a way that they are independent of the
reciprocal lattice vector $\mathbf{Q}$ and depend only on the distance to the
zone center $\tilde{\mathbf{q}}$.
We may therefore use Eqs. (\ref{eq:p+corr})-(\ref{eq:p-zcorr}),
with the replacement 
\begin{eqnarray}
\mathbf{q} \to \tilde{\mathbf{q}}
\end{eqnarray}
 to describe the correlations
at small $\tilde{\mathbf{q}}$ in the vicinity of {\it all}
reciprocal lattice vectors $\mathbf{Q}$, i.e. near all Brillouin zone
centers.


To demonstrate that the constraints derived in the continuum theory
[Eqs. (\ref{eq:p+constraint})-(\ref{eq:p-constraint})] are
equivalent to the constraints in the lattice theory [Eq. (\ref{eq:tauconstraints}) ]
in the limit $\tilde{q} \to 0$
we consider the following linear combinations
of the  vectors $\tau(\mathbf{q})=\tau(\mathbf{Q}+\tilde{\mathbf{q}})$ 
\begin{eqnarray}
&&\tau_{AC+}(\mathbf{Q},\tilde{\mathbf{q}})= 
\frac{1}{\sqrt{2}} 
(\exp(i \mathbf{Q} \cdot \mathbf{r}_A)\tau_A(\mathbf{Q}+\tilde{\mathbf{q}})
\nonumber \\
&& \qquad \qquad \qquad
+
(\exp(i \mathbf{Q} \cdot \mathbf{r}_C)\tau_C(\mathbf{Q}+\tilde{\mathbf{q}})) 
\label{eq:tauAC+} 
\\
&&\tau_{AC-}(\mathbf{Q},\tilde{\mathbf{q}})
= \frac{1}{\sqrt{2}} 
(
\exp(i \mathbf{Q} \cdot \mathbf{r}_A)\tau_A(\mathbf{Q}+\tilde{\mathbf{q}})
\nonumber
\\ 
&& \qquad \qquad \qquad
-
\exp(i \mathbf{Q} \cdot \mathbf{r}_C)
\tau_C(\mathbf{Q}+\tilde{\mathbf{q}})) \nonumber \\
\label{eq:tauAC-}\\
&&\tau_{BD+}(\mathbf{Q},\tilde{\mathbf{q}})=
 \frac{1}{\sqrt{2}} 
(\exp(i \mathbf{Q} \cdot \mathbf{r}_B)\tau_B(\mathbf{Q}+\tilde{\mathbf{q}})+ \nonumber \\
&&
\qquad \qquad \qquad
\exp(i \mathbf{Q} \cdot \mathbf{r}_D)\tau_D(\mathbf{Q}+\tilde{\mathbf{q}}))  \nonumber \\
\label{eq:tauBD+}\\
&&\tau_{BD-}(\mathbf{Q},\tilde{\mathbf{q}})= 
 \frac{1}{\sqrt{2}} 
(\exp(i \mathbf{Q} \cdot \mathbf{r}_B)\tau_B(\mathbf{Q}+\tilde{\mathbf{q}})- \nonumber \\
&&
\qquad \qquad \qquad
\exp(i \mathbf{Q} \cdot \mathbf{r}_D)\tau_D(\mathbf{Q}+\tilde{\mathbf{q}}))  \nonumber \\
\label{eq:tauBD-} \; ,
\end{eqnarray}
where ${\bf r}_{A}, {\bf r}_B, {\bf r}_C$ and ${\bf r}_D$ are the positions of each oxygen vertex
in the unit cell.
We can then express the constraints on $\tilde{\sigma}(\mathbf{q})$ 
[Eq.~(\ref{eq:tauconstraints})] as 
\begin{eqnarray}
(\tau_{AC+}(\mathbf{Q},\tilde{\mathbf{q}}), \tilde{\sigma}(\mathbf{Q}+\tilde{\mathbf{q}}))&=&0   
\label{eq:tauAC+constraint}\\
(\tau_{AC-}(\mathbf{Q},\tilde{\mathbf{q}}), \tilde{\sigma}(\mathbf{Q}+\tilde{\mathbf{q}}))&=&0   
\label{eq:tauAC-constraint}\\
(\tau_{BD+}(\mathbf{Q},\tilde{\mathbf{q}}), \tilde{\sigma}(\mathbf{Q}+\tilde{\mathbf{q}}))&=&0  
\label{eq:tauBD+constraint}\\
(\tau_{BD-}(\mathbf{Q},\tilde{\mathbf{q}}), \tilde{\sigma}(\mathbf{Q}+\tilde{\mathbf{q}}))&=&0 \; .
\label{eq:tauBD-constraint} 
\end{eqnarray}
Since
\begin{eqnarray}
(\tau_{AC+}(\mathbf{Q},\tilde{\mathbf{q}}), \tilde{\sigma}(\mathbf{Q}+\tilde{\mathbf{q}})) &=& m_{+}(\tilde{\mathbf{q}}) \\
(\tau_{AC-}(\mathbf{Q},\tilde{\mathbf{q}}), \tilde{\sigma}(\mathbf{Q}+\tilde{\mathbf{q}})) &=& m_{-}(\tilde{\mathbf{q}}) \; ,
\end{eqnarray}
the constraints Eq.~(\ref{eq:tauAC+constraint}) and Eq.~(\ref{eq:tauAC-constraint}) 
are satisfied if
\begin{eqnarray}
m_{+}(\tilde{\mathbf{q}})=m_{-}(\tilde{\mathbf{q}})=0 \quad \forall \ \  \mathbf{q} \; .
\end{eqnarray}


The constraints on $\tau(\mathbf{Q},\tilde{\mathbf{q}})$, Eq.~(\ref{eq:tauBD+constraint}) 
and Eq.~(\ref{eq:tauBD-constraint}),
 give rise to constraints on the $\tilde{{\bf q}}$ dependence of 
${\bf P}_+ (\tilde{\mathbf{q}})$ and ${\bf P}_-(\tilde{\mathbf{q}})$.
Reexpressing Eq.~(\ref{eq:tauBD+constraint}) in terms of 
${\bf P}_+ (\tilde{\mathbf{q}})$ [Eq.~(\ref{eq:p+def})] 
and ${\bf P}_-(\tilde{\mathbf{q}})$ [Eq.~(\ref{eq:p-def})], 
and expanding to linear order in $\mathbf{q}$ we find
\begin{eqnarray}
\tilde{\mathbf{q}} \cdot {\bf P}_+(\tilde{\mathbf{q}}) +
\mathcal{O}(a_{\sf O}^2)=0
\label{eq:p+longwavelength}
\end{eqnarray}
This precisely the Fourier transform of Eq.~(\ref{eq:p+constraint}),
i.e. the required condition on ${\bf P}_+$.
Similarly, Eq.~(\ref{eq:tauBD-constraint}) becomes
\begin{eqnarray}
P^z_-(\tilde{\mathbf{q}})+ ia_{\sf O} \left( \frac{2}{3} \mathbf{q} \cdot {\bf P}^-(\tilde{\mathbf{q}}) -  q_z P^z_-(\tilde{\mathbf{q}}) \right)
+
\mathcal{O}(a_{\sf O}^2)
=0 \nonumber \\
\label{eq:p-longwavelength}.
\end{eqnarray}
which is the Fourier transform of Eq.~(\ref{eq:p-constraint}).


Having established that the lattice--based theory is equivalent to the continuum field theory 
at long wavelength, we now turn to the problem of calculating the structure 
factors which describe proton--proton correlations for arbitrary ${\bf q}$.
Within the lattice--based theory, this reduces to constructing a
matrix $\mathcal{P}$ which projects states into the subspace 
of proton configurations {\it orthogonal} to the set of vectors 
$\tau_{\alpha}(\mathbf{q})$.
Explicit construction of $\mathcal{P}$ is messy, 
and the final expression for the structure factor must
be evaluated numerically.


We proceed by using Gram-Schmidt orthogonalisation \cite{arfken95} to construct 
from $\{ \tau_{\alpha} (\mathbf{q})\}$ an orthogonal basis set $\{ \tau_{\alpha} (\mathbf{q})\}'$
\begin{eqnarray}
\tau_A'&=&\tau_A \\
\tau_B'&=&\tau_B-\frac{(\tau_B, \tau_A')}{(\tau_A', \tau_A')}\tau_A' \\
\tau_C'&=&\tau_C-\frac{(\tau_C, \tau_A')}{(\tau_A', \tau_A')}\tau_A'-\frac{(\tau_C, \tau_B')}{(\tau_B', \tau_B')}\tau_B' \\
\tau_D'&=&\tau_D-\frac{(\tau_D, \tau_A')}{(\tau_A', \tau_A')}\tau_A'-\frac{(\tau_D, \tau_B')}{(\tau_B', \tau_B')}\tau_B'-\frac{(\tau_D, \tau_C')}{(\tau_C', \tau_C')}\tau_C' \nonumber \\
\end{eqnarray}
The projection matrix $\mathcal{P}(\mathbf{q})$ is then given by
\begin{eqnarray}
\mathcal{P}_{ij}(\mathbf{q})=\delta_{ij}-\sum_{\alpha} \frac{\tau_{\alpha i}' \tau_{\alpha j}^{\prime \ast}}{(\tau_{\alpha}', \tau_{\alpha}')}
\end{eqnarray}


The structure factor is obtained by acting on the Fourier transform of some
general (non-ice rule obeying) state $\tilde{\sigma}'(\mathbf{q})$
\begin{eqnarray}
&&\langle \sigma_i (-\mathbf{q}) \sigma_j(\mathbf{q}) \rangle=
  \nonumber \\
&& \qquad \qquad=\sum_{mn} \mathcal{P}_{in}(-\mathbf{q}) \mathcal{P}_{jm}(\mathbf{q})
\langle \sigma_n'(-\mathbf{q})  \sigma_m'(\mathbf{q}) \rangle \nonumber \\
&& \qquad \qquad=\sum_{mn} \mathcal{N}_0 \delta_{mn}
\mathcal{P}_{in}(-\mathbf{q}) \mathcal{P}_{jm}(\mathbf{q})
 \nonumber \\
&& \qquad \qquad =\mathcal{N}_0
\mathcal{P}_{ij}(\mathbf{q})
\end{eqnarray}
where $\mathcal{N}_0$ is a normalisation constant and in the last
step we have used the identities
\begin{eqnarray}
\mathcal{P}(-\mathbf{q})&=&\mathcal{P}(\mathbf{q})^{\ast} \\
\mathcal{P}(\mathbf{q})^{\ast}&=&\mathcal{P}(\mathbf{q})^T \\
\mathcal{P}(\mathbf{q})^2&=&\mathcal{P}(\mathbf{q}).
\end{eqnarray}

The constant $\mathcal{N}_0$ is fixed by Eq. (\ref{eq:avenorm}).
This method was used to calculate the scattering patterns
in Figs. \ref{fig:Sq.Ising.classical} and \ref{fig:Sq.proton.classical}.

\section{Relating structure factors to correlations of ${\bf P}_+(\mathbf{q})$ and ${\bf P}_-(\mathbf{q})$}
\label{appendix:structure.factors.and.fields}

In this Appendix we derive the relationship between the structure factors
$S_{\sf Ising}(\mathbf{q})$ 
and $S^{\sf diffuse}_{\sf proton}(\mathbf{q})$, 
discussed in Section~\ref{subsec:classical.lattice.theory} 
and Section~\ref{subsec:classical.predictions} of the main text, 
and the correlation functions of the fields 
${\bf P}_{\pm}$, introduced in Section~\ref{subsec:classical.fields}.  
In so doing we establish the necessary conditions for a zone centre to exhibit 
singular, pinch--point scattering. 


The structure factors $S_{\sf Ising}(\mathbf{q})$~[Eq.~(\ref{eq:S.of.q.Ising})] 
and $S^{\sf diffuse}_{\sf proton}(\mathbf{q})$~[Eq.~(\ref{eq:SHdiffuse})]
can both be expressed in terms of the correlations of the Ising variable 
$\sigma_{\nu}(\mathbf{q})$~[Eq.~(\ref{eq:sigma.q})], as
\begin{eqnarray}
&&S_{\sf \chi}(\mathbf{q})=
 \sum_{\nu \nu'} \langle \sigma_{\nu}(\mathbf{q})
\sigma_{\nu'}(-\mathbf{q}) \rangle 
\ \eta^{\chi}_{\nu}(\mathbf{q}) 
\ \eta^{\chi}_{\nu'}(-\mathbf{q}) 
\label{eq:s.chi.q.definition}
\end{eqnarray}
where 
\begin{eqnarray}
{\chi}={\sf Ising} \; , \; {\sf proton} \, , 
\end{eqnarray}
the sum on $\nu$ runs over all bonds within the unit cell, 
and the coefficients $\eta^{\chi}_{\nu}(\mathbf{q})$ depend 
on which structure factor is being calculated. 
In the case of $S_{\sf Ising}(\mathbf{q})$ 
\begin{eqnarray}
\eta_{\nu}^{\sf Ising}(\mathbf{q})=1 \ \ \forall \ \nu, \mathbf{q}  \,\, ,
\end{eqnarray}
while for $S^{\sf diffuse}_{\sf proton}(\mathbf{q})$
\begin{eqnarray}
\eta^{\sf proton}_{\nu} ( \mathbf{q} )
    = i \sin( \mathbf{q} \cdot {\bf a}_{\nu} )  \, ,
\end{eqnarray}
where the vectors ${\bf a}_{\nu}$ describe the displacement of the
protons from their bond midpoints, as defined in Eq.~(\ref{eq:dispvectors}).


The expression for $S_{\sf \chi}(\mathbf{q})$ [Eq.~(\ref{eq:s.chi.q.definition})] 
can be factorised to give
\begin{eqnarray}
S_{\chi}(\mathbf{q}) = \langle \big| \sum_{\nu} 
\eta^{\chi}_{\nu}(\mathbf{q}) \sigma_{\nu}(\mathbf{q}) \big|^2 \rangle
\label{eq:S.chi.q.factorised}
\end{eqnarray}
We are interested in understanding the behaviour of this structure factor 
near to a given reciprocal lattice vector $\mathbf{Q}$, in terms of 
the continuum field theory developed in Section~\ref{subsec:classical.fields}.  
To this end, we write
\begin{eqnarray}
\mathbf{q} = \mathbf{Q} + \tilde{\mathbf{q}}
\end{eqnarray}
and express $S_{\chi}(\mathbf{q})$ in terms of the fields 
$m_{\pm}(\tilde{\mathbf{q}})$ [Eq.~(\ref{eq:FT.m})] 
and ${\bf P}_{\pm}(\tilde{\mathbf{q}})$ [Eq.~(\ref{eq:FT.P})], 
to obtain
\begin{eqnarray}
&&\sum_{\nu} 
\eta^{\chi}_{\nu}(\mathbf{Q}+\tilde{\mathbf{q}}) \ 
\sigma_{\nu}(\mathbf{Q}+\tilde{\mathbf{q}})
\nonumber \\
&& \quad =
\mu_{+}^{\chi}(\mathbf{Q},\tilde{\mathbf{q}})\ m_{+}(\tilde{\mathbf{q}})
+
\mu_{-}^{\chi}(\mathbf{Q},\tilde{\mathbf{q}})\ m_{-}(\tilde{\mathbf{q}}) \nonumber \\
&& \qquad 
+ \; 
\boldsymbol{\lambda}_{+}^{\chi}(\mathbf{Q},\tilde{\mathbf{q}}) \cdot {\bf P}_{+}(\tilde{\mathbf{q}})
+ \; 
\boldsymbol{\lambda}_{-}^{\chi}(\mathbf{Q},\tilde{\mathbf{q}}) \cdot {\bf P}_{-}(\tilde{\mathbf{q}}) \; ,
\nonumber \\
\end{eqnarray}
where the vectors $\boldsymbol{\lambda}^{\sf Ising}_{\pm}(\mathbf{q})$ are defined
in Eq.~(\ref{eq:defn-Ising-lambda}) and Eq.~(\ref{eq:defn-proton-lambda}), below, 
and the scalar functions $\mu_{\pm}^{\chi}(\mathbf{Q},\tilde{\mathbf{q}})$ 
play no part in our subsequent discussion.


In any state obeying the ice rules
\begin{eqnarray}
m_{+}(\tilde{\mathbf{q}})=m_{-}(\tilde{\mathbf{q}})=0  
\end{eqnarray}
so terms in these fields can safely be dropped
\begin{eqnarray}
&&\sum_{\nu} 
\eta^{\chi}_{\nu}(\mathbf{Q}+\tilde{\mathbf{q}}) 
\sigma_{\nu}(\mathbf{Q}+\tilde{\mathbf{q}})
\nonumber \\
&& \qquad =  
    \boldsymbol{\lambda}_{+}^{\chi}(\mathbf{Q},\tilde{\mathbf{q}}) \cdot {\bf P}_{+}(\tilde{\mathbf{q}})
+  \boldsymbol{\lambda}_{-}^{\chi}(\mathbf{Q},\tilde{\mathbf{q}}) \cdot {\bf P}_{-}(\tilde{\mathbf{q}}) \; .
\nonumber \\
\end{eqnarray}
It follows that, for $| \tilde{\mathbf{q}} | \ll 1$, the structure factor 
$S_{\chi}(\mathbf{q})$~[Eq.~\ref{eq:S.chi.q.factorised}]  can be written as 
\begin{eqnarray}
S_{\chi} (\mathbf{Q} + \tilde{\mathbf{q}}) 
\approx 
     \sum_{\upsilon = \pm}
          F^\chi_\upsilon ( \mathbf{Q} )
         \langle 
              | \boldsymbol{\hat{\lambda}}_{\mathbf{Q}, \upsilon}^\chi
              \cdot {\bf P}_\upsilon ( \tilde{\mathbf{q}} ) |^2 
         \rangle \; ,
\label{eq:S.chi.q}
\end{eqnarray}
where $\boldsymbol{\hat{\lambda}}_{\mathbf{Q}, \upsilon}^\chi$ is a unit 
vector in the direction of
\begin{eqnarray}
\boldsymbol{\lambda}_{\mathbf{Q}, \upsilon}^\chi 
   = \boldsymbol{\lambda}^\chi_\upsilon (\mathbf{Q}, \tilde{\mathbf{q}} = 0 ) \; ,
   \label{eq:lambda.Q.upsilon}
\end{eqnarray}
[cf. Eq.~(\ref{eq:defn-Ising-lambda}) and Eq.~(\ref{eq:defn-proton-lambda})], and the form factor 
\begin{eqnarray}
F^\chi_\upsilon ( \mathbf{Q} ) = | \boldsymbol{\lambda}_{\mathbf{Q}, \upsilon}^\chi  |^2 \; .
\label{eq.F.Q.upslion}
\end{eqnarray}


The final result for the structure factor, Eq.~(\ref{eq:S.chi.q}), is strikingly simple.
We learn that, evaluated near to a reciprocal lattice 
vector $\mathbf{Q}$, the structure factor $S_{\chi} (\mathbf{q})$ 
measures correlations of both ${\bf P}_+ (\tilde{\mathbf{q}})$ and ${\bf P}_- (\tilde{\mathbf{q}})$, 
where each is projected onto the vector $\boldsymbol{\hat{\lambda}}_{\mathbf{Q}, +}^\chi$ 
and $\boldsymbol{\hat{\lambda}}_{\mathbf{Q}, -}^\chi$, respectively.
These correlations are mixed with a weight controlled by the form factors
$F^\chi_+ ( \mathbf{Q} )$ and $F^\chi_- ( \mathbf{Q} )$.   
The form factors are, in turn, fully determined by the vectors 
$\boldsymbol{\lambda}_{\mathbf{Q}, +}^\chi $ and 
$\boldsymbol{\lambda}_{\mathbf{Q}, -}^\chi $.
These vectors 
therefore control both the ``selection rules'' which determine which of the 
fields ${\bf P}_+$ and ${\bf P}_-$ is manifest in the scattering around 
a given reciprocal lattice vector, and the way in which these fields 
couple to a given experimental probe.


All that remains is to substitute the appropriate 
$\boldsymbol{\lambda}_{\mathbf{Q}, \upsilon}^\chi$, 
in the expression for  $S_{\chi} (\mathbf{q})$~[Eq.~(\ref{eq:S.chi.q})], 
using Eq.~(\ref{eq:lambda.Q.upsilon}).
In the specific case of the Ising structure factor 
$S_{\sf Ising}(\mathbf{q})$~[Eq.~(\ref{eq:S.of.q.Ising})], 
we have 
\begin{eqnarray}
&&\boldsymbol{\lambda}^{\sf Ising}_{\pm}(\mathbf{Q}, \tilde{\mathbf{q}})
\nonumber \\
&& \quad = 
   \exp(-i\mathbf{Q} \cdot \mathbf{r}_A)
      	\sum_{i=0}^3 
 	\frac{3}{4 a_{\sf O}} {\bf d}_i^{A} 
 	\exp[-i(\mathbf{Q} + \tilde{\mathbf{q}})\cdot{\bf d}_i^2/2]
 \nonumber \\
&& \quad \quad
   \pm \exp(-i\mathbf{Q} \cdot \mathbf{r}_C)
	\sum_{i=0}^3 \frac{3}{4 a_{\sf O}} {\bf d}_i^{C} 
	\exp[-i( \mathbf{Q} + \tilde{\mathbf{q}} )\cdot{\bf d}_i^C/2]  \; ,
\nonumber \\
\label{eq:defn-Ising-lambda}
\end{eqnarray}
where the vectors $\mathbf{d}_i^{A}, \mathbf{d}_i^{C}$ 
are given in Eqs. (\ref{eq:dA0})--(\ref{eq:dC3}).


Meanwhile, for the proton structure factor 
$S^{\sf diffuse}_{\sf proton}(\mathbf{q})$ [Eq.~(\ref{eq:SHdiffuse})], 
we have 
\begin{eqnarray}
&& \boldsymbol{\lambda}^{\sf proton}_{\pm} ( \mathbf{Q}, \tilde{\mathbf{q}} )
\nonumber \\
&& \quad = \exp(-i\mathbf{Q} \cdot \mathbf{r}_A)
 	\sum_{i=0}^3 \frac{3}{4 a_{\sf O}} {\bf d}_i^{A} 
	\exp[-i(\mathbf{Q} + \tilde{\mathbf{q}}) \cdot{\bf d}_i^2/2] 
\nonumber \\
&& \qquad \qquad \qquad \qquad \qquad \qquad
	\times i\sin [ \phi \ (\mathbf{Q} +\tilde{\mathbf{q}} \cdot{\bf d}_i^A )]
\nonumber \\
&& \quad \pm  \exp(-i\mathbf{Q} \cdot \mathbf{r}_C)
	\sum_{i=0}^3 \frac{3}{4 a_{\sf O}} {\bf d}_i^{C} 
	\exp[-i( \mathbf{Q} +\tilde{\mathbf{q}} ) \cdot{\bf d}_i^C/2] 
\nonumber \\
&& \qquad \qquad \qquad \qquad \qquad \qquad
	\times i\sin [ \phi \  (\mathbf{Q} +\tilde{\mathbf{q}}\cdot{\bf d}_i^C )]
\nonumber \\
\label{eq:defn-proton-lambda}
\end{eqnarray}
where the parameter 
\begin{eqnarray}
\phi = 0.15
\end{eqnarray}
expresses the relative displacement of the protons from the 
midpoint of the bond, as defined in Eq.~(\ref{eq:displacement.parameter}).

\section{Calculation of the dispersion of emergent photons on the ice Ih lattice}
\label{appendix:photons}

The Hamiltonian of the $U(1)$ gauge theory on the pyrochlore
lattice is
\begin{eqnarray}
&&\mathcal{H}_{\sf U(1)}
 = \frac{\mathcal{U}}{2} \sum_{\langle \mathbf{r} \mathbf{r}' \rangle \in  {\sf CS}}
E_{\mathbf{r} \mathbf{r}'}^2 
 +\frac{\mathcal{U^{\prime}}}{2} \sum_{\langle \mathbf{r} \mathbf{r}' \rangle \in {\sf MS}}
E_{\mathbf{r} \mathbf{r}'}^2  \nonumber \\
 && \qquad + \frac{\mathcal{K}}{2} \sum_{\hexagon \in I}
   \left[  \nabla_{\scriptsize\hexagon} \times A  \right]^2 
   + \frac{\mathcal{K^{\prime}}}{2} \sum_{\hexagon \in II}
   \left[ \nabla_{\scriptsize\hexagon} \times A \right]^2 
\label{eq:HU1-app}
\end{eqnarray}
where the sum $\sum_{\langle \mathbf{r} \mathbf{r}' \rangle \in  {\sf CS}}$ runs over centre symmetric
oxygen--oxygen bonds, $\sum_{\langle \mathbf{r} \mathbf{r}' \rangle \in  {\sf MS}}$ runs over mirror symmetric 
oxygen--oxygen bonds [cf. Fig. \ref{fig:bondsymmetries}],
 $ \sum_{\mathbf{r}_p \in I}$ runs over plaquettes normal to the optical axis
and  $ \sum_{\mathbf{r}_p \in II}$ runs over plaquettes parallel to the optical 
axis [cf. Eq. (\ref{eq:htunnelling2})].
We may condense this as
\begin{eqnarray}
\label{eq:HIh-condensed}
\mathcal{H}_{\sf Ih} 
   = \frac{\mathcal{U}}{2} \sum_{\mathbf{r}} \sum_m \alpha_{m}
 E_{\mathbf{r} m}^2 
   + \frac{\mathcal{K}}{2} \sum_{\mathbf{r}} \sum_{p} \beta_{p}
   \left[ \left( \nabla_{\scriptsize\hexagon} \times A \right)_{\mathbf{r}, p} \right]^2 
\nonumber \\
\end{eqnarray}
where $\sum_{\mathbf{r}}$ is a sum over primitive unit cells and the sums over $m$ and $p$
are over bond and plaquette midpoints in a single unit cell respectively.
For the eight component objects $\boldsymbol{\alpha}$ and $\boldsymbol{\beta}$ we have
\begin{eqnarray}
\boldsymbol{\alpha}&=&\bigg( 1, 1, 1, \frac{\mathcal{U}^{\prime}}{\mathcal{U}}, 
1, 1,1, \frac{\mathcal{U}^{\prime}}{\mathcal{U}} \bigg) \\
\boldsymbol{\beta}&=&\bigg(  \frac{\mathcal{K}^{\prime}}{\mathcal{K}},  \frac{\mathcal{K}^{\prime}}{\mathcal{K}}, 
 \frac{\mathcal{K}^{\prime}}{\mathcal{K}}, 1,
\frac{\mathcal{K}^{\prime}}{\mathcal{K}},
 \frac{\mathcal{K}^{\prime}}{\mathcal{K}},  \frac{\mathcal{K}^{\prime}}{\mathcal{K}}, 
 1 \bigg).
\end{eqnarray}


We will use the notation 
\begin{eqnarray}
E_{(\mathbf{r}, m)}=E_{\mathbf{r}, \mathbf{r}+{\bf d}_m} \\
A_{(\mathbf{r}, m)}=A_{\mathbf{r}, \mathbf{r}+{\bf d}_m}
\end{eqnarray}
where ${\bf d}_m$ is a bond vector in the direction of one of the eight
inequivalent bonds in the unit cell.


We need now to write down an decomposition of these fields in terms of
photon operators.
This is
\begin{eqnarray}
\label{eq:AquantisedIh}
&& A_{(\mathbf{r}, m)}
   =\sqrt{\frac{4}{N}} \sum_{\mathbf{k}} \sum_{\lambda=1}^{8}
  \sqrt{\frac{\alpha_m \mathcal{U}}{\omega_{\lambda}(\mathbf{k})}} 
    \times
 \nonumber \\
&& \quad \qquad
          \left( \exp\left[-i \mathbf{k} \cdot (\mathbf{r}+\mathbf{d}_m/2)\right] 
          \eta_{m \lambda}(\mathbf{k}) a^{\phantom\dagger}_{\lambda}(\mathbf{k}) \right. \nonumber \\
&&
\left. \quad \qquad
+  \exp\left[i \mathbf{k} \cdot (\mathbf{r}+\mathbf{d}_m/2)\right] 
             \eta^{\ast}_{\lambda m}(\mathbf{k}) a^{\dagger}_{\lambda}(\mathbf{k}) \right) \\
\label{eq:EquantisedIh}
&& E_{(\mathbf{r}, m)}
    = i  \sqrt{\frac{4}{N}} \sum_{\mathbf{k}} \sum_{\lambda=1}^{8} 
            \sqrt{\frac{\omega_{\lambda}(\mathbf{k})}{\alpha_m \mathcal{U}}} 
    \times \nonumber \\
&& \quad \qquad \left( \exp\left[-i \mathbf{k} \cdot (\mathbf{r}+\mathbf{d}_m/2)\right] 
           \eta_{m \lambda}(\mathbf{k}) a^{\phantom\dagger}_{\lambda}(\mathbf{k}) 
\right. \nonumber\\
&& \left. \quad \qquad
         - \exp\left[i \mathbf{k} \cdot (\mathbf{r}+\mathbf{d}_m/2)\right] 
           \eta^{\ast}_{\lambda m}(\mathbf{k}) a^{\dagger}_{\lambda}(\mathbf{k}) \right)
\end{eqnarray}

It is easy to check that that Eqs. (\ref{eq:AquantisedIh})-(\ref{eq:EquantisedIh}) fulfil the commutation relationship
\begin{eqnarray}
[E_{({\bf r}, m)}, A_{({\bf r}, m)}]=i.
\end{eqnarray}


\begin{figure*}
\centering
\subfigure[\ \ $\langle {\bf P}_{+}(-\mathbf{q}, -\omega)
\cdot {\bf P}_{+}(\mathbf{q}, \omega) \rangle$]{
\includegraphics[width=.45\textwidth]{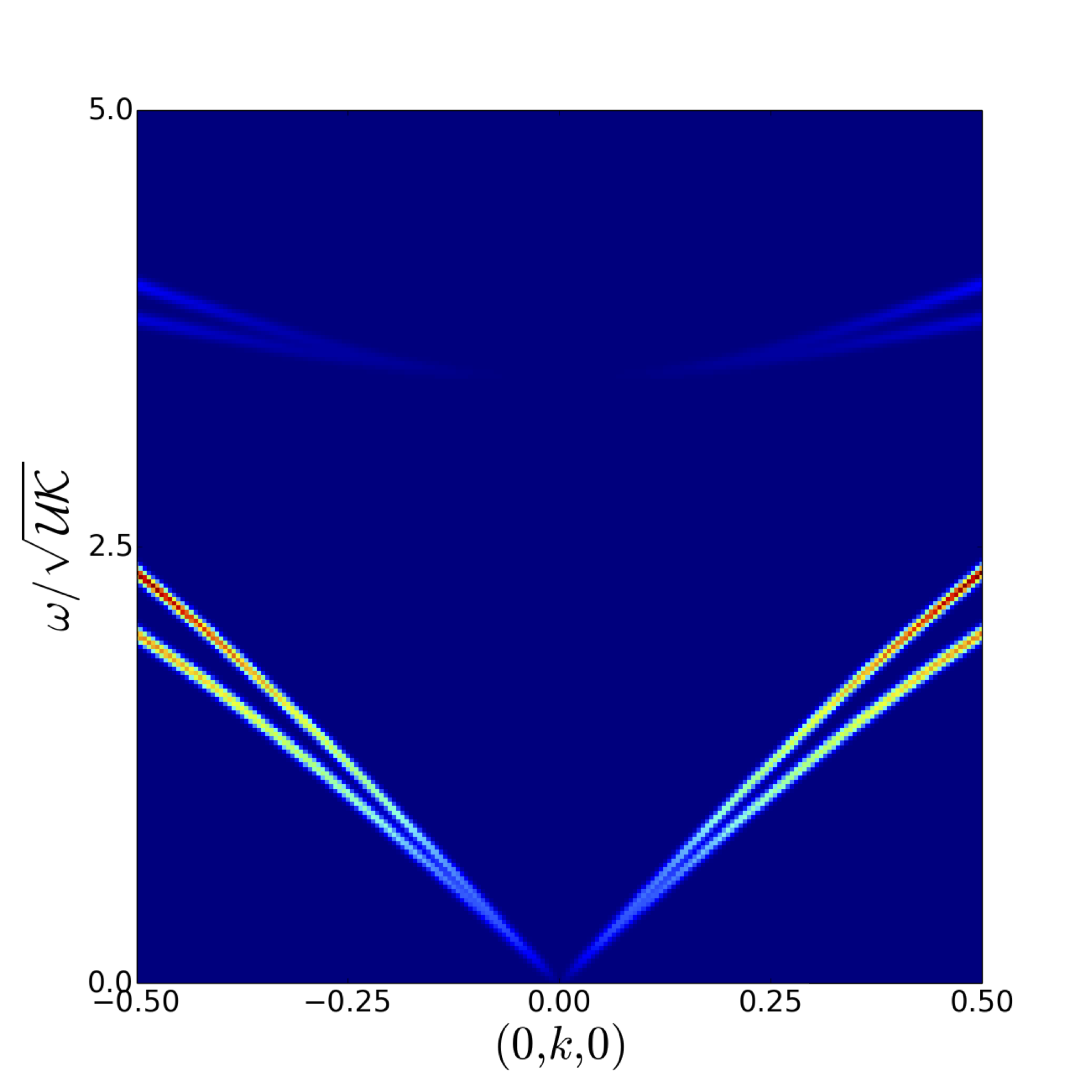}}%
    \qquad
\subfigure[\ \  $\langle {\bf P}_{-}(-\mathbf{q}, -\omega)
\cdot {\bf P}_{-}(\mathbf{q}, \omega) \rangle$]{%
\includegraphics[width=.45\textwidth]{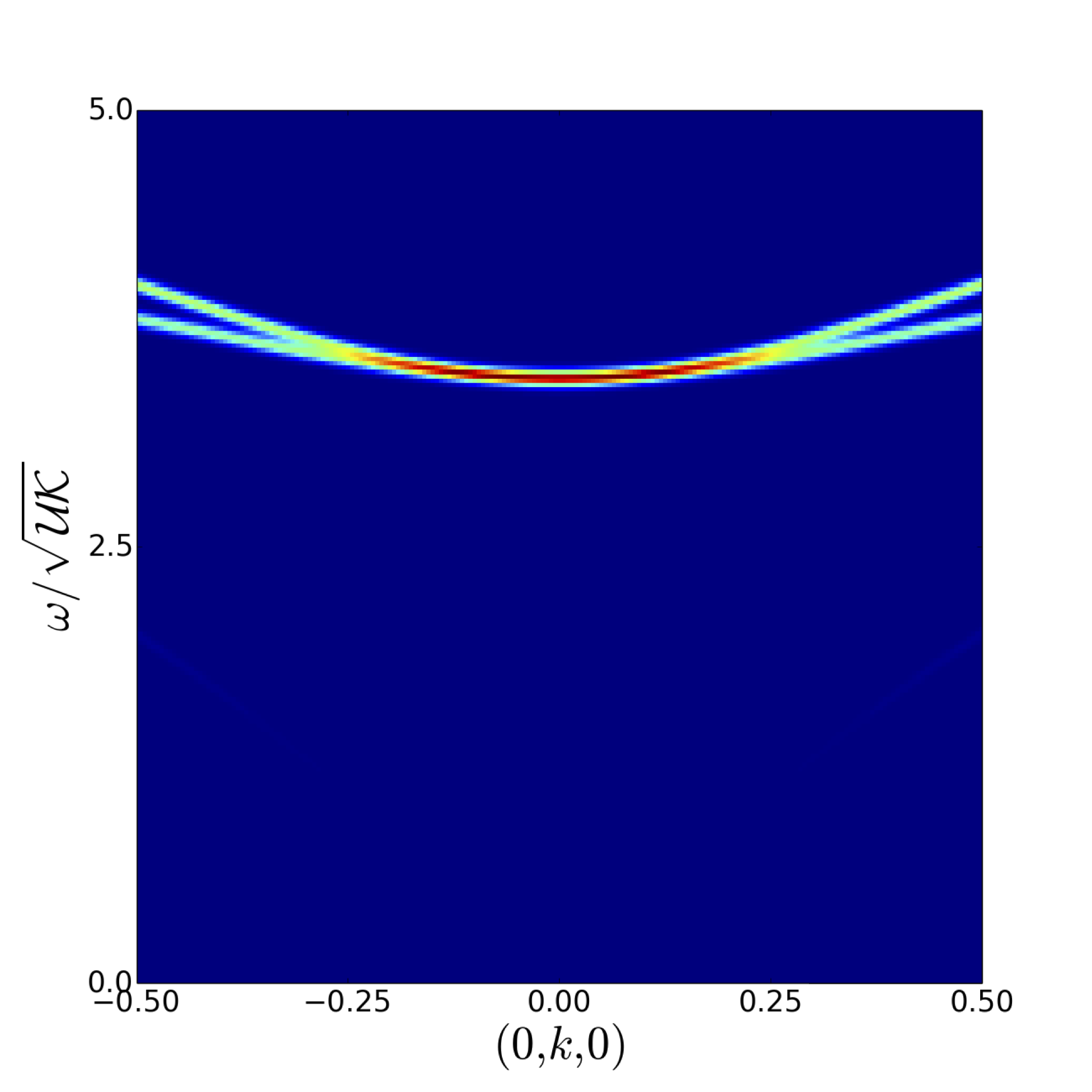}}%
\caption{
(Color online).
Identification of the excitations of the lattice gauge theory, 
$\mathcal{H}_{\sf U(1)}$~[Eq.~(\ref{eq:HU1-app})], 
with quantised fluctuations of the fields ${\bf P}_{+}$ and ${\bf P}_{-}$.
(a)~Dynamical structure factor of the uniform polarisation, ${\bf P}_{+}$
[Eq.~(\ref{eq:FT.P.quantum})], calculated within the lattice gauge theory.
(b)~Equivalent results for the dynamical structure factor of the staggered polarisation, ${\bf P}_{-}$.
For $q \to 0$, the correlations of ${\bf P}_+$ are directly associated with the gapless, 
birefringent photons of the lattice gauge theory, while the correlations of ${\bf P}_-$ 
are associated with its gapped optical modes.
This identification breaks down at shorter wavelengths, where the correlations of 
${\bf P}_+$ and ${\bf P}_-$ have contributions from all modes of the lattice gauge theory.
Calculations were carried out for the symmetric choice of parameters,
$\mathcal{U}=\mathcal{U}'$ and $\mathcal{K}=\mathcal{K}'$.
} 
\label{fig:p+p-omega}
\end{figure*}


As in Ref. \cite{benton12}, we may
 write $\nabla_{\scriptsize \hexagon} \times A_{\mathbf{r}m}$ in terms
of our Fourier decomposition as a matrix $\uuline{Z}(\mathbf{k})$ acting on the vectors 
$\uline{\eta}_{\lambda}(\mathbf{k})$.
\begin{widetext}
We have
\begin{eqnarray}
&&\sqrt{\beta_p} \left( \nabla_{\scriptsize\hexagon} \times A \right)_{(\mathbf{r}, p)}  
= \sqrt{\frac{4}{N}} \sum_{\mathbf{k}} \sum_{\lambda=1}^{4} 
    \sqrt{\frac{\mathcal{U}}{\omega_{\lambda}(\mathbf{k})}} 
  \bigg\{ \exp[-i \mathbf{k} \cdot (\mathbf{r}_p)] a^{\phantom\dagger}_{\lambda}(\mathbf{k})
   \sum_{m} \sqrt{\beta_p} Z_{pm}(\mathbf{k}) \sqrt{\alpha_m} \eta_{m \lambda}(\mathbf{k})  \nonumber \\
 && \qquad \qquad \qquad
+ \exp\left[ i \mathbf{k} \cdot (\mathbf{r}_p) \right] a^{\dagger}_{\lambda}(\mathbf{k})
  \sum_{m}  
       \eta^{\ast}_{\lambda m}  \sqrt{\beta_p}  Z_{pm}(\mathbf{k})^{\ast} \sqrt{\alpha_m} \bigg\}. 
\label{eq:curl1hA}
\end{eqnarray}


The matrix $\uuline{Z}(\mathbf{k})$ is
\begin{eqnarray}
\label{eq:Z}
\uuline{Z}(\mathbf{k})= \qquad 
 \qquad  \qquad  \qquad  \qquad  \qquad  \qquad  \qquad  \qquad  \qquad  \qquad  \qquad 
\qquad  \qquad  \qquad  \qquad  \qquad  \qquad  \qquad  \qquad  \qquad  \qquad  \qquad 
\qquad
\nonumber \\
  \begin{pmatrix} 
  0 & 2i \sin(\mathbf{k} \cdot \mathbf{c}_{01}) & 2i \sin(\mathbf{k} \cdot \mathbf{c}_{02}) & 2i \sin(\mathbf{k} \cdot \mathbf{c}_{03}) 
  & 0 & 0 & 0 & 0 \\  
  2i \sin(\mathbf{k} \cdot \mathbf{c}_{10}) & 0 &  e^{- i \mathbf{k} \cdot \mathbf{C}_{12}}  &  -e^{- i \mathbf{k} \cdot \mathbf{C}_{13}}  
  & 0 & 0 &  e^{- i \mathbf{k} \cdot \mathbf{C}_{16}}  &  -e^{- i \mathbf{k} \cdot \mathbf{C}_{17}}   \\
  2i \sin(\mathbf{k} \cdot \mathbf{c}_{20}) &  -e^{- i \mathbf{k} \cdot \mathbf{C}_{21}}  & 0 & e^{- i \mathbf{k} \cdot \mathbf{C}_{23}}
   & 0 & -e^{- i \mathbf{k} \cdot \mathbf{C}_{25}} & 0 & e^{- i \mathbf{k} \cdot \mathbf{C}_{27}} \\
  2i \sin(\mathbf{k} \cdot \mathbf{c}_{30}) & e^{- i \mathbf{k} \cdot \mathbf{C}_{31}} & -e^{- i \mathbf{k} \cdot \mathbf{C}_{32}} & 0 
  & 0  & e^{- i \mathbf{k} \cdot \mathbf{C}_{35}} & -e^{- i \mathbf{k} \cdot \mathbf{C}_{36}} & 0 \\
    0 & 0 & 0 & 0 & 0 &  2i \sin(\mathbf{k} \cdot \mathbf{c}_{45})  &  2i \sin(\mathbf{k} \cdot \mathbf{c}_{46})  &  2i \sin(\mathbf{k} \cdot \mathbf{c}_{47})  \\  
  0 & 0 & -e^{-i \mathbf{k} \cdot \mathbf{C}_{52}} &e^{-i \mathbf{k} \cdot \mathbf{C}_{53}} & 2i \sin(\mathbf{k} \cdot \mathbf{c}_{54}) & 0 & 
  -e^{-i \mathbf{k} \cdot \mathbf{C}_{56}} & e^{-i \mathbf{k} \cdot \mathbf{C}_{57}}  \\
  0 & e^{-i \mathbf{k} \cdot \mathbf{C}_{61}} & 0 & -e^{-i \mathbf{k} \cdot \mathbf{C}_{63}} & 2i \sin(\mathbf{k} \cdot \mathbf{c}_{64}) 
  & e^{-i \mathbf{k} \cdot \mathbf{C}_{65}} & 0 & -e^{-i \mathbf{k} \cdot \mathbf{C}_{67}} \\
  0 & -e^{-i \mathbf{k} \cdot \mathbf{C}_{71}} & e^{-i \mathbf{k} \cdot \mathbf{C}_{72}} & 0 & 2i \sin(\mathbf{k} \cdot \mathbf{c}_{74}) & 
  -e^{-i \mathbf{k} \cdot \mathbf{C}_{75}} & e^{-i \mathbf{k} \cdot \mathbf{C}_{76}} & 0 
  \end{pmatrix} \nonumber \\
  \nonumber 
  \end{eqnarray}
\end{widetext}
Where the vectors $\mathbf{c}_n$, $\mathbf{C}_{nm}$ are vectors joining the central points of plaquettes
to the bond midpoints around the outside.
These vectors are defined in Appendix \ref{appendix:lattice} along with the numbering convention for sites and plaquettes.

Defining
\begin{eqnarray}
Y_{mp}(\mathbf{k})=\sqrt{\alpha_m} Z_{mp} (\mathbf{k}) \sqrt{\beta_p}
\end{eqnarray}
we may write
\begin{eqnarray}
&&\sqrt{\beta_p} \left( \nabla_{\scriptsize\hexagon} \times A \right)_{(\mathbf{r}, p)}  
= \sqrt{\frac{4}{N}} \sum_{\mathbf{k}} \sum_{\lambda=1}^{4} 
    \sqrt{\frac{\mathcal{U}}{\omega_{\lambda}(\mathbf{k})}}  \nonumber \\
&& \qquad
 \times \bigg\{ \exp[-i \mathbf{k} \cdot (\mathbf{r}_p)] a^{\phantom\dagger}_{\lambda}(\mathbf{k}) 
  \times \sum_{m}  Y_{pm}(\mathbf{k})  \eta_{m \lambda}(\mathbf{k})  \nonumber \\
 && \qquad
+ \exp\left[ i \mathbf{k} \cdot (\mathbf{r}_p) \right] a^{\dagger}_{\lambda}(\mathbf{k}) 
 \times \sum_{m}  
       \eta^{\ast}_{\lambda m} Y^{\dagger}_{mp}(\mathbf{k})  \bigg\}.
\label{eq:curl1hB}
\end{eqnarray}
The matrix $\uuline{Y} (\mathbf{k})$ is not guaranteed to be Hermitian, so we cannot
necessarily form an orthonormal set of vectors $\uline{\eta}_{\lambda}(\mathbf{k})$
from its eigenvectors.
However, the matrix $\uuline{Y}^{\dagger} (\mathbf{k}) \cdot \uuline{Y} (\mathbf{k})$ is Hermitian
and has positive semi-definite eigenvalues, so we may write
\begin{eqnarray}
\uuline{Y}^{\dagger} (\mathbf{k}) \cdot \uuline{Y} (\mathbf{k}) \uline{\eta}_{\lambda}(\mathbf{k})
=\zeta_{\lambda}(\mathbf{k})^2 \uline{\eta}_{\lambda}(\mathbf{k}).
\end{eqnarray}
Squaring and summing over $\mathbf{r}$ and $p$ we have
\begin{eqnarray}
&& \frac{\mathcal{K}}{2} \sum_{\mathbf{r}} \sum_{p} \beta_{p}
   \left[ \left( \nabla_{\scriptsize\hexagon} \times A \right)_{\mathbf{r}, p} \right]^2 =
 \frac{\mathcal{U} \mathcal{K}}{4} \sum_{\mathbf{k}} \sum_{\lambda} 
\nonumber \\
&& \qquad \qquad
 \frac{\zeta_{\lambda}(\mathbf{k})^2}{\omega_{\lambda}(\mathbf{k})}
  \bigg\{ 
a_{\lambda}(\mathbf{k}) a_{\lambda}(-\mathbf{k}) 
   + a_{\lambda}^{\dagger}(\mathbf{k})a_{\lambda}^{\dagger}(-\mathbf{k}) 
\nonumber \\
&& \qquad \qquad
 + a_{\lambda}^{\dagger}(\mathbf{k}) a_{\lambda}(\mathbf{k}) 
   + a_{\lambda}(\mathbf{k})a_{\lambda}^{\dagger}(\mathbf{k})
    \bigg\}.
\end{eqnarray}
Similarly
\begin{eqnarray}
&&\frac{\mathcal{U}}{2}\sum_{(\mathbf{e}, m)}  
 \alpha_m  \left(
E_{\mathbf{r} m}
   \right)^2 
   = \frac{1}{4} \sum_{\mathbf{k}}  \sum_{\lambda=1}^{4} 
    \omega_{\lambda}(\mathbf{k})
 \nonumber \\
&& \qquad \qquad
 \bigg\{-a_{\lambda}(\mathbf{k}) a_{\lambda}(-\mathbf{k}) 
   - a_{\lambda}^{\dagger}(\mathbf{k})a_{\lambda}^{\dagger}(-\mathbf{k}) 
\nonumber \\
&& \qquad \qquad
 +a_{\lambda}^{\dagger}(\mathbf{k}) a_{\lambda}(\mathbf{k}) 
   + a_{\lambda}(\mathbf{k})a_{\lambda}^{\dagger}(\mathbf{k}) \bigg\}.
\end{eqnarray}
Inserting this into the Hamiltonian [Eq. (\ref{eq:HU1-app})]
results in 
\begin{eqnarray}
\mathcal{H}_{\sf Ih}=\sum_{\mathbf{k}} \sum_{\lambda=1}^{8} \omega_{\lambda}(\mathbf{k})
\left( a_{\lambda}^{\dagger} (\mathbf{k}) a_{\lambda} (\mathbf{k}) +\frac{1}{2}\right)
\end{eqnarray}
where the dispersion $\omega_{\lambda}(\mathbf{k})$ is fixed by requiring the off-diagonal
terms to vanish
\begin{eqnarray}
\omega_{\lambda}(\mathbf{k})=\sqrt{\mathcal{U} \mathcal{K}} |\zeta_{\lambda}(\mathbf{k}) |.
\end{eqnarray}
The functions $\zeta_{\lambda}(\mathbf{k})$ are found by numerical diagonalization  of
$\uuline{Y}^{\dagger} (\mathbf{k}) \cdot \uuline{Y} (\mathbf{k})$.
Four of the modes $\lambda$ are unphysical, zero-energy modes (cf. the two unphysical
zero energy modes which occur in the pyrochlore case \cite{benton12}).
The remaining four modes are now non-degenerate, which is again in contrast to the 
cubic symmetry case.
Two of these modes are gapless and linearly-dispersing, and therefore recognisable as
photons, while the other two modes are gapped and are associated with
quantised fluctuations of the classical field ${\bf P}_-$ (see Section \ref{section:classical}).
The identification of the gapless modes with fluctuations of ${\bf P}_+$ and the gapped
modes with fluctuations of ${\bf P}_-$ is illustrated in Fig. \ref{fig:p+p-omega}.

The time evolution of the electromagnetic fields is given by the time
evolution of the photon operators $a_{\lambda}(\mathbf{q}), a_{\lambda}^{\dagger}(\mathbf{q})$,
which, since the photons are eigenstates of $\mathcal{H}_{\sf U(1)}$, is simply
\begin{eqnarray}
a^{\dagger}_{\lambda}(\mathbf{q})(t)=e^{i \omega_{\lambda} (\mathbf{q}) t} a^{\dagger}_{\mathbf{q}}(0) \quad
a^{\phantom\dagger}_{\lambda}(\mathbf{q})(t)=e^{-i  \omega_{\lambda} (\mathbf{q})  t} a^{\phantom \dagger}_{\mathbf{q}}(0). \nonumber \\
\end{eqnarray}
Therefore
\begin{eqnarray}
 &&E_m(\mathbf{q}, t)=
\frac{i}{\sqrt{2}}
\sum_{\lambda}
\sqrt{\frac{\omega_{\lambda}(\mathbf{q})}{\alpha_m\mathcal{U}}}
\left(
\eta_{n \lambda}(\mathbf{q}) a^{\phantom\dagger}_{\lambda}(\mathbf{q}) e^{-i \omega_{\lambda} (\mathbf{q}) t} \right.\nonumber \\
&& \qquad \qquad\left.+ \eta_{\lambda n}^{\ast} (\mathbf{q})e^{i \omega_{\lambda} (\mathbf{q}) t}a^{\dagger}_{\lambda}(\mathbf{q}) 
\right) 
\end{eqnarray}
and the dynamical
structure factor for the electric fields is 
\begin{eqnarray}
S^{mn}_E(\mathbf{q}, \omega) &\equiv& \int dt e^{-i \omega t} \langle E_m(-\mathbf{q}, 0) E_n(-\mathbf{q}, t) \rangle \nonumber \\
&=&
\frac{1}{2} \sum_{\lambda}
\left[\delta(\omega-\omega_{\lambda}(\mathbf{q})) 
(1+n_B(\omega))
+ \right. \nonumber \\
&&  \left.
\delta(\omega+\omega_{\lambda}(\mathbf{q})) 
n_B(-\omega) 
\right]  
\frac{\omega_{\lambda}(\mathbf(q))}{\alpha_m \alpha_n \mathcal{U}} 
\eta_{m \lambda} \eta_{\lambda n}^{\ast}  \; . 
\nonumber \\
\label{eq:Ecorr-3}
\end{eqnarray}


\begin{figure*}
\centering
\subfigure[\ \ Monte Carlo- hk0 plane]{%
\includegraphics[width=.22\textwidth]{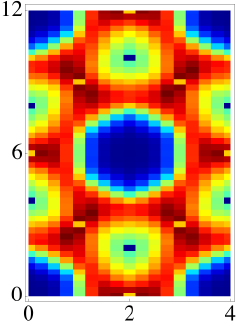}}%
\quad
\subfigure[\ \ Monte Carlo- 0kl plane]{%
\includegraphics[width=.43\textwidth]{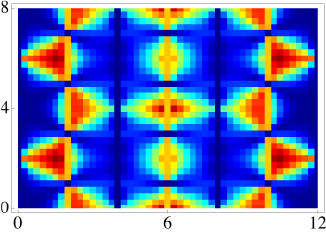}}%
\quad
\subfigure[\ \ Monte Carlo- h0l plane]{%
\includegraphics[width=.29\textwidth]{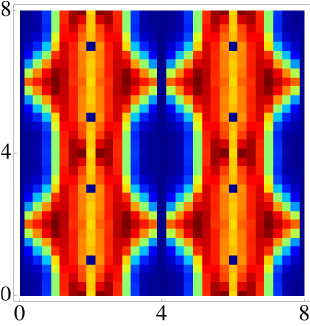}}  \\
\subfigure[\ \ Theory- hk0 plane]{%
\includegraphics[width=.22\textwidth]{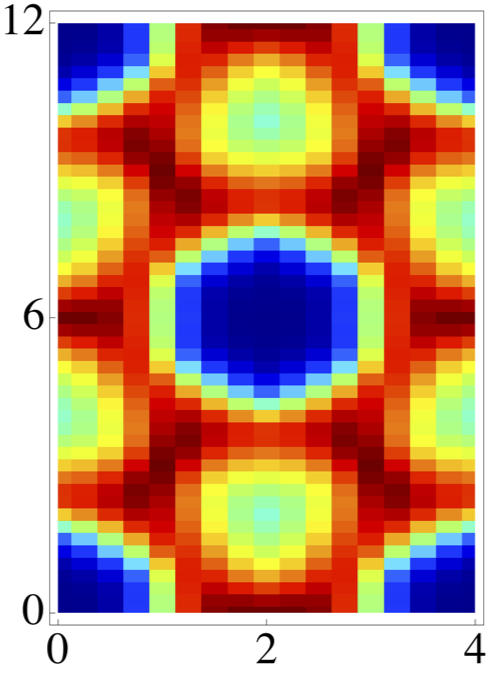}}%
\quad
\subfigure[\ \ Theory- 0kl plane]{%
\includegraphics[width=.43\textwidth]{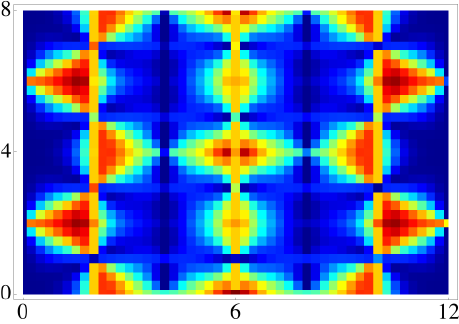}}%
\quad
\subfigure[\ \ Theory- h0l plane]{%
\includegraphics[width=.29\textwidth]{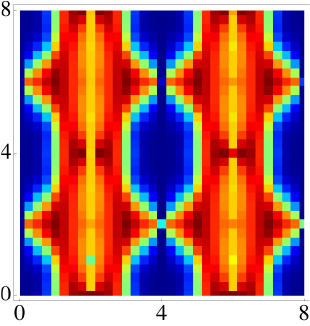}}%
\caption{
(Color online).
Ising structure factor $S_{\sf Ising}(\mathbf{q})$~[Eq. (\ref{eq:S.of.q.Ising})]
for  a classical model of ice Ih, calculated from classical Monte Carlo 
simulation [(a)-(c)] and from lattice theory [(d)-(f)].
The theory calculation is performed using
the method outlined in Appendix \ref{appendix:projection}, which is based
on the method described for spin ice in Ref. \cite{henley05}.
This method consists in writing the ice rule constraints as orthogonality
conditions in Fourier space.
In the long wavelength limit these conditions reduce to those obtained
from the continuum field--theory presented in Section \ref{subsec:classical.fields}
[Eqs. (\ref{eq:p+constraint})-(\ref{eq:p-constraint})].
Near the zone centres the correlations are well decribed by
a combination of pinch point singularities, reflecting the algebraic correlations of ${\bf P}_+(\mathbf{q})$
and smooth features reflecting the short ranged correlations of ${\bf P}_-(\mathbf{q})$.
} 
\label{fig:Sq.Ising.classical-MC}
\end{figure*}


\begin{figure*}
\centering
\subfigure[\ \ VMC-1024 bonds]{%
  \includegraphics[width=.3\textwidth]{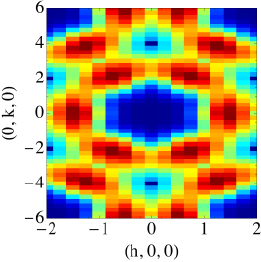}}%
    \qquad
\subfigure[\ \ Lattice Gauge Theory (1024 bonds)]{%
  \includegraphics[width=.3\textwidth]{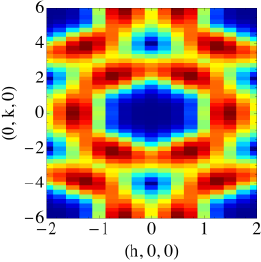}}%
    \qquad
\subfigure[\ \ Lattice Gauge Theory (thermodynamic limit)]{%
  \includegraphics[width=.3\textwidth]{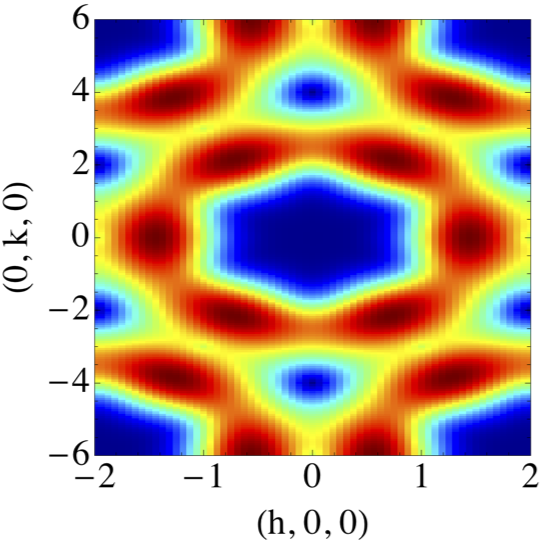}}%
\caption{
(Color online).
Comparison between the predictions of the lattice gauge theory 
$\mathcal{H}_{\sf U(1)}$~[Eq. (\ref{eq:HU1})], and variational Monte Carlo 
simulation of $\mathcal{H}_{\sf tunnelling}^{\sf hexagonal}$~[Eq. (\ref{eq:htunnelling2})] 
for the ``Ising'' structure--factor $S_{\sf Ising}(\mathbf{q}, t=0)$~[Eq.~(\ref{eq:equaltimeSzz})],
for the parameter sets given in Eq.~(\ref{eq:plain-vanilla-parameters}) and
Eq.~(\ref{eq:microscopic-plain-vanilla-parameters}).
There is excellent agreement between the two methods, validating our 
description of the proton--liquid in water ice Ih in terms of a quantum 
${\sf U(1)}$ lattice gauge theory.
} 
\label{fig:isingSq-quantum}
\end{figure*}


Since the electric field is in one--to--one correspondence with the proton
configuration, the dynamical structure factor $S^{mn}_E(\mathbf{q}, \omega)$  
also determines the scattering of neutrons or X--rays from protons in water ice.
The relevant structure factor for the coherent scattering of neutrons is given 
by~\cite{squires12}
\begin{eqnarray}
&&S_{\sf coh}(\mathbf{q}, \omega)= \frac{1}{2\pi}
\sum_j \sum_{j'}
\int_{-\infty}^{\infty} dt e^{- i \omega t} \times \nonumber \\
&& \qquad \qquad
\langle \exp(-i \mathbf{q} \cdot \mathbf{R}_j(0))
\exp(i \mathbf{q} \cdot \mathbf{R}_{j'}(t)) \; .
\rangle
\label{eq:Scoh-definition}
\end{eqnarray}
Within our treatment, the position of the proton
on bond $j$ of the lattice, $\mathbf{R}_j (t)$, is given by
\begin{eqnarray}
\mathbf{R}_j (t) = \mathbf{r}^0_j+\mathbf{a}_j \sigma_j(t)
\label{eq:protonpos}
\end{eqnarray}
where ${\bf r}^0_j$ is the bond midpoint, $\mathbf{a}_j $ describes 
the displacement of the two proton sites on the bond from the bond midpoint 
and $\sigma_j(t) = \pm 1$ is the Ising variable describing the bond polarisations.


Using Eqs.  (\ref{eq:Scoh-definition})-(\ref{eq:protonpos}) and the relationship between
the Ising variables $\sigma_j(t)$ and the electric fields $E_j(t)$ 
\begin{eqnarray}
\sigma_j(t)=2 E_j(t)
\end{eqnarray}
we find that the diffuse contribution to $S_{\sf coh}(\mathbf{q}, \omega)$ is given by
\begin{eqnarray}
&& S_{\sf coh}^{\sf H^+}(\mathbf{q}, \omega)
=4 \sum_{m,n}
\sin(\mathbf{q}\cdot \mathbf{a}_m)
\sin(\mathbf{q}\cdot \mathbf{a}_n)
S^{mn}_E(\mathbf{q}, \omega) \nonumber \\
\label{eq:SH-definition}
\end{eqnarray}
where $m, n$ index the eight sublattices of
bonds and the displacement
of a proton from the bond midpoint of bond $m$
is $\pm {\bf a}_{m}$.


We have also calculated the structure factor of the fields ${\bf P}_{+}(\mathbf{r})$
and ${\bf P}_{-}(\mathbf{r})$ within the lattice gauge theory 
$\mathcal{H}_{\sf U(1)}$~[Eq.~(\ref{eq:HU1-app})].  
To do this, by analogy with Eq. (\ref{eq:FT.P}), we 
introduce the Fourier transform of the fields ${\bf P}_{\pm}(\mathbf{q})$
for wavevector $\mathbf{q}=\mathbf{Q}+\tilde{\mathbf{q}}$, in a Brillouin 
zone centred on reciprocal lattice vector ${\bf Q}$
\begin{eqnarray}
&&{\bf P}_{\pm}(\tilde{\mathbf{q}}, \omega)=\sqrt{\frac{4}{N}}
\int_{-\infty}^{\infty} \exp(-i \omega t) \times \nonumber \\
&&\qquad \qquad \left[ \exp(i \mathbf{Q \cdot \mathbf{r}_A})
\sum_{\mathbf{r} \in A} 
\exp(-i \mathbf{q} \cdot \mathbf{r}) {\bf P}_{\pm}(\mathbf{r}_A, t) +
\right. \nonumber \\
&&\qquad \qquad
\left.
 \exp(i \mathbf{Q \cdot \mathbf{r}_C})
\sum_{\mathbf{r} \in C}
\exp(-i \mathbf{q} \cdot \mathbf{r}) {\bf P}_{\pm}(\mathbf{r}_C, t)
 \right]. \nonumber \\
\label{eq:FT.P.quantum}
\end{eqnarray}
where we have chosen a definition such that ${\bf P}_{\pm}(\tilde{\mathbf{q}}, \omega)$
is the same in all Brillouin zones. 
The vectors ${\bf r}_A, {\bf r}_C$ are the positions of the 
$A$ and $C$ oxygen vertices within a primitive unit cell.
Restricting the sum to the $A$ and $C$ sublattices of oxygen ions
means that each bond only contributes to the sum once.
In Fig.~\ref{fig:p+p-omega} we plot \mbox{$\langle{\bf P}_{+}(-\mathbf{q}, -\omega)
\cdot {\bf P}_{+}(\mathbf{q}, \omega) \rangle$}
and  \mbox{$\langle{\bf P}_{-}(-\mathbf{q}, -\omega)
\cdot {\bf P}_{-}(\mathbf{q}, \omega) \rangle$}.


The results of this analysis are shown in Fig. \ref{fig:p+p-omega}.
At long wavelengths, the correlations of ${\bf P}_+$ are directly associated with the gapless 
emergent photons, while the fluctuations of ${\bf P}_-$ are associated with the gapped
optical modes.
This confirms the conclusions of the symmetry--based analysis given in 
Section~\ref{subsection:phenomenology}.  

\section{Comparison with numerics}
\label{appendix:numerics}

\subsection{Classical Monte Carlo simulation}

In this Appendix we compare our calculations of the correlation functions
in the classical ice problem with the results of Monte Carlo simulations for a cluster of 1024
oxygen--oxygen bonds.


To obtain numerical results for the correlations in the absence of quantum tunnelling 
it is necessary to take an equally weighted average over ice rule configurations.
This is accomplished numerically by starting from a configuration with
zero total electric polarisation and then acting randomly with the hexagonal ring
exchange operators [cf. Eq. (\ref{eq:htunnelling2})]
to generate new configurations.
Since these operations preserve total electric polarisation
the average obtained in this way only includes states of vanishing
total polarisation.
However, this set of states is representative of the manifold as 
a whole, as in the case of spin ices.
%


In Fig. \ref{fig:Sq.Ising.classical-MC} we compare results for the Ising structure factor
$S_{\sf Ising}(\mathbf{q})$  [Eq. (\ref{eq:S.of.q.Ising})] between these Monte Carlo
calculations [(a)-(c)] and the projection method outlined in Appendix~\ref{appendix:projection}
[(d)-(f)].
Noticeable differences are visible for $\mathbf{q}$ located exactly at Brillouin zone centers.
This is caused by the restriction to states of vanishing
total polarisation in the simulations. 
However, for any $\mathbf{q}$ not exactly at a Brillouin zone center 
there is very good agreement between the theory calculation and Monte Carlo,
strongly validating our understanding
of the classical ice problem. 

\subsection{Quantum Monte Carlo simulation}
\label{subsection:QMC}

In order to validate our description of quantum ice Ih,
we have compared the predictions of the lattice gauge theory 
$\mathcal{H}_{\sf U(1)}$~[Eq. (\ref{eq:HU1})] for equal--time correlation functions, 
with the results of variational quantum Monte Carlo (VMC) simulations of the 
microscopic model $\mathcal{H}_{\sf tunnelling}^{\sf hexagonal}$~[Eq.~(\ref{eq:htunnelling2})].


VMC simulations were carried out for an orthorhombic
 cluster of 1024 oxygen--oxygen bonds
with a one-parameter variational 
wavefunction of the form
\begin{eqnarray}
| \psi^{\sf var}_{\alpha} \rangle=\exp[ \alpha \mathcal{N}_f] 
| \psi_0 \rangle
\label{eq:VMCwavefunction}
\end{eqnarray}
where the operator $\mathcal{N}_f$ measures the number
``flippable'' plaquettes (of both type $I$ and $II$), $\alpha$ is a
variational parameters, and $| \psi_0 \rangle$ is an equal weight superposition of 
all ice configurations within a given flux--sector of the Hilbert space.
This variational wave funtion correctly describes the physics of quantum liquids 
based on a lattice ${\sf U}(1)$ gauge theory~\cite{sikora09,sikora11,shannon12,benton12}, 
but is heavily biased toward liquid, rather than ordered ground states~\cite{sikora11}.
Since our intention here is to investigate the properties of a proton--liquid in water ice, 
rather than to investigate possible ordered ground states, the wave function 
Eq.~(\ref{eq:VMCwavefunction}) is sufficient for our purposes.


In Fig. \ref{fig:isingSq-quantum} we show the equal-time ``Ising'' structure factor
\begin{eqnarray}
S_{zz}(\mathbf{q}, t=0)=\langle {\sf S}^z(-\mathbf{q}, t=0) {\sf S}^z(\mathbf{q}, t=0)  \rangle
\label{eq:equaltimeSzz}
\end{eqnarray}
in the $(h, k, 0)$ plane of reciprocal space.
Results were calculated 
from the lattice gauge theory \mbox{$\mathcal{H}_{\sf U(1)}$~[Eq. (\ref{eq:HU1})]}, 
for parameters 
\begin{eqnarray}
\mathcal{U} = \mathcal{U}' = \mathcal{K} = \mathcal{K}' = 1
\label{eq:plain-vanilla-parameters}
\end{eqnarray}
and from the microscopic model 
$\mathcal{H}_{\sf tunnelling}^{\sf hexagonal}$~[Eq. (\ref{eq:htunnelling2})], for parameters 
\begin{eqnarray}
g_1 = g_2 = 1
\label{eq:microscopic-plain-vanilla-parameters}
\end{eqnarray}
using VMC, as described in \cite{sikora11,benton12}.
A field-renormalization of $S^z$, allowed in principle within the lattice gauge 
theory~\cite{banerjee08,benton12}, has also been set equal to one.


The agreement between the results of the two methods is excellent,
confirming that the lattice gauge theory correctly describes the liquid phase of the
microscopic model.

\section{Calculation of the incoherent scattering cross section}
\label{appendix:incoherent}

The structure factor for inelastic, incoherent, 
neutron scattering from a set of nuclei at 
located at positions ${\bf R}_j$ is \cite{squires12}
\begin{eqnarray}
&&S_{\sf inc}(\mathbf{q}, \omega)= \frac{1}{2\pi}
\sum_j
\int_{-\infty}^{\infty} dt e^{- i \omega t}
 \nonumber \\
&&
\qquad \qquad 
\langle \exp(-i \mathbf{q} \cdot \mathbf{R}_j(0))
\exp(i \mathbf{q} \cdot \mathbf{R}_j(t))
\rangle
\label{eq:Sinc-definition}
\end{eqnarray}


We approximate the position of the proton
on bond $j$ of the lattice, $\mathbf{R}_j (t)$,
to be given by
\begin{eqnarray}
\mathbf{R}_j (t) = \mathbf{r}^0_j+\mathbf{a}_j \sigma_j(t)
\end{eqnarray}
where
${\bf r}^0_j$ is the bond midpoint,
 $\mathbf{a}_j $ describes the displacement
of the two proton sites on the bond from the bond midpoint
and $\sigma_j(t)=\pm 1$ is the Ising variable
describing the bond polarisations.


Since $\sigma_j(t)=\pm 1$ 
and $\bf{r}_j^0$ and ${\bf a}_j$ are constants
we can write
\begin{eqnarray}
&&\exp(i \mathbf{q} \cdot \mathbf{R}_j(t)) 
\nonumber \\
&& \quad = \exp(i \mathbf{q} \cdot \mathbf{r}_j^0)
\left[ \cos(\mathbf{q} \cdot \mathbf{a}_j)
+
i \sigma_j (t)
\sin(\mathbf{q} \cdot \mathbf{a}_j)
\right] \nonumber \\
\label{eq:exponentsimplified}
\end{eqnarray}


Since 
\begin{eqnarray}
\langle \sigma_j (t) \rangle=0
\end{eqnarray}
on inserting Eq. (\ref{eq:exponentsimplified})
into Eq. (\ref{eq:Sinc-definition}) we obtain
two terms, corresponding to the elastic
and inelastic contributions
to the incoherent scattering
\begin{eqnarray}
S_{\sf inc}(\mathbf{q}, \omega)
=S_{\sf inc, el}(\mathbf{q}, \omega)
+S_{\sf inc, inel}(\mathbf{q}, \omega)
\end{eqnarray}


The elastic contribution is simply
\begin{eqnarray}
S_{\sf inc, el}(\mathbf{q}, \omega)=
\delta(\omega)
\sum_j \cos^2(\mathbf{q} \cdot \mathbf{a}_j)
\label{eq:sqel}
\end{eqnarray}
while the inelastic contribution is
\begin{eqnarray}
S_{\sf inc, inel}(\mathbf{q}, \omega)=
\sum_j \sin^2(\mathbf{q} \cdot \mathbf{a}_j)
\langle \sigma_j(-\omega) \sigma_j(\omega) \rangle
\label{eq:sqinel}
\end{eqnarray}


For the purposes of comparison with
experiments we need to integrate the
momentum transfer dependence
over angle,
which gives
\begin{eqnarray}
&&S_{\sf inc, el}^{\sf pow}(Q, \omega)=
\delta(\omega)
\sum_j \frac{1}{2} \left( 1+\frac{\sin(2 Q |a_j|)}{2 Q |a_j|} \right)
\label{eq:sqelpow} \\
&&S_{\sf inc, inel}^{\sf pow}(Q, \omega)=
\sum_j  \frac{1}{2}  \left( 1-\frac{\sin(2 Q |a_j|)}{2 Q |a_j|} \right)
\langle \sigma_j(-\omega) \sigma_j(\omega) \rangle \nonumber \\
\label{eq:sqinelpow} 
\end{eqnarray}
where $Q=|\mathbf{q}|$.


The local correlation function $\langle \sigma_j(-\omega) \sigma_j(\omega) \rangle$
can be rewritten as a sum over Fourier space, so we have
\begin{eqnarray}
&&S_{\sf inc, inel}(\mathbf{q}, \omega)=
\sum_{i=1}^8 \sin^2(\mathbf{q} \cdot \mathbf{a}_i)
\sum_{\mathbf{q}'}
\langle \sigma_i(-\mathbf{q}', -\omega) \sigma_i(\mathbf{q}', \omega) \rangle
\nonumber \\
\label{eq:incoherent-strcuture-factor}
&&\\
&&S_{\sf inc, inel}^{\sf pow}(Q, \omega)= \nonumber \\
&&\sum_{i=1}^8  \frac{1}{2}  \left( 1-\frac{\sin(2 Q |a_j|)}{2 Q |a_j|} \right) 
\sum_{\mathbf{q}'}
\langle \sigma_i(-\mathbf{q}', -\omega) \sigma_i(\mathbf{q}', \omega) \rangle
\nonumber \\
\label{eq:incoherent-strcuture-factor-pow}
\end{eqnarray}
where the sum over $i$ now runs over the eight
sublattices of bonds. 


Within the lattice gauge theory description the 
correlations of the Ising variables $\sigma$
is directly related to the correlations of the
electric field $E$
\begin{eqnarray}
\langle \sigma_i(-\mathbf{q}', -\omega) \sigma_i(\mathbf{q}', \omega) \rangle
=
4 \langle E_i(-\mathbf{q}', -\omega) E_i(\mathbf{q}', \omega) \rangle \nonumber \\
\end{eqnarray}
so calculating the correlations of $E$
in the gauge theory enables us to calculate the
incoherent scattering
from protons, as shown in Fig. \ref{fig:powderfig}.
The effects of finite temperature, which lead to the thermal excitation 
of photons, can also be included, as described in [\onlinecite{benton12}].


Calculations were carried out for the symmetric parameter set given in 
\begin{eqnarray}
\mathcal{U}=\mathcal{U}' \; ,  \; \mathcal{K}=\mathcal{K}'
\nonumber
\end{eqnarray}
[cf. Eq.~(\ref{eq:plain-vanilla-parameter-set})], 
with energy scale
\begin{eqnarray}
\sqrt{\mathcal{U}\mathcal{K}} = 0.018 \  \text{meV} \; ,
\nonumber
\end{eqnarray}
[cf. Eq.~(\ref{eq:symmetric-parameter-set})].   
The one remaining free parameter of the theory, $\sqrt{\frac{U}{K}}$,
is used to fix the average 
normalisation of the electric fields of the gauge theory [cf. Eq. (\ref{eq:Econstraint})].
At $T=5$K this gives
\begin{eqnarray}
\sqrt{\frac{U}{K}}=47.8.
\end{eqnarray}


Experimental measurements are carried out with finite energy resolution.
To make a comparison with experiment, it is therefore necessary to convolute both the 
elastic response (a delta--function in energy) and the prediction for incoherent inelastic 
scattering with a representation of the experimental line--shape.
We adopt the simplest representative line shape, a Gaussian 
\begin{eqnarray}
F_{\sf exp}(\omega)=
\frac{1}{\sqrt{2 \pi} \delta} \exp\left( -\frac{1}{2} \frac{\omega^2}{\delta^2}\right) \; ,
\end{eqnarray}
with energy width
\begin{eqnarray}
\delta &=& 0.03 \ \ \text{meV} \; ,
\nonumber 
\end{eqnarray}
chosen as representative of the elastic line shown in Ref.~\onlinecite{bove09}.




\begin{thebibliography}{10}
\bibitem{footnote1}
Neither Bjerrum nor ionic defects play a direct role in the theories 
constructed in this article, which assume that the ice rules are
perfectly obeyed.   
%
However the quantum tunnelling processes shown in Fig.~\ref{fig:loops} 
could be achieved through the virtual excitation of a pair of defects which 
annihilated one another, after traversing a closed path on the lattice.
\end{thebibliography}
\end{document}